\documentclass[twocolumn,english]{revtex4-2}
\usepackage{lmodern}
\usepackage{lmodern}

\usepackage[T1]{fontenc}
\usepackage[latin9]{inputenc}
\usepackage{babel}
\usepackage{amsmath}
\usepackage{amssymb}
\usepackage{graphicx}
\usepackage[unicode=true,pdfusetitle,
 bookmarks=true,bookmarksnumbered=false,bookmarksopen=false,
 breaklinks=false,pdfborder={0 0 1},backref=false,colorlinks=false]
 {hyperref}

\makeatletter

\providecommand{\tabularnewline}{\\}
\newcommand{\lyxdot}{.}

\usepackage{babel}

\makeatother

\begin{document}
\title{Solutions of the imploding shock problem in a medium with varying
density}
\author{Itamar Giron}
\author{Shmuel Balberg}
\author{Menahem Krief}
\email{menahem.krief@mail.huji.ac.il}

\address{Racah Institute of Physics, The Hebrew University, 9190401 Jerusalem,
Israel}
\begin{abstract}
We consider the solutions of the Guderley problem, consisting of an
imploding strong shock wave in an ideal gas with a power law initial
density profile. The self-similar solutions, and specifically the
similarity exponent which determines the behavior of the accelerating
shock, are studied in detail, for cylindrical and spherical symmetries
and for a wide range of the adiabatic index and the spatial density
exponent. We then demonstrate how the analytic solutions can be reproduced
in Lagrangian hydrodynamic codes, thus demonstrating their usefulness
as a code validation and verification test problem. 
\end{abstract}
\maketitle

\section{Introduction \label{sec:introduction}}

The imploding shock problem, studied first by Guderly \cite{guderley1942starke},
involves a strong converging shock in a uniform ideal gas. It has
long been considered one of the fundamental one-dimensional problems
in compressible hydrodynamic flow \cite{zel2002physics,Landau1987Fluid}.
In particular, it is well known that the flow structure settles on
a self-similar solution \cite{lazarus1981self,lazarus1977similarity,ramsey2012guderley,zel2002physics,sakurai1960problem,ponchaut2006imploding,ramu1993converging,welsh1967imploding,meyer1982selfsimilar}.

From a practical perspective, an imploding shock is a possible driving
mechanism for initiating detonation in combustional material, since
the shock velocity increases as it propagates inward. This possibility
has several interesting prospects, such as in astrophysics \cite{ArnettLivne1994a,ArnettLivne1994b,ShenBildsten2014}
and in inertial confinement fusion \cite{Valletetal2013}. In fact,
as demonstrated by Kushnir et al. \cite{kushnir2012imploding}, initiation
strongly depends on the ratio of shock radius to shock velocity during
implosion, creating a threshold that can be assessed from the pure
hydrodynamic solution.

A thorough investigation and solution of the Guderly problem was carried
out by Lazarus \cite{lazarus1977similarity,lazarus1981self}. The
general approach to the solution of the shock velocity, as well as
the downstream density profile, is to transform the system of partial
differential equations governing the motion (the Euler equations)
into a system of ordinary differential equations (the self-similar
equations), by introducing a self-similar variable. The similarity
exponent describing the flow is then found by requiring the solution
to pass through a sonic point without generating a singularity. This
is an example of a "second-type" self-similar solution, which is
also reflected by the fact that the exponents of the flow cannot be
deduced by dimensional analysis (as opposed, for example, to the well
known Sedov-Taylor explosion problem \cite{Sedov1946,Taylor1950,coggeshall1986lie,coggeshall1991analytic,pakula1985self,reinicke1991point,krief2021analytic,yalinewich2017analytic,faran2021non}).

The extensive survey \cite{lazarus1977similarity,lazarus1981self}
on an imploding shock in a uniform medium examined the solutions for
a wide range of values of the adiabatic index of the medium, $\gamma$,
and the dimensionality (cylindrical or spherical). While other works
\cite{sakurai1960problem,sharma1995similarity,toque2001self} have
considered some aspects of an imploding shock wave in a power law
density profile of the form $\rho_{0}r^{\mu}$ for $\mu\geq0$, a
comprehensive survey of this generalized Guderley problem has not
been carried out, and this is one of our goals here. We systematically
generalize the algorithms developed by Lazarus for numerical calculations
of the similarity exponent of an imploding shock in a spatial power
law density profile. This algorithm is then used to conduct a semi-analytic
survey of the flow for a wide range of values for $\gamma$ and for
both positive and negative values of $\mu$.

A highly attractive aspect of the Guderley problem is that it offers
a nontrivial test problem for hydrodynamic code verification. This
was demonstrated recently in several studies \cite{ramsey2012guderley,ramsey2012simulation,ramsey2012surrogate,ramsey2017verification,ramsey2018converging,ramsey2019piston,ruby2019boundary,singh2020kinematics},
for a uniform density medium. The converging nature of the flow, coupled
with compression and shock discontinuity, present unique subtleties,
and is therefore very useful for code validation and verification,
given that the results can be compared to solutions obtained by self-similar
methods. Here we expand on this point and present numerical methods
and analysis of the imploding shock in a power-law density profile.
In particular we present comparisons between the semi-analytic solutions
and the numerical simulations for a variety of $\gamma$ and $\mu$.

We note that the full Guderley problem is actually two-fold, and also
includes the reflected shock from the center. This shock can be described
by a second set of self-similar solutions \cite{lazarus1977similarity,lazarus1981self},
which can again be useful in code test problems \cite{ramsey2012guderley},
thus combining converging and diverging flow.

The structure of the article is as follows. In section \ref{sec:The-Guderley-Problem}
we review the Guderley problem, by presenting the equations, notation
and conventions we use. In section \ref{sec:The-Self-Similar-Formulation}
we describe the self-similar representation of the problem, derive
the self-similar equations and analyze the their singular points.
A robust algorithm for the calculation of the similarity exponent
and the self-similar profiles for general values of $\mu$ is developed.
We present the resulting similarity exponent for a wide range of values
of $\gamma$ and $\mu$, and also compare the results to previously
published works when such exist. Turning our focus to numerical simulation
of imploding shocks, in section \ref{sec:Comparison-to-numerical}
a test problem for hydrodynamic codes is developed by properly defining
the initial and boundary condition of a piston with velocity given
from the analytic solution \cite{ramsey2017verification}, and examine
the accessible accuracy of the solution. We conclude in section \ref{sec:summary}.
The main text is followed by several appendices. Details on the iterative
algorithm for calculating the similarity exponent are discussed in
Appendix \ref{app:Numerical-Calculation-of}, which in turn depends
on a critical value of the adiabatic exponents, described in Appendix
\ref{app:gamma_crit}. For completeness, the calculation of approximate
similarity exponents are given in Appendix \ref{app:Approximated-Calculation-of},
and the numerical hydrodynamic Lagrangian scheme is laid out in Appendix
\ref{app:The-hydrodynamic-Scheme}.

\section{The Guderley Problem \label{sec:The-Guderley-Problem}}

We summarize the general setting of the Guderley problem, involving
a strong shock wave propagating from $r=\infty$ through an ideal
gas medium, which is initially cold and at rest. The shock wave converges
on an axis (in cylindrical symmetry) or a point (in spherical symmetry).
The Euler equations, which govern the flow variables of the gas are:

\begin{equation}
\frac{\partial\rho}{\partial t}+\frac{\partial\left(\rho u\right)}{\partial r}+\left(n-1\right)\frac{\rho u}{r}=0,\label{continuity equation}
\end{equation}

\begin{equation}
\frac{\partial u}{\partial t}+u\frac{\partial u}{\partial r}+\frac{c^{2}}{\gamma\rho}\frac{\partial\rho}{\partial r}+2c\frac{\partial c}{\partial r}=0,
\end{equation}
\begin{equation}
\frac{\partial c}{\partial t}+u\frac{\partial c}{\partial r}+\frac{\gamma-1}{2}c\left(\frac{\partial u}{\partial r}+\frac{\left(n-1\right)u}{r}\right)=0,\label{energy equation}
\end{equation}
where $\rho$ denotes the fluid density, $u$ the velocity, $c$ the
sound speed,and $n$ the symmetry constant ($n=2$ for cylindrical
symmetry and $n=3$ for spherical symmetry). The material is assumed
to be an ideal gas with an adiabatic index $\gamma$, so that the
equation of state is:
\begin{equation}
p=\left(\gamma-1\right)\rho e,
\end{equation}
relating the pressure $p$ to the specific internal energy $e$ and
density $\rho$. The sound speed is then:
\begin{equation}
c^{2}=\frac{\gamma p}{\rho}.
\end{equation}

Across the strong shock wave, the Euler equations \eqref{continuity equation}-\eqref{energy equation}
are not valid, and the strong-shock Rankine--Hugoniot relations \cite{zel2002physics}
set boundary conditions for the flow variables in the shocked region:
\begin{equation}
\frac{\rho_{\text{shocked}}}{\rho_{0}r_{\text{shock}}^{\mu}}=\frac{\gamma+1}{\gamma-1},\label{hugonoit den}
\end{equation}
\begin{equation}
u_{\text{shocked}}=\frac{2}{\gamma+1}D,
\end{equation}
\begin{equation}
c_{\text{shocked}}=-\frac{\sqrt{2\gamma\left(\gamma-1\right)}}{\gamma+1}D,\label{hugonoit c_s}
\end{equation}
where $r_{\text{shock}}$ is the shock position, and $D$ is the shock
velocity.

\section{The Self-Similar Formulation\label{sec:The-Self-Similar-Formulation}}

In deriving the self-similar solution we follow the notation and formulation
set forth by Lazarus \cite{lazarus1977similarity,lazarus1981self}.
The shock position is assumed to have a power law temporal dependence:
\begin{equation}
r_{\text{shock}}(t)=A\left(-t\right)^{\frac{1}{\lambda}},\label{eq:rshock}
\end{equation}
where $\lambda$ is the similarity exponent. It is customary to set
$A=1$, so that $r_{\text{shock}}=1$ at $t=-1$, so that convergence,
$r_{\text{shock}}=0$, occurs at $t=0$. The dimensionless independent
variable is set as:

\begin{equation}
x=\frac{t}{r^{\lambda}},
\end{equation}
so that the shock position is given by $x=-1$.

In this work we study the solutions for a medium with an initial spatial
power-law density profile of the form:
\[
\rho(r,t=-\infty)=\rho_{0}r^{\mu}.
\]
A subtlety exists for sufficiently negative values of $\mu$. We consider
here the range $\mu\geq-n$, for which the mass enclosed by the shock
is finite, even though the density diverges for $r\rightarrow0$.
As shown in Ref. \cite{modelevsky2021revisiting}, self-similar solutions
do exist also for $\mu<-n$, and in fact include a transition at some
$\mu_{b}<-n$ to shocks that converge at infinite times as they propagate
through a steep density gradient.

The self-similar nature of the flow variables is postulated in the
form: 
\begin{equation}
u(r,t)=-\frac{r}{\lambda t}V(x),\label{Similarity V}
\end{equation}

\begin{equation}
c(r,t)=-\frac{r}{\lambda t}C(x),\label{Similarity C}
\end{equation}

\begin{equation}
\rho(r,t)=\rho_{0}r^{\mu}R(x),\label{Similarity R}
\end{equation}
where $V(x),C(x),R(x)$ are the similarity functions. Inserting equations
\eqref{Similarity V}-\eqref{Similarity R} into the Euler equations
\eqref{continuity equation}-\eqref{energy equation} and employing
the relations $\partial f/\partial r=-\lambda xr^{-1}\partial f/\partial x$,
$\ \ \ \partial f/\partial t=xt^{-1}\partial f/\partial x$, results
in the a system of nonlinear ordinary differential equations (ODEs)
for $V(x),\ C(x)$ and $R(x)$: 
\begin{equation}
\boldsymbol{A}\frac{d}{dx}\left[\begin{array}{c}
R\\
V\\
C
\end{array}\right]=\left[\begin{array}{c}
V(\mu+n)\\
R\left(C^{2}(2+\mu)+\gamma V(\lambda+V)\right)\\
\frac{C\left(2\left(V+\lambda\right)+n(\gamma-1)\right)}{2\gamma(1+V)}
\end{array}\right],\label{eq:linsys}
\end{equation}
with $\boldsymbol{A}$ being the matrix: 
\[
\boldsymbol{A}=\lambda x\left[\begin{array}{ccc}
\frac{1+V}{R} & 1 & 0\\
C^{2} & \gamma R(V+1) & 2RC\\
0 & \frac{C(\gamma-1)}{2\gamma(V+1)} & \frac{1}{\gamma}
\end{array}\right].
\]
The derivatives can be readily inverted by employing Kramers' law.
The result is commonly written as:

\begin{equation}
\lambda xR'=\frac{\Delta_{1}}{\Delta},\ \ \lambda xV'=\frac{\Delta_{2}}{\Delta},\ \ \lambda xC'=\frac{\Delta_{3}}{\Delta},\label{ODES}
\end{equation}
with the following discriminants,
\begin{equation}
\Delta(V,C)=(V+1)^{2}-C^{2},\label{det}
\end{equation}
\begin{align}
\Delta_{1}(R,V,C,\lambda,\mu) & =R\Bigg[\frac{2(1-\lambda)+\mu(\gamma V+1)}{\gamma(V+1)}C^{2}\nonumber \\
 & +V(V+\lambda)-(n+\mu)V(V+1)\Bigg],
\end{align}

\begin{align}
\Delta_{2}(V,C,\lambda,\mu) & =C^{2}\left(nV+\frac{2(\lambda-1)-\mu}{\gamma}\right)\nonumber \\
 & \ \ \ \ -V(V+1)(V+\lambda),\label{eq:delta2}
\end{align}
\begin{align}
\Delta_{3}(V,C,\lambda,\mu) & =C\Bigg[C^{2}\left(1+\frac{2(\lambda-1)+\mu(\gamma-1)}{2\gamma(1+V)}\right)\nonumber \\
 & \ \ \ \ -(V+1)^{2}-(n-1)(\gamma-1)\frac{V(1+V)}{2}\nonumber \\
 & \ \ \ \ -(\lambda-1)\frac{(3-\gamma)V+2}{2}\Bigg].\label{eq:delta3}
\end{align}
A single differential equation for $C\left(V\right)$ is then obtained
from the set of equations \eqref{ODES}:

\begin{equation}
\frac{dC}{dV}=\frac{\Delta_{3}(V,C,\lambda,\mu)}{\Delta_{2}(V,C,\lambda,\mu)}.\label{eq:dCdV}
\end{equation}
This equation \eqref{eq:dCdV} lends to integration from the known
values at the shock front, for which the self-similar functions \eqref{Similarity V}-\eqref{Similarity R}
are determined through the strong shock jump relations \eqref{hugonoit den}-\eqref{hugonoit c_s}:

\begin{equation}
R_{s}\equiv R\left(-1\right)=\frac{\gamma+1}{\gamma-1},\label{eq:R_shock}
\end{equation}

\begin{equation}
V_{s}\equiv V\left(-1\right)=-\frac{2}{\gamma+1},\label{eq:V_shock}
\end{equation}

\begin{equation}
C_{s}\equiv C\left(-1\right)=\frac{\sqrt{2\gamma(\gamma-1)}}{\gamma+1}.\label{eq:C_shock}
\end{equation}
Since for $r\rightarrow\infty$ (for which $x\rightarrow0$) the pressure
and velocity are finite, it is evident from equations \eqref{Similarity V}-\eqref{Similarity C},
that:

\begin{equation}
V_{\infty}\equiv V\left(0\right)=0,\label{eq:v_infinity}
\end{equation}
\begin{equation}
C_{\infty}\equiv C\left(0\right)=0.\label{eq:c_infinity}
\end{equation}
Hence, the two points $(V_{s},C_{s})$ and $(V_{\infty},C_{\infty})$,
must be connected continuously in the $C-V$ plane by the $C\left(V\right)$
curve, with the full path obtained by integrating eq. \eqref{eq:dCdV}.
Since $\gamma>1$, the point $\left(V_{s},C_{s}\right)$ lies above
the line $C=V+1$, and the solution of eq. \eqref{eq:dCdV} must intersect
with this line as the integration advances to $\left(V_{\infty},C_{\infty}\right)$
which corresponds to the origin, as shown in Fig. \ref{fig:Illustration-of-the-plane}
(and also in Figs. \eqref{fig:CV_example}-\eqref{fig:not_true_lam}
below). Given that, at some point along the profile $C\left(V\right)$,
we must have $C=V+1$, for which $\Delta=0$, and equations \eqref{ODES}-\eqref{eq:dCdV}
are singular. This singularity is removable only if the intersection
point satisfies $\Delta_{1}=\Delta_{2}=\Delta_{3}=0$ as well. This
requirement is a constraint which enables the determination of the
similarity exponent $\lambda$.

\begin{figure}[h]
\begin{centering}
\includegraphics[scale=0.38]{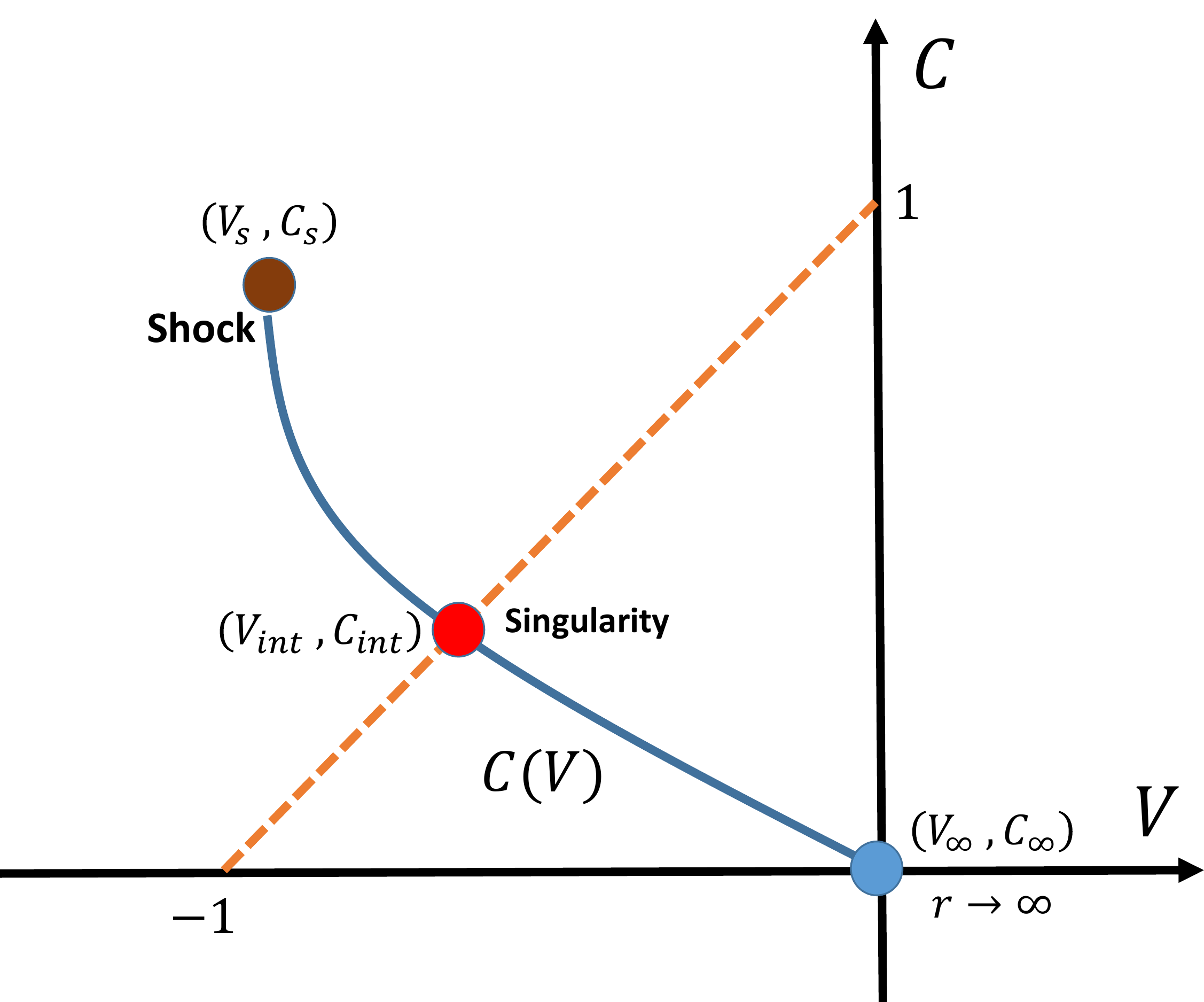} 
\par\end{centering}
\caption{Illustration of the integration of eq. \eqref{eq:dCdV} in the $C-V$
plane. As explained in the text, the integration of the $C(V)$ curve
(blue line) starts from the shock (brown point, equations \eqref{eq:V_shock}-\eqref{eq:C_shock},
passes through the singular point (in red), which is the intersection
point of $C(V)$ with the line $C=V+1$ (orange dotted line), and
continues to the origin (blue point, equations \eqref{eq:v_infinity}-\eqref{eq:c_infinity}).
\label{fig:Illustration-of-the-plane}}
\end{figure}

\subsection{The Critical Points of the Similarity Equations\label{subsec:The-Critical-Points}}

In order to obtain an expression for the removable singular point,
which we denote by $(V_{int},C_{int})$, we set $\Delta_{2}=0$ and
substitute the relation $C^{2}=(V+1)^{2}$ in eq. \eqref{eq:delta2}.
The result is the following cubic equation for the intersection point:

\begin{align}
 & \Bigg[\left(1+\frac{(\gamma-1)(1-\lambda)-\mu}{\gamma(n-1)}\right)V_{int}\nonumber \\
 & \ \ +V_{int}^{2}+\frac{2(\lambda-1)-\mu}{\gamma(n-1)}\Bigg]\left(V_{int}+1\right)=0.
\end{align}
The root $V_{int}=-1$ is not physical since it corresponds to $C_{int}=0$,
which implies zero pressure. Therefore, the two possible roots are:

\begin{align}
V_{\mp}= & \mp\left[\frac{1}{4}\left(\frac{\gamma-2-\frac{\mu}{\lambda-1}}{\gamma(n-1)}-1\right)^{2}-\frac{\left(2-\frac{\mu}{\lambda-1}\right)\left(\lambda-1\right)}{\gamma\left(n-1\right)}\right]^{1/2}\nonumber \\
 & +\frac{\gamma-2-\frac{\mu}{\lambda-1}}{2\gamma(n-1)}-\frac{1}{2}.\label{eq:roots}
\end{align}
It can be shown \cite{lazarus1981self} that the correct root should
be chosen according to:

\begin{equation}
V_{int}=\begin{cases}
V_{-}, & \gamma<\gamma_{crit}\\
V_{+}, & \gamma\geq\gamma_{crit}
\end{cases}\label{eq:V_int}
\end{equation}
with the corresponding value of $C$ being: 
\begin{equation}
C_{int}=V_{int}+1.\label{eq:C_int}
\end{equation}
The critical adiabatic constant, $\gamma_{crit}=\gamma_{crit}\left(\mu,n\right)$,
is defined as the value of $\gamma$ for which the discriminant of
the quadratic equation \eqref{eq:roots} is zero, that is, when $V_{+}=V_{-}$.
In Appendix \ref{app:gamma_crit}, we describe in detail the numerical
method for calculating $\gamma_{crit}\left(\mu,n\right)$.

As shown below, assessing the $L_{int}\equiv C'\left(V_{int}\right)$
at the singular point serves as a useful quantity for identifying
the correct solution to the differential equation \eqref{eq:dCdV}.
To this end, we define $c=C-C_{int}$, $v=V-V_{int}$ and expand $\Delta_{2}$
and $\Delta_{3}$ around the singularity (where $\Delta_{2}\left(V_{int},C_{int}\right)=\Delta_{3}\left(V_{int},C_{int}\right)=0$)
in eq. \eqref{eq:dCdV}:

\begin{equation}
L_{int}\approx\frac{v\frac{\partial\Delta_{3}}{\partial V}+c\frac{\partial\Delta_{3}}{\partial C}}{v\frac{\partial\Delta_{2}}{\partial V}+c\frac{\partial\Delta_{2}}{\partial C}}.\label{eq:Lintap}
\end{equation}
Since $L_{int}\approx c/v$, eq. \eqref{eq:Lintap} reduces to a quadratic
form:

\begin{equation}
\frac{\partial\Delta_{2}}{\partial C}L_{int}^{2}+\left(\frac{\partial\Delta_{2}}{\partial V}-\frac{\partial\Delta_{3}}{\partial C}\right)L_{int}-\frac{\partial\Delta_{3}}{\partial V}=0.
\end{equation}
As shown in \cite{lazarus1981self}, only one of the roots of this
quadratic relation corresponds to the imploding shock problem, so
that the slope at the intersection point is:

\begin{equation}
L_{int}=\frac{\frac{\partial\Delta_{3}}{\partial C}-\frac{\partial\Delta_{2}}{\partial V}-\sqrt{\left(\frac{\partial\Delta_{3}}{\partial C}-\frac{\partial\Delta_{2}}{\partial V}\right)^{2}+4\frac{\partial\Delta_{2}}{\partial C}\frac{\partial\Delta_{3}}{\partial V}}}{2\frac{\partial\Delta_{2}}{\partial C}},\label{eq:L_int}
\end{equation}
where the partial derivatives are evaluated at $\left(V_{int},C_{int}\right)$,
and can be calculated analytically by differentiating equations \eqref{eq:delta2}-\eqref{eq:delta3}
as follows:

\begin{equation}
\frac{\partial\Delta_{2}}{\partial V}=nC^{2}-V(V+1)-V(V+\lambda)-(V+1)(V+\lambda),\label{2dV}
\end{equation}
\begin{equation}
\frac{\partial\Delta_{2}}{\partial C}=2C\left(\frac{2(\lambda-1)-\mu}{\gamma}+nV\right),
\end{equation}
\begin{align}
\frac{\partial\Delta_{3}}{\partial V} & =-\Bigg[2(1+V)+\frac{(3-\gamma)(\lambda-1)}{2}\nonumber \\
 & \ \ \ \ \ \ \ +\frac{2(\lambda-1)+(\gamma-1)\mu}{2\gamma\left(V+1\right)^{2}}C^{2}\nonumber \\
 & \ \ \ \ \ \ \ +(n-1)(\gamma-1)\left(V+\frac{1}{2}\right)\Bigg]C,
\end{align}
\begin{align}
\frac{\partial\Delta_{3}}{\partial C} & =3C^{2}\left(1+\frac{2(\lambda-1)+\mu(\gamma-1)}{2\gamma(1+V)}\right)\nonumber \\
 & \ \ \ +(V+1)^{2}+(\lambda-1)\frac{(3-\gamma)V+2}{2}\nonumber \\
 & \ \ \ -(\gamma-1)(n-1)\frac{V(V+1)}{2}.\label{3dC}
\end{align}

\subsection{Calculation of the similarity exponent\label{subsec:Calculation-lambda}}

\begin{figure}[t]
\begin{centering}
\includegraphics[scale=0.5]{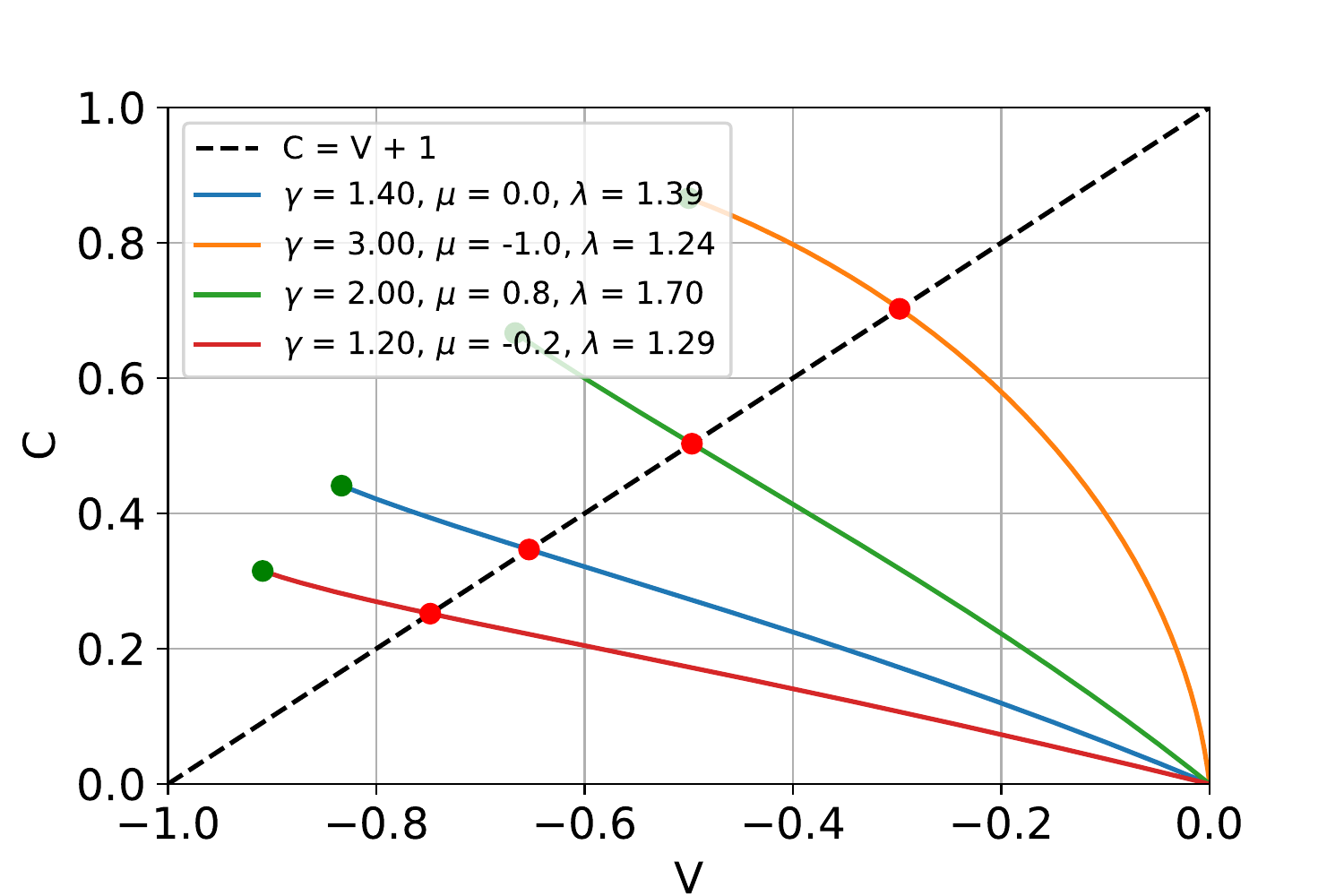}
\par\end{centering}
\caption{$C\left(V\right)$ curves obtained from integration of eq. \eqref{eq:dCdV}
with the correct value of the similar exponent $\lambda$ for various
cases with $n=3$, as detailed in the legend, where the exact value
of $\lambda$ is also given. The integration starts from the shock
front $\left(V_{s},C_{s}\right)$ (in green), and continues smoothly
through the singularity point $\left(V_{int},C_{int}\right)$ (in
red), which is the intersection with the $C=V+1$ line (dashed black
line), until it reaches the origin (at $\left(V_{\infty},C_{\infty}\right)$
which corresponds to $r\rightarrow\infty$). \label{fig:CV_example}}
\end{figure}

\begin{figure*}[t]
\begin{centering}
\includegraphics[scale=0.47]{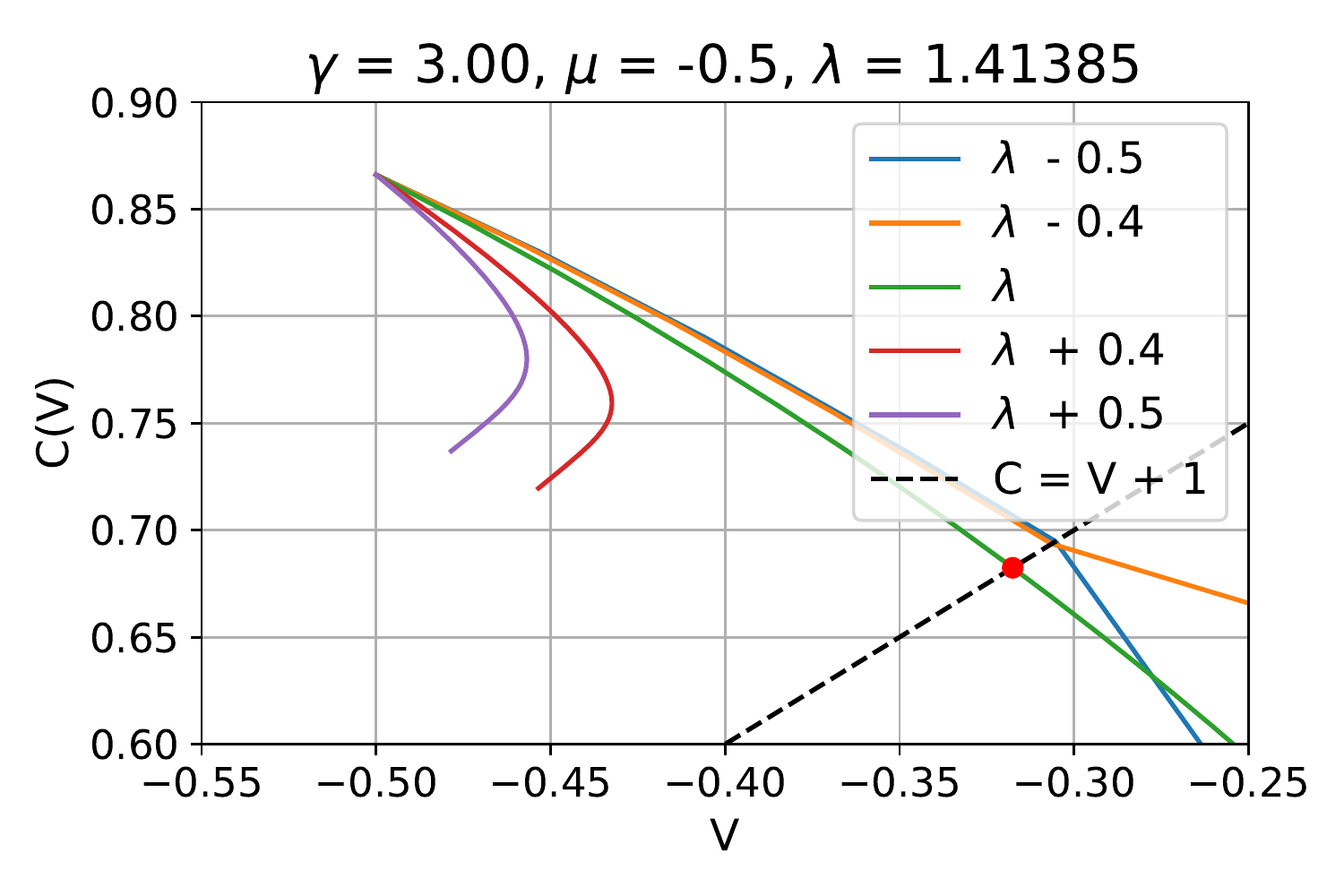}\includegraphics[scale=0.47]{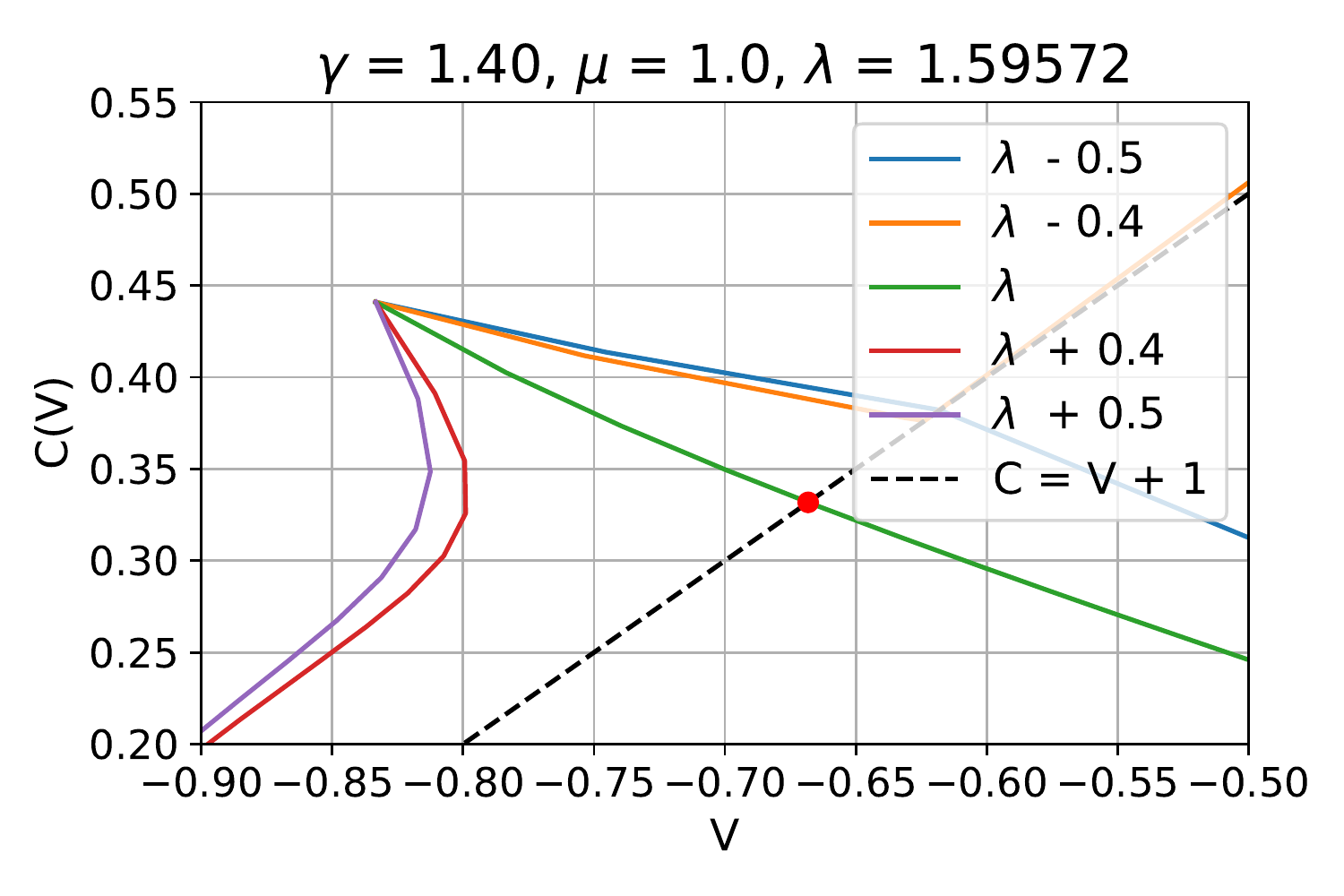}
\par\end{centering}
\begin{centering}
\includegraphics[scale=0.47]{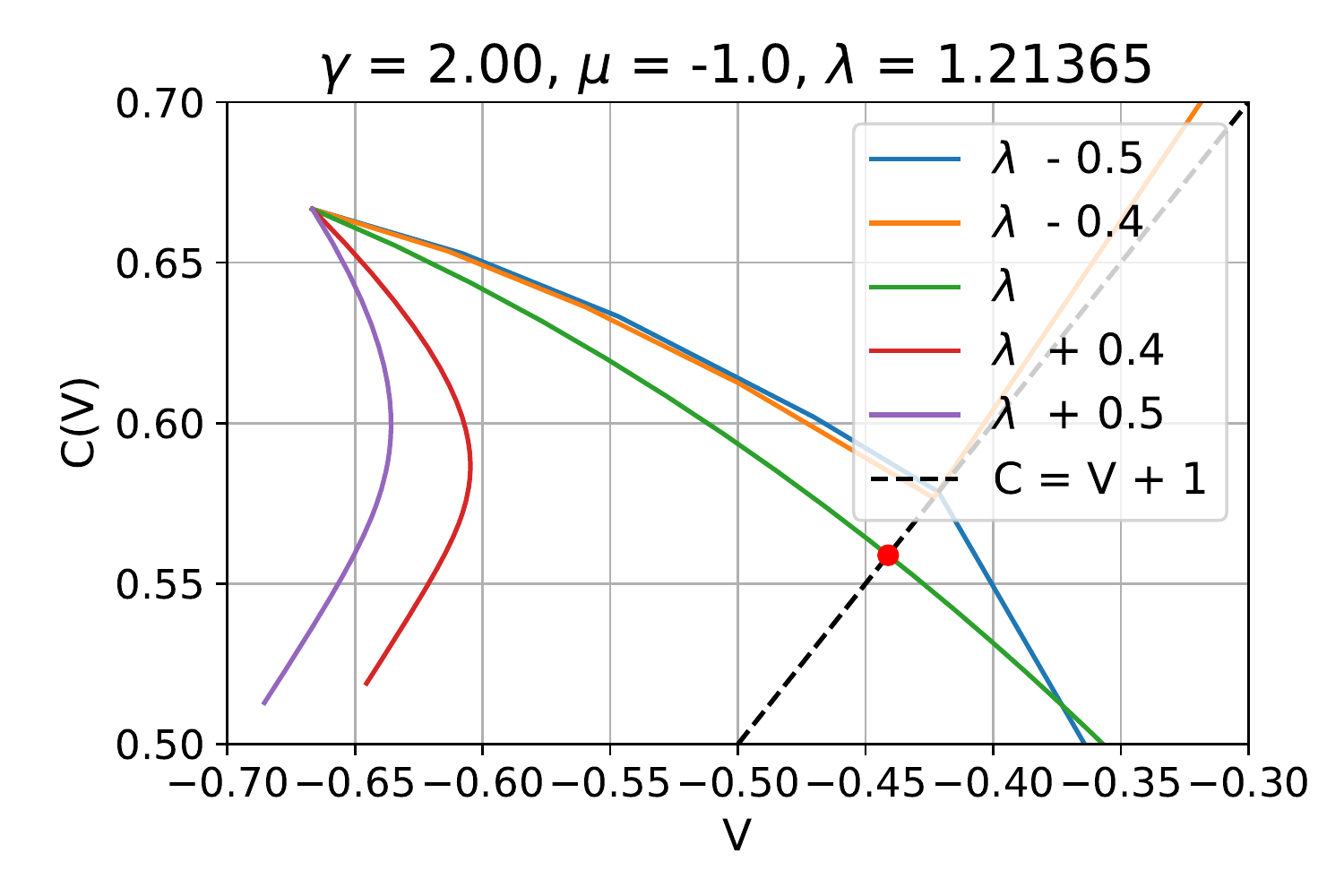}\includegraphics[scale=0.47]{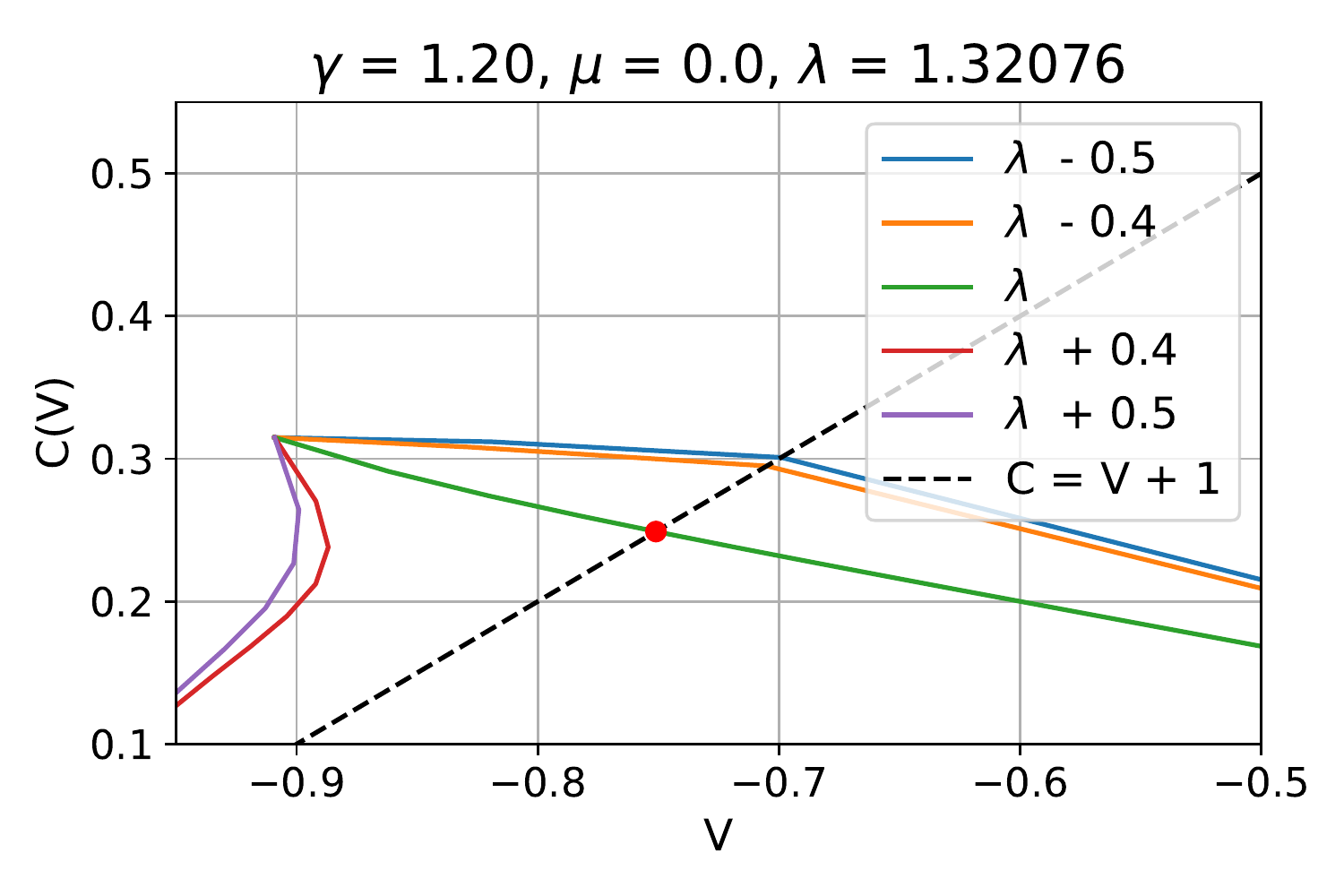}
\par\end{centering}
\caption{$C\left(V\right)$ curves for several cases (for $n=3$, and parameters
which are given in the titles), each calculated with the exact similarity
exponent $\lambda$ (green line), as well incorrect exponents $\lambda\pm0.5$
(blue and purple lines) and $\lambda\pm0.4$ (orange and red lines).
It is evident that only the curve integrated with the correct value
of $\lambda$, intersects the line $C=V+1$ (black dashed line) in
a smooth fashion. \label{fig:not_true_lam}}
\end{figure*}

\begin{figure}[t]
\begin{centering}
\includegraphics[scale=0.6]{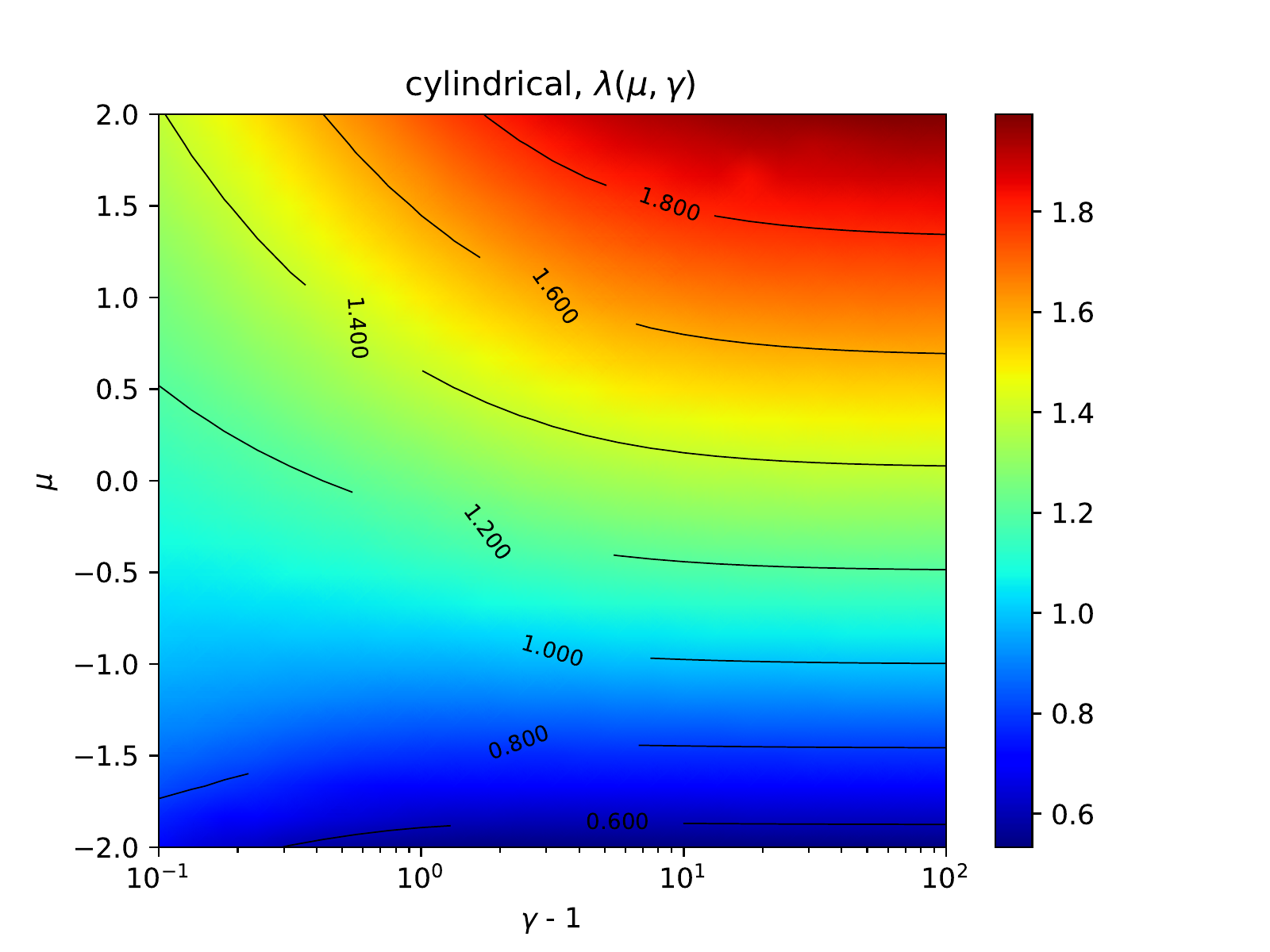}
\par\end{centering}
\begin{centering}
\includegraphics[scale=0.6]{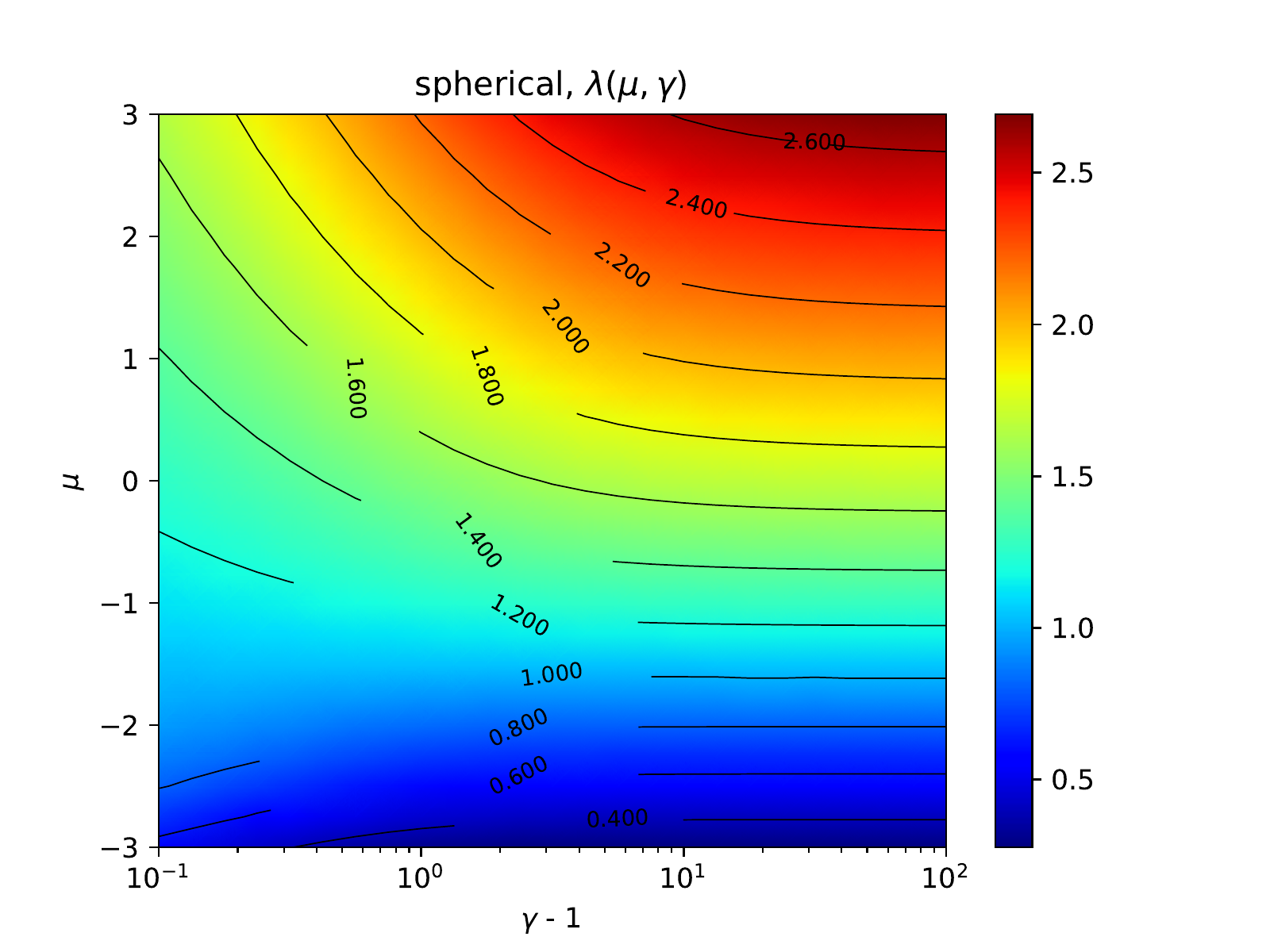}
\par\end{centering}
\caption{A color plot for the similarity exponent $\lambda$ as a function
of $\gamma$ and $\mu$ for cylindrical (upper figure) and spherical
(lower figure) symmetries.\label{fig:lambda-gamma-mu}}
\end{figure}

\begin{figure}[h]
\begin{centering}
\includegraphics[scale=0.5]{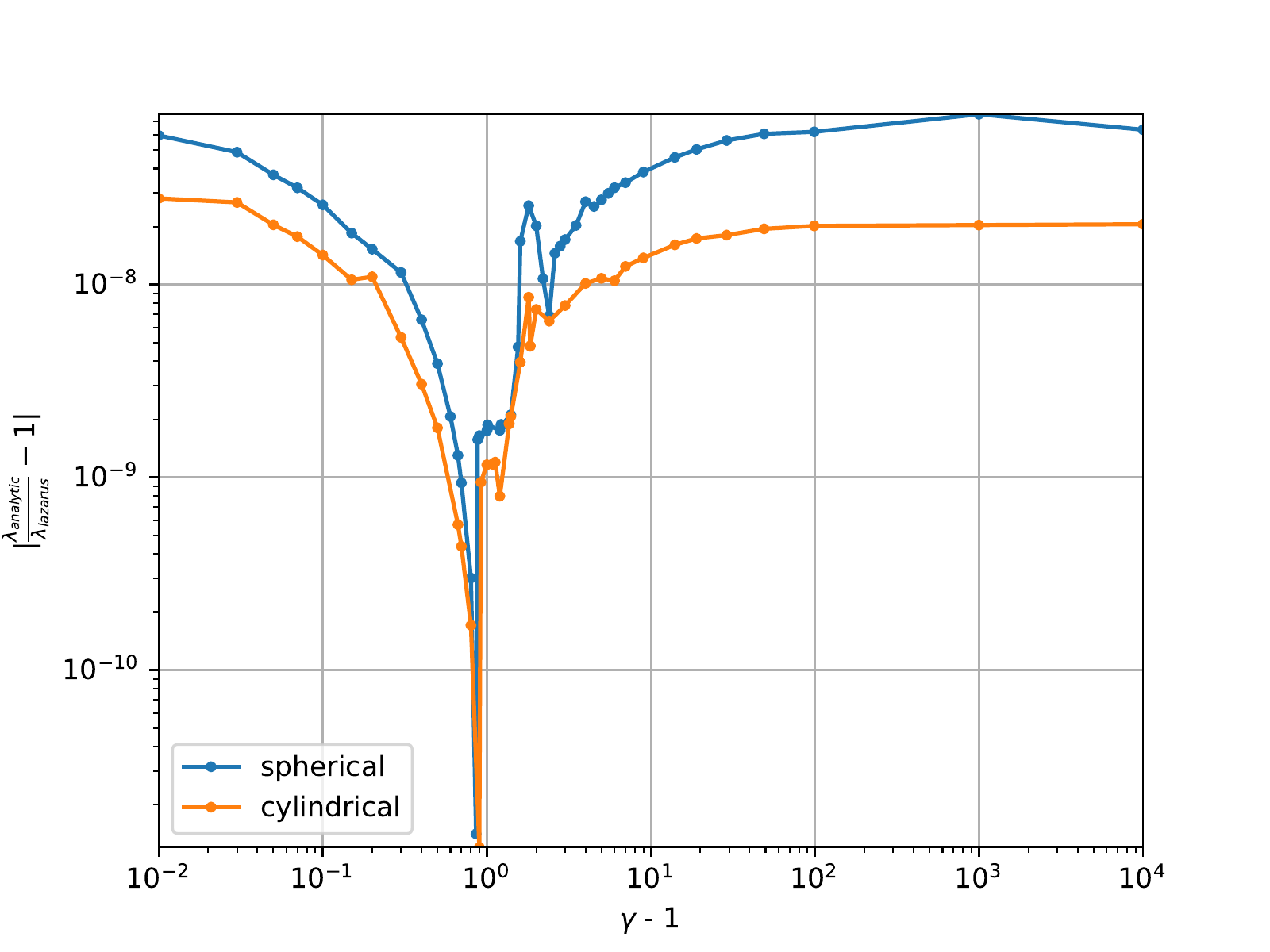}
\par\end{centering}
\caption{The relative error between the similarity exponents $\lambda$ calculated
in this work and the standard results of Lazarus \cite{lazarus1981self}
(see tables \ref{tab:Comparison-cylindrical}-\ref{tab:Comparison-spherical}),
as a function of $\gamma$ for spherical (blue) and cylindrical (orange)
symmetric flows, and $\mu=0$.\label{fig:error-lazarus}}
\end{figure}

Solving the similarity exponent $\lambda$ is the main required step
for calculating the similarity profiles through the ODEs in eq. \eqref{eq:linsys}
and the resulting flow variables via equations \eqref{Similarity V}-\eqref{Similarity R}.

As discussed above, the set of equations \eqref{Similarity V}-\eqref{Similarity R}
is non singular only if integration of the curve $C\left(V\right)$
(eq. \eqref{eq:dCdV}) intersects the line $C=V+1$ precisely at $V=V_{int}$
(eq. \eqref{eq:V_int}), which will correspond to a slope $C'\left(V_{int}\right)=L_{int}$
(eq. \eqref{eq:L_int}). In Appendix \ref{app:Numerical-Calculation-of}
we present an iterative numerical algorithm for the calculation of
$\lambda$ utilizing the fact that these conditions must be satisfied
simultaneously.

Fig. \ref{fig:CV_example} presents integrated $C\left(V\right)$
curves and the corresponding intersection points for various cases,
calculated with the correct values of $\lambda.$ It is evident that
the integration crosses the singularity in a smooth manner and reaches
the origin $\left(V_{\infty},C_{\infty}\right)$ as required. In contrast,
Fig. \ref{fig:not_true_lam} compares the integration of $C\left(V\right)$
with the correct value of the similar exponent $\lambda$ to integrations
with the incorrect similarity exponents $\lambda\pm0.4$ and $\lambda\pm0.5$.
It is evident that only integrations using the correct similarity
exponent pass the singular points smoothly, while integrations with
incorrect similarity exponents either do not reach the $C=V+1$ line
at all, or result with unphysical behaviour such as a discontinuous
change of slope across the intersection.

Tables \ref{tab:Comparison-cylindrical}-\ref{tab:Comparison-spherical}
list our results of the calculated values of $\lambda$ using the
algorithm presented the special case $\mu=0$ in cylindrical and spherical
symmetry, for a wide range of adiabatic exponents. The results are
compared with those of published by Lazarus \cite{lazarus1981self}.
The relative disparity between the results is presented as a function
of $\gamma$ in Fig. \ref{fig:error-lazarus} and it is evident that
an agreement to seven significant digits or better is achieved.

To the best of our knowledge, similar results were published in the
past only for for some values of $\mu>0$, notably by Sharma et al.
\cite{sharma1995similarity} and Toque \cite{toque2001self}, which
we compare to our calculations in table \ref{tab:compare_toque}.
Finally, in table \ref{tab:canonical} we present numerical results
of $\lambda$ for spherical and cylindrical symmetries, the typical
values $\gamma=5/3,1.4$ and various different values of $\mu$, including
$\mu<0$. In Fig. \ref{fig:lambda-gamma-mu}, color plots for $\lambda\left(\gamma,\mu\right)$
for a wide range of $\gamma,\mu$ are shown for spherical and cylindrical
symmetries. Further support for our results was found after this work
had been completed by \cite{modelevsky2021revisiting}.

As can be expected, $\lambda$ is a monotonically increasing function
of both $\gamma$ and $\mu$. We note in passing that for $\gamma\gg1$,
the similarity exponent approaches a constant value (which depends
on $\mu$), as was already noted by Lazarus \cite{lazarus1977similarity}.

\begin{table*}[t]
\begin{centering}
\begin{tabular}{|c|c|c|}
\hline 
$\gamma$  & $\lambda_{\text{Lazarus}}$  & $\lambda$\tabularnewline
\hline 
$1.01$  & $1.05539838$  & $1.05539835$\tabularnewline
\hline 
$1.03$  & $1.08507376$  & $1.08507379$\tabularnewline
\hline 
$1.05$  & $1.10238925$  & $1.10238927$\tabularnewline
\hline 
$1.07$  & $1.11506921$  & $1.11506923$\tabularnewline
\hline 
$1.1$  & $1.12962686$  & $1.12962688$\tabularnewline
\hline 
$1.15$  & $1.14757733$  & $1.14757734$\tabularnewline
\hline 
$1.2$  & $1.16122032$  & $1.16122033$\tabularnewline
\hline 
$1.3$  & $1.18172136$  & $1.18172136$\tabularnewline
\hline 
$1.4$  & $1.19714143$  & $1.19714143$\tabularnewline
\hline 
$1.5$  & $1.20955913$  & $1.20955913$\tabularnewline
\hline 
$\frac{5}{3}$  & $1.22605379$  & $1.22605379$\tabularnewline
\hline 
\end{tabular}%
\begin{tabular}{|c|c|c|}
\hline 
$\gamma$  & $\lambda_{\text{Lazarus}}$  & $\lambda$\tabularnewline
\hline 
$1.7$  & $1.22889310$  & $1.22889310$\tabularnewline
\hline 
$1.8$  & $1.23670552$  & $1.23670552$\tabularnewline
\hline 
$1.9$  & $1.24362784$  & $1.24362784$\tabularnewline
\hline 
$1.92$  & $1.24492082$  & $1.24492082$\tabularnewline
\hline 
$2$  & $1.24982448$  & $1.24982448$\tabularnewline
\hline 
$2.0863$  & $1.25468301$  & $1.25468301$\tabularnewline
\hline 
$2.0883$  & $1.25479079$  & $1.25479079$\tabularnewline
\hline 
$2.125$  & $1.25673437$  & $1.25673437$\tabularnewline
\hline 
$2.2$  & $1.26049898$  & $1.26049898$\tabularnewline
\hline 
$2.3676$  & $1.26806432$  & $1.26806432$\tabularnewline
\hline 
$2.3678$  & $1.26807272$  & $1.26807272$\tabularnewline
\hline 
\end{tabular}%
\begin{tabular}{|c|c|c|}
\hline 
$\gamma$  & $\lambda_{\text{Lazarus}}$  & $\lambda$\tabularnewline
\hline 
$2.4$  & $1.26940764$  & $1.26940764$\tabularnewline
\hline 
$2.6$  & $1.27698161$  & $1.27698162$\tabularnewline
\hline 
$2.8$  & $1.28351397$  & $1.28351398$\tabularnewline
\hline 
$2.8392$  & $1.28469123$  & $1.28469124$\tabularnewline
\hline 
$2.83929$  & $1.28469390$  & $1.28469391$\tabularnewline
\hline 
$3$  & $1.28921366$  & $1.28921367$\tabularnewline
\hline 
$3.4$  & $1.29869509$  & $1.29869510$\tabularnewline
\hline 
$4$  & $1.30952673$  & $1.30952674$\tabularnewline
\hline 
$5$  & $1.32204998$  & $1.32204999$\tabularnewline
\hline 
$6$  & $1.33056278$  & $1.33056279$\tabularnewline
\hline 
$7$  & $1.33673018$  & $1.33673020$\tabularnewline
\hline 
\end{tabular}%
\begin{tabular}{|c|c|c|}
\hline 
$\gamma$  & $\lambda_{\text{Lazarus}}$  & $\lambda$\tabularnewline
\hline 
$8$  & $1.34140548$  & $1.34140549$\tabularnewline
\hline 
$10$  & $1.34802513$  & $1.34802515$\tabularnewline
\hline 
$15$  & $1.35699098$  & $1.35699100$\tabularnewline
\hline 
$20$  & $1.36153562$  & $1.36153564$\tabularnewline
\hline 
$30$  & $1.36612239$  & $1.36612242$\tabularnewline
\hline 
$50$  & $1.36982259$  & $1.36982261$\tabularnewline
\hline 
$100$  & $1.37261589$  & $1.37261592$\tabularnewline
\hline 
$1000$  & $1.37514328$  & $1.37514331$\tabularnewline
\hline 
$9999$  & $1.37539672$  & $1.37539675$\tabularnewline
\hline 
\multicolumn{1}{c}{} & \multicolumn{1}{c}{} & \multicolumn{1}{c}{}\tabularnewline
\multicolumn{1}{c}{} & \multicolumn{1}{c}{} & \multicolumn{1}{c}{}\tabularnewline
\end{tabular}
\par\end{centering}
\caption{Comparison of the values of the similarity exponent $\lambda$ calculated
in this work and those calculated by Lazarus \cite{lazarus1981self},
for different values of $\gamma$ and $\mu=0$, in cylindrical symmetry.
\label{tab:Comparison-cylindrical}}
\end{table*}

\begin{table*}[t]
\begin{centering}
\begin{tabular}{|c|c|c|}
\hline 
$\gamma$  & $\lambda_{\text{Lazarus}}$  & $\lambda$\tabularnewline
\hline 
$1.01$  & $1.10881007$  & $1.10881001$\tabularnewline
\hline 
$1.03$  & $1.16716916$  & $1.16716922$\tabularnewline
\hline 
$1.05$  & $1.20156643$  & $1.20156647$\tabularnewline
\hline 
$1.07$  & $1.22695814$  & $1.22695818$\tabularnewline
\hline 
$1.1$  & $1.25632911$  & $1.25632914$\tabularnewline
\hline 
$1.15$  & $1.29284049$  & $1.29284052$\tabularnewline
\hline 
$1.2$  & $1.32075654$  & $1.32075656$\tabularnewline
\hline 
$1.3$  & $1.36281235$  & $1.36281237$\tabularnewline
\hline 
$1.4$  & $1.39436078$  & $1.39436079$\tabularnewline
\hline 
$1.5$  & $1.41959135$  & $1.41959136$\tabularnewline
\hline 
$1.6$  & $1.44052881$  & $1.44052882$\tabularnewline
\hline 
$\frac{5}{3}$  & $1.45269272$  & $1.45269272$\tabularnewline
\hline 
\end{tabular}%
\begin{tabular}{|c|c|c|}
\hline 
$\gamma$  & $\lambda_{\text{Lazarus}}$  & $\lambda$\tabularnewline
\hline 
$1.7$  & $1.45832858$  & $1.45832858$\tabularnewline
\hline 
$1.8$  & $1.47372274$  & $1.47372274$\tabularnewline
\hline 
$1.86$  & $1.48201847$  & $1.48201847$\tabularnewline
\hline 
$1.88$  & $1.48464620$  & $1.48464620$\tabularnewline
\hline 
$1.9$  & $1.48720971$  & $1.48720972$\tabularnewline
\hline 
$2$  & $1.49914683$  & $1.49914683$\tabularnewline
\hline 
$2.01$  & $1.50026616$  & $1.50026616$\tabularnewline
\hline 
$2.012$  & $1.50048851$  & $1.50048851$\tabularnewline
\hline 
$2.2$  & $1.51937505$  & $1.51937505$\tabularnewline
\hline 
$2.2215$  & $1.51130884$  & $1.51130884$\tabularnewline
\hline 
$2.2217$  & $1.52132663$  & $1.52132664$\tabularnewline
\hline 
$2.4$  & $1.53589867$  & $1.53589867$\tabularnewline
\hline 
\end{tabular}%
\begin{tabular}{|c|c|c|}
\hline 
$\gamma$  & $\lambda_{\text{Lazarus}}$  & $\lambda$\tabularnewline
\hline 
$2.55194$  & $1.54657142$  & $1.54657143$\tabularnewline
\hline 
$2.6$  & $1.54966637$  & $1.54966640$\tabularnewline
\hline 
$2.8$  & $1.56131989$  & $1.56131993$\tabularnewline
\hline 
$3$  & $1.57131262$  & $1.57131266$\tabularnewline
\hline 
$3.2$  & $1.57997558$  & $1.57997560$\tabularnewline
\hline 
$3.4$  & $1.58755678$  & $1.58755679$\tabularnewline
\hline 
$3.6$  & $1.59424597$  & $1.59424599$\tabularnewline
\hline 
$3.8$  & $1.60019098$  & $1.60019100$\tabularnewline
\hline 
$4$  & $1.60550871$  & $1.60550874$\tabularnewline
\hline 
$4.5$  & $1.61663097$  & $1.61663100$\tabularnewline
\hline 
$5$  & $1.62542433$  & $1.62542437$\tabularnewline
\hline 
$5.5$  & $1.63254761$  & $1.63254766$\tabularnewline
\hline 
\end{tabular}%
\begin{tabular}{|c|c|c|}
\hline 
$\gamma$  & $\lambda_{\text{Lazarus}}$  & $\lambda$\tabularnewline
\hline 
$6$  & $1.63843333$  & $1.63843337$\tabularnewline
\hline 
$6.5$  & $1.64337694$  & $1.64337699$\tabularnewline
\hline 
$7$  & $1.64758710$  & $1.64758715$\tabularnewline
\hline 
$8$  & $1.65437385$  & $1.65437391$\tabularnewline
\hline 
$10$  & $1.66375840$  & $1.66375846$\tabularnewline
\hline 
$15$  & $1.67605129$  & $1.67605136$\tabularnewline
\hline 
$20$  & $1.68210044$  & $1.68210053$\tabularnewline
\hline 
$30$  & $1.68808305$  & $1.68808315$\tabularnewline
\hline 
$50$  & $1.69282046$  & $1.69282056$\tabularnewline
\hline 
$100$  & $1.69634476$  & $1.69634486$\tabularnewline
\hline 
$1000$  & $1.69949536$  & $1.69949549$\tabularnewline
\hline 
$9999$  & $1.69980930$  & $1.69980941$\tabularnewline
\hline 
\end{tabular}
\par\end{centering}
\caption{Comparison of the values of the similarity exponent $\lambda$ calculated
in this work and those calculated by Lazarus \cite{lazarus1981self},
for different values of $\gamma$ and $\mu=0$, in spherical symmetry.
\label{tab:Comparison-spherical}}
\end{table*}

\begin{table}[t]
\begin{centering}
\begin{tabular}{|c|c|c|c|}
\hline 
$\gamma$  & $\mu$  & $\lambda_{\text{Sharma}}$  & $\lambda$\tabularnewline
\hline 
$1.2$  & $0.5$  & $1.2458754$  & $1.2458735$\tabularnewline
\hline 
$\frac{5}{3}$  & $0.5$  & $1.3415641$  & $1.3415624$\tabularnewline
\hline 
$2.0$  & $0.5$  & $1.3749395$  & $1.3749381$\tabularnewline
\hline 
$1.2$  & $1.0$  & $1.3281454$  & $1.3280348$\tabularnewline
\hline 
$\frac{5}{3}$  & $1.0$  & $1.4527030$  & $1.4527000$\tabularnewline
\hline 
$2.0$  & $1.0$  & $1.4949825$  & $1.4949798$\tabularnewline
\hline 
$1.1$  & $2.0$  & $1.3932402$  & $1.3932370$\tabularnewline
\hline 
$1.2$  & $2.0$  & $1.4886240$  & $1.4886198$\tabularnewline
\hline 
$1.4$  & $2.0$  & $1.5914957$  & $1.5914907$\tabularnewline
\hline 
$\frac{5}{3}$  & $2.0$  & $1.6682122$  & $1.6682070$\tabularnewline
\hline 
$2.0$  & $2.0$  & $1.7270734$  & $1.7270682$\tabularnewline
\hline 
$3.0$  & $2.0$  & $1.8181848$  & $1.8176319$\tabularnewline
\hline 
$6.0$  & $2.0$  & $1.9065023$  & $1.9064994$\tabularnewline
\hline 
\end{tabular}%
\begin{tabular}{|c|c|c|c|}
\hline 
$\gamma$  & $\mu$  & $\lambda_{\text{Sharma}}$  & $\lambda$\tabularnewline
\hline 
$1.2$  & $0.5$  & $1.4041700$  & $1.4041673$\tabularnewline
\hline 
$\frac{5}{3}$  & $0.5$  & $1.4964265$  & $1.4964238$\tabularnewline
\hline 
$2.0$  & $0.5$  & $1.5691222$  & $1.5691202$\tabularnewline
\hline 
$1.2$  & $1.0$  & $1.4859164$  & $1.4859128$\tabularnewline
\hline 
$\frac{5}{3}$  & $1.0$  & $1.5957192$  & $1.5957154$\tabularnewline
\hline 
$2.0$  & $1.0$  & $1.6816074$  & $1.6816040$\tabularnewline
\hline 
$1.1$  & $2.0$  & $1.5177710$  & $1.5177653$\tabularnewline
\hline 
$1.2$  & $2.0$  & $1.6465296$  & $1.6465223$\tabularnewline
\hline 
$1.4$  & $2.0$  & $1.7895314$  & $1.7895229$\tabularnewline
\hline 
$\frac{5}{3}$  & $2.0$  & $1.8997554$  & $1.8997468$\tabularnewline
\hline 
$2.0$  & $2.0$  & $1.9865095$  & $1.9865016$\tabularnewline
\hline 
$3.0$  & $2.0$  & $2.1228456$  & $2.1228406$\tabularnewline
\hline 
$6.0$  & $2.0$  & $2.2571023$  & $2.2571027$\tabularnewline
\hline 
\end{tabular}
\par\end{centering}
\caption{Comparison of the of the similarity exponent $\lambda$ calculated
in this work and those calculated by Sharma et al. \cite{sharma1995similarity},
for different values $\gamma$ and $\mu>0$, in cylindrical (left
table) and spherical (right table) symmetry. \label{tab:compare_toque}}
\end{table}

\begin{table}[t]
\begin{centering}
\begin{tabular}{|c|c|c|}
\hline 
\multicolumn{3}{|c|}{$\gamma=\frac{5}{3}$}\tabularnewline
\hline 
$\mu$  & $\lambda_{n=2}$  & $\lambda_{n=3}$\tabularnewline
\hline 
$-1$  & $0.96265849$  & $1.19582757$\tabularnewline
\hline 
$-0.25$  & $1.16563261$  & $1.39227335$\tabularnewline
\hline 
$0.5$  & $1.34156241$  & $1.56912017$\tabularnewline
\hline 
$1.25$  & $1.50723161$  & $1.73682914$\tabularnewline
\hline 
$2$  & $1.66820698$  & $1.89974683$\tabularnewline
\hline 
\end{tabular}%
\begin{tabular}{|c|c|c|}
\hline 
\multicolumn{3}{|c|}{$\gamma=1.4$}\tabularnewline
\hline 
$\mu$  & $\lambda_{n=2}$  & $\lambda_{n=3}$\tabularnewline
\hline 
$-1$  & $0.96426155$  & $1.17286279$\tabularnewline
\hline 
$-0.25$  & $1.14366554$  & $1.34177491$\tabularnewline
\hline 
$0.5$  & $1.29970718$  & $1.49642378$\tabularnewline
\hline 
$1.25$  & $1.44745345$  & $1.64464959$\tabularnewline
\hline 
$2$  & $1.59149071$  & $1.78952289$\tabularnewline
\hline 
\end{tabular}
\par\end{centering}
\caption{The similarity exponent for canonical values $\gamma=\frac{5}{3}$
(left table) and $\gamma=1.4$ (right table) and various values of
$\mu$, in cylindrical and spherical symmetries. \label{tab:canonical}}
\end{table}

\subsection{Calculation of the similarity profiles}

\begin{figure*}[t]
\begin{centering}
\includegraphics[scale=0.47]{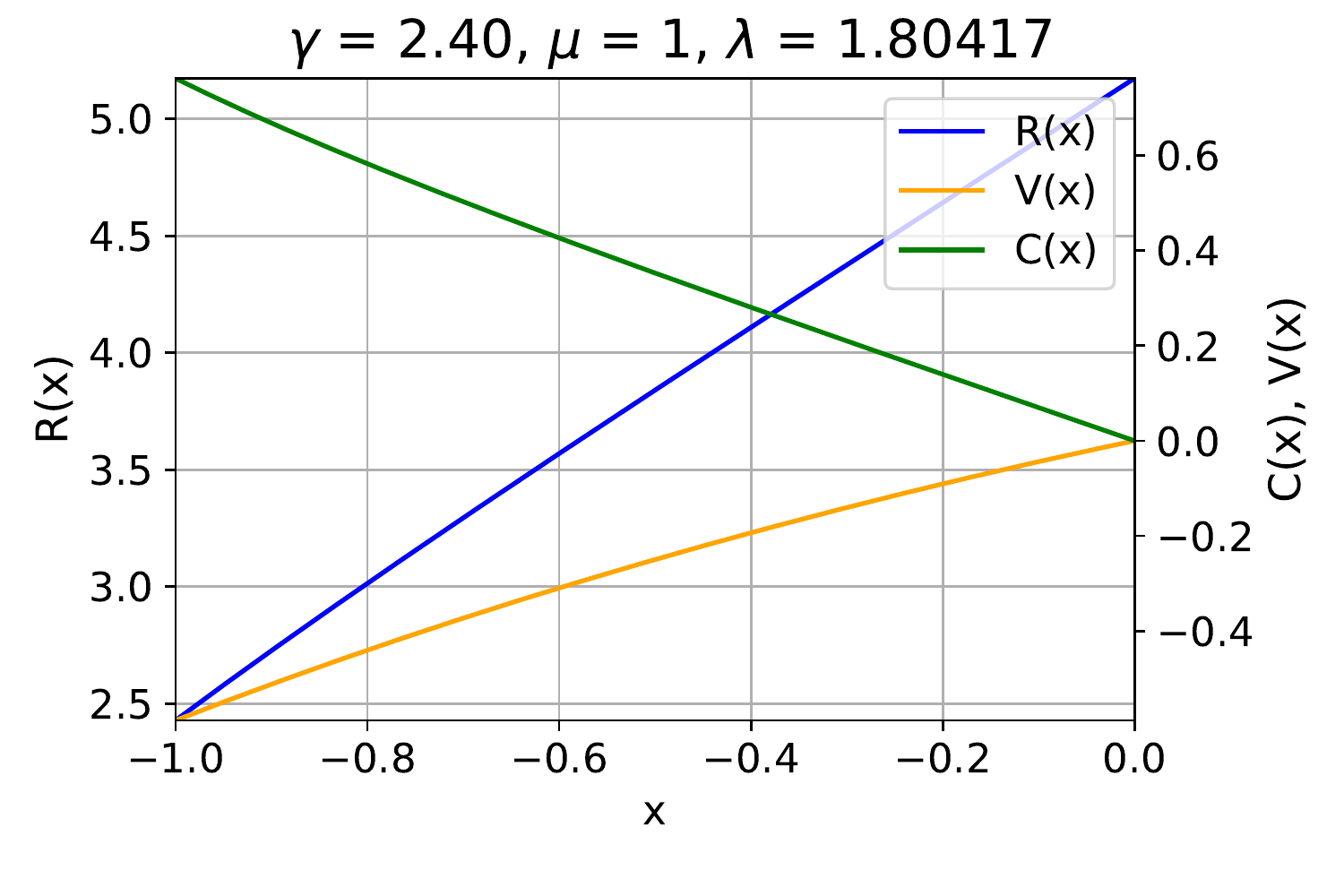}\includegraphics[scale=0.47]{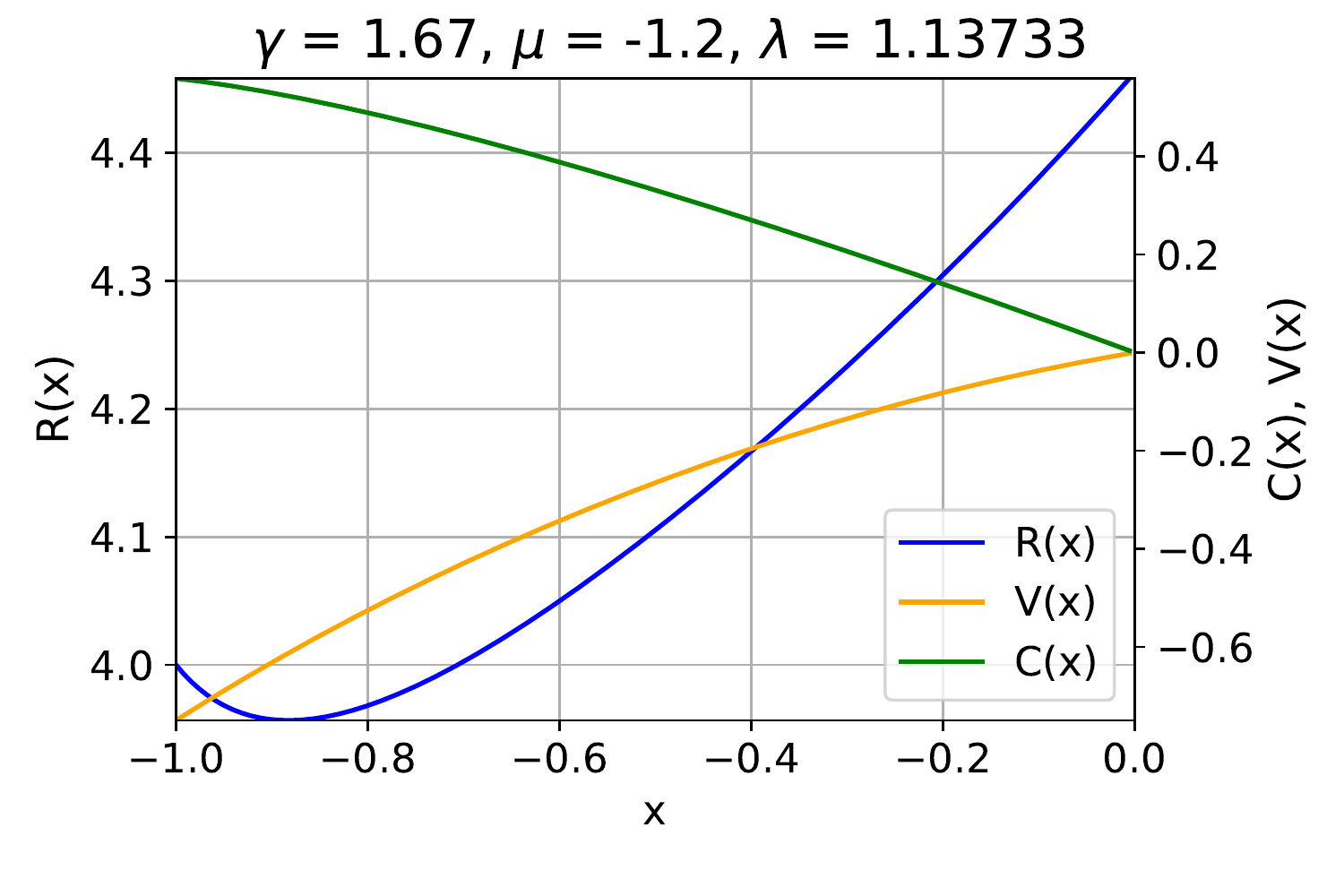}
\par\end{centering}
\begin{centering}
\includegraphics[scale=0.47]{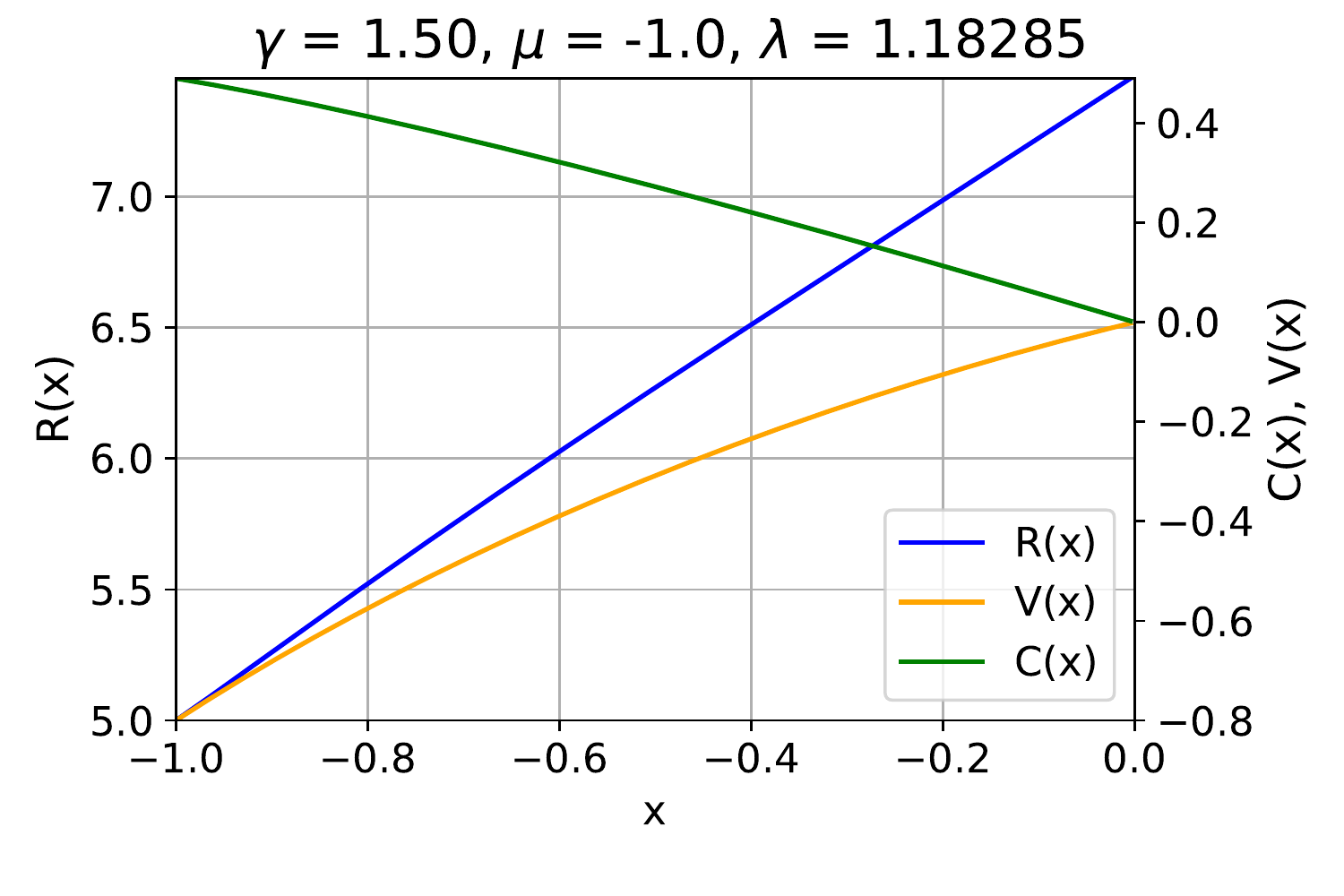}\includegraphics[scale=0.47]{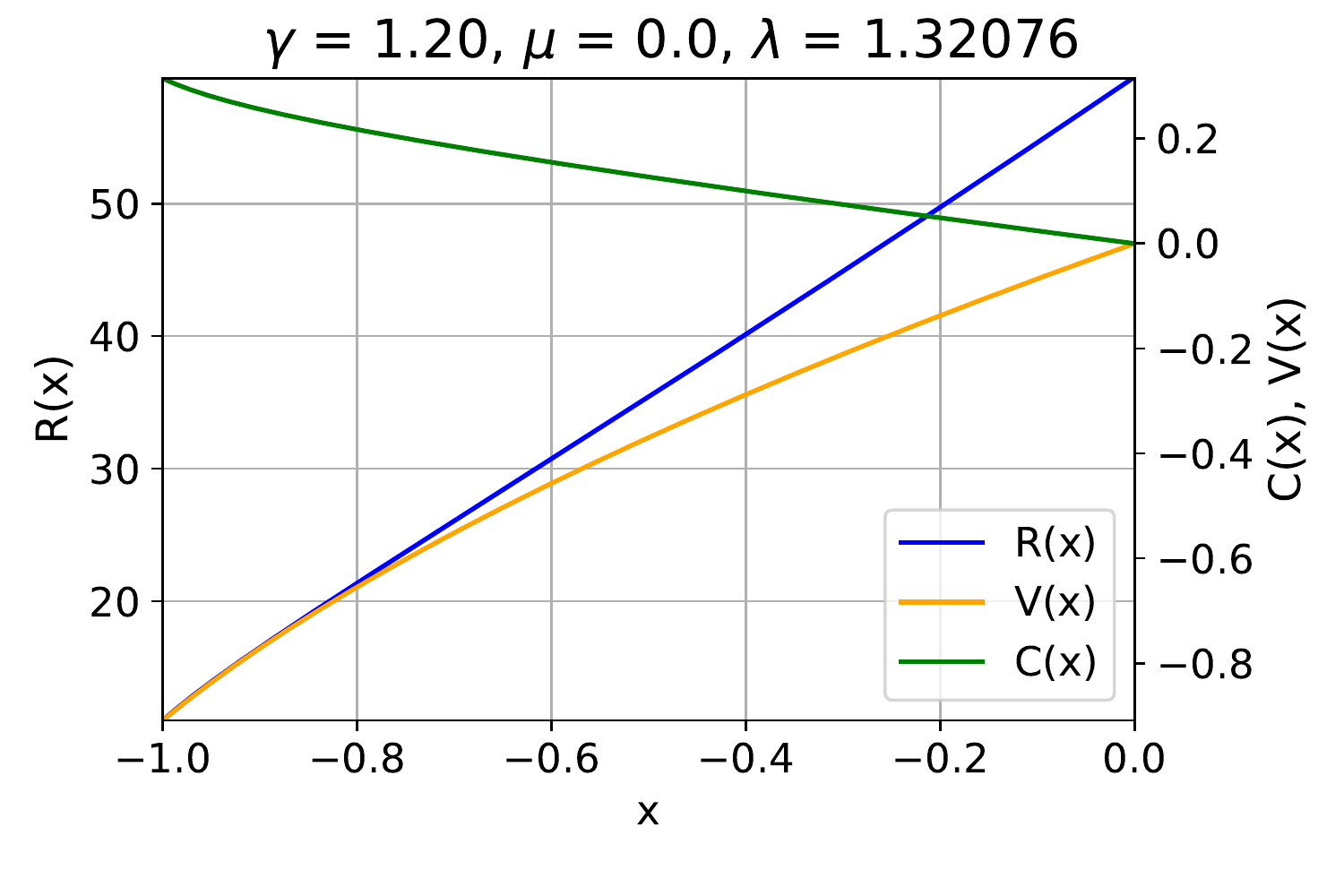}
\par\end{centering}
\caption{The similarity profiles $R\left(x\right)$ (blue line, left $y$ axis),
$V\left(x\right)$ (orange line, right $y$ axis) and $C\left(x\right)$
(green line, right $y$ axis), in spherical symmetry, for different
values of $\mu$ and $\gamma$, as specified in the title of each
sub-figure (where the value of the similarity exponent $\lambda$
is also given).\label{fig:self-similar-profiles}}
\end{figure*}

Once the correct value of the similarity exponent is obtained, the
integration of the system of differential equations \eqref{ODES}
for the similarity profiles $R(x),V(x)$ and $C(x)$ can be performed.
Integration of eq. \eqref{ODES} from the shock front at $x=-1$ with
the initial values (see equations \eqref{eq:R_shock}-\eqref{eq:C_shock})
can, in principle be completed all the way to $x=0$ (which corresponds
to $r\rightarrow\infty$). The ODE integration is performed via the
LSODA integrator \cite{hindmarsh1983odepack}. Numerical examples
of the similarity profiles $R(x),V(x),C(x)$ for several values of
$\gamma$ and $\mu$ in spherical symmetry are presented in Fig. \ref{fig:self-similar-profiles}.

\section{Comparison to numerical simulations \label{sec:Comparison-to-numerical}}

\begin{figure*}[t]
\begin{centering}
\includegraphics[scale=0.47]{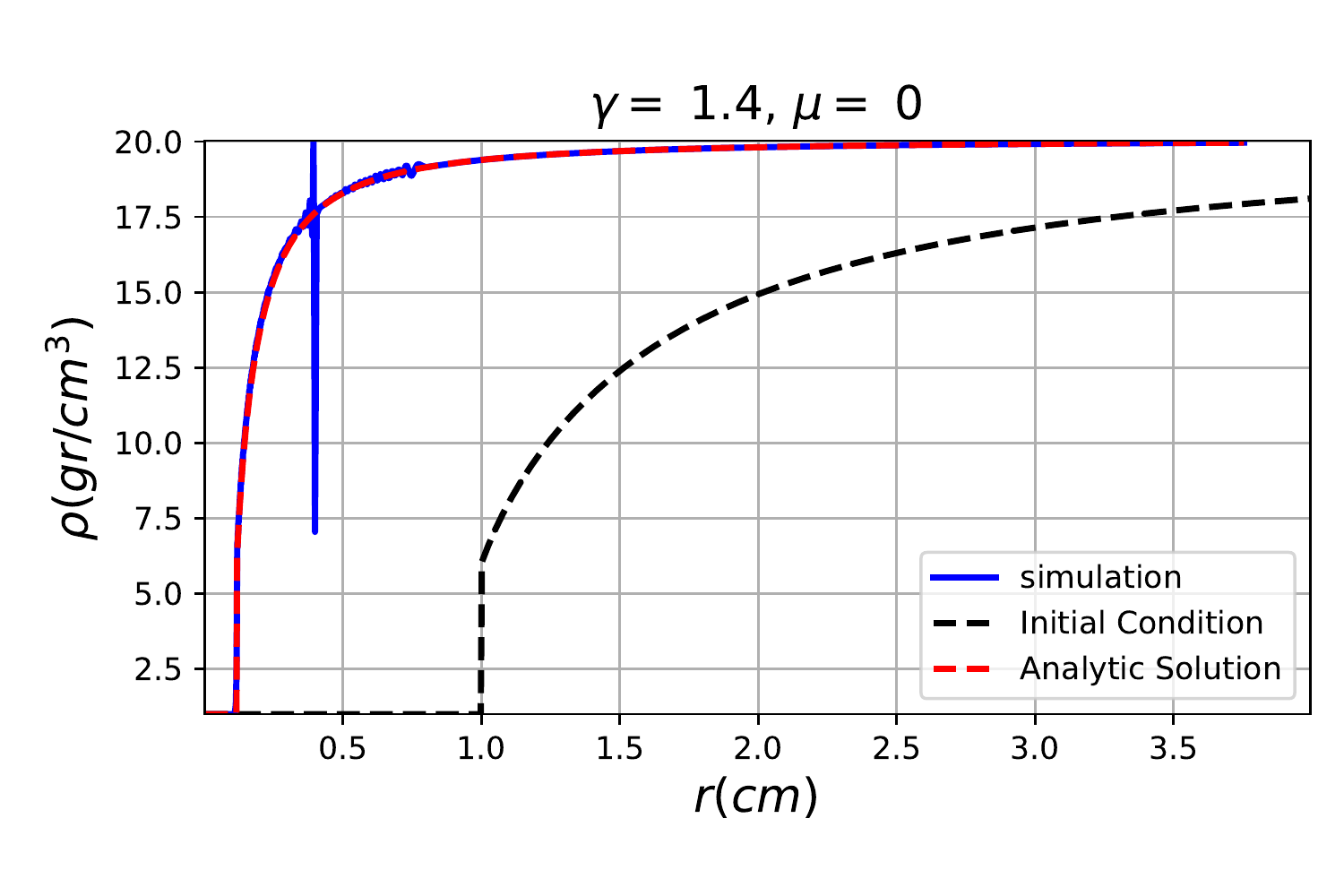}\includegraphics[scale=0.47]{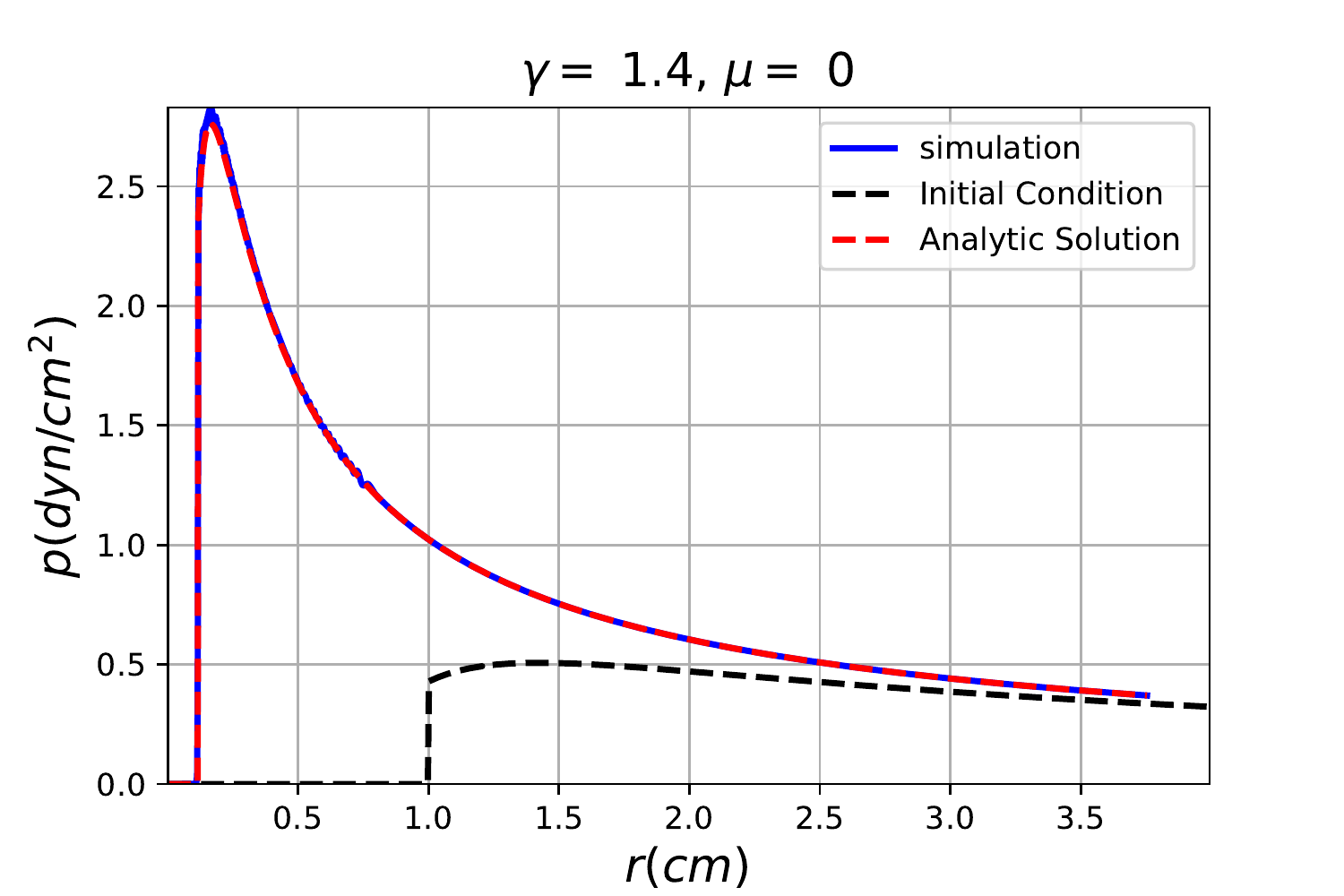}
\par\end{centering}
\begin{centering}
\includegraphics[scale=0.47]{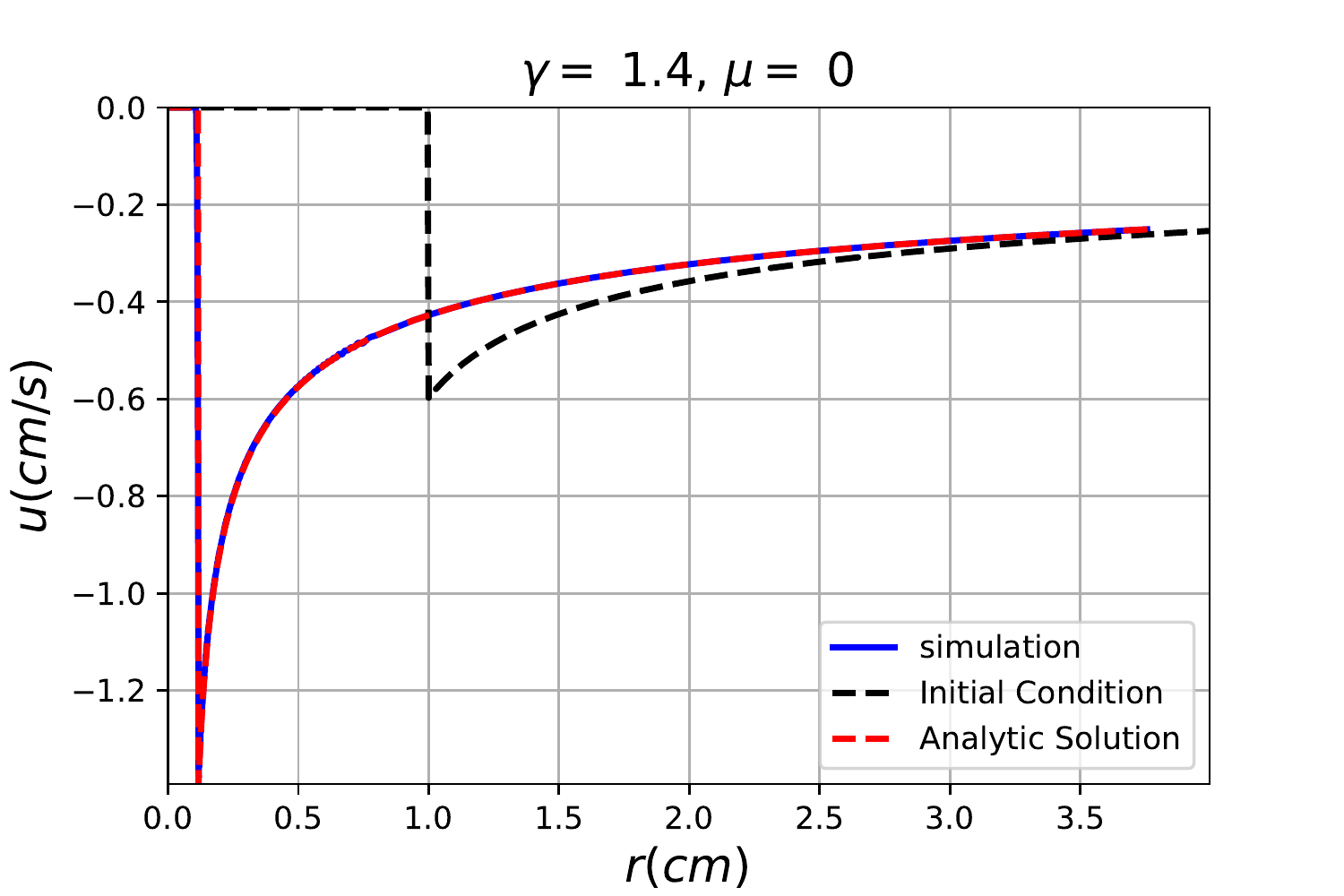}\includegraphics[scale=0.47]{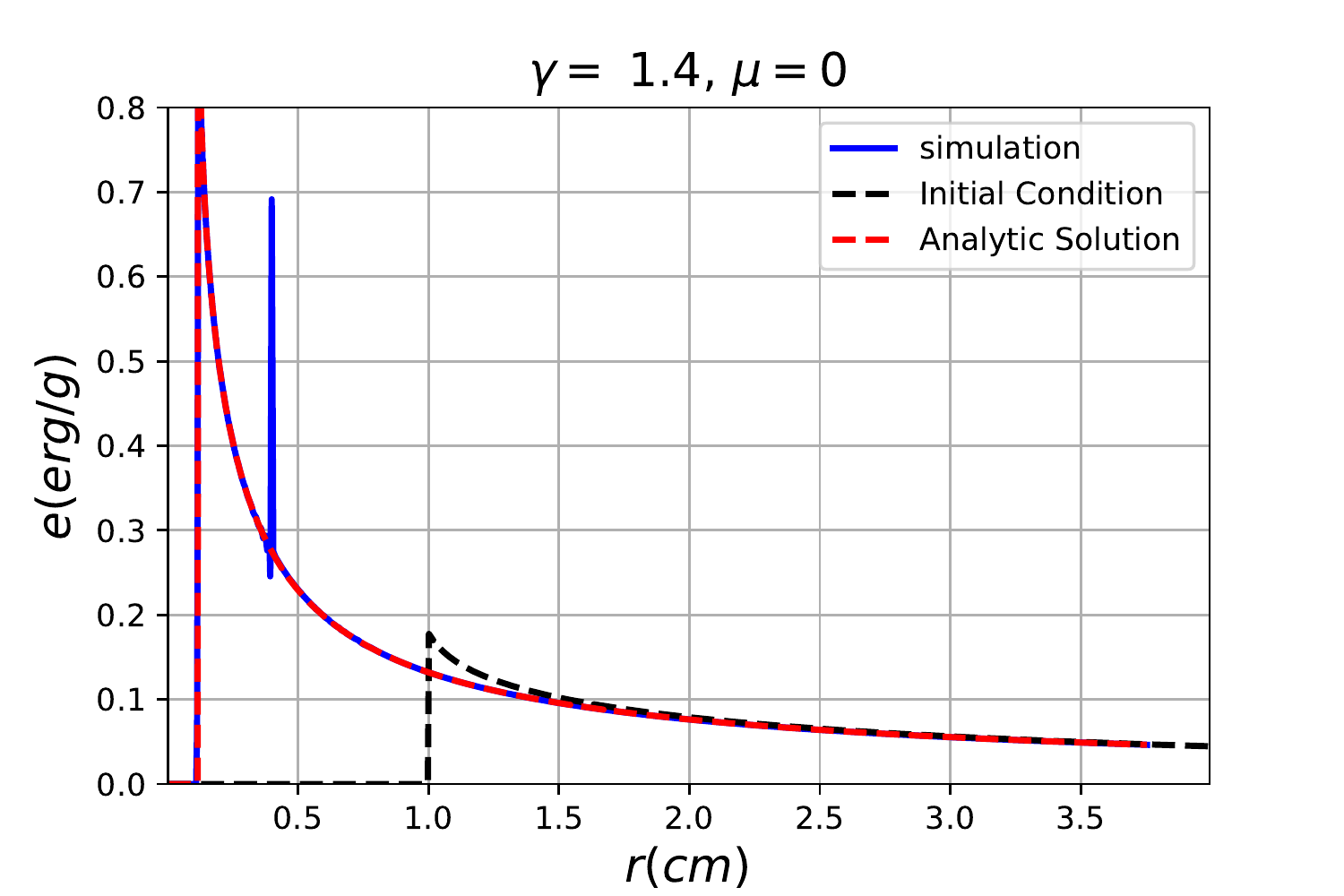}
\par\end{centering}
\caption{A comparison of hydrodynamic profiles (density, pressure, velocity
and specific internal energy), resulting from numerical simulations
(in blue) and the corresponding analytic solutions (in red) for $\gamma=1.4,\ \mu=0$.
The simulations were performed with $N=1000$ cells. The dashed black
lines describe the profiles that were used to initialize the simulations.
\label{fig:profiles}}
\end{figure*}

\begin{figure*}[t]
\begin{centering}
\includegraphics[scale=0.47]{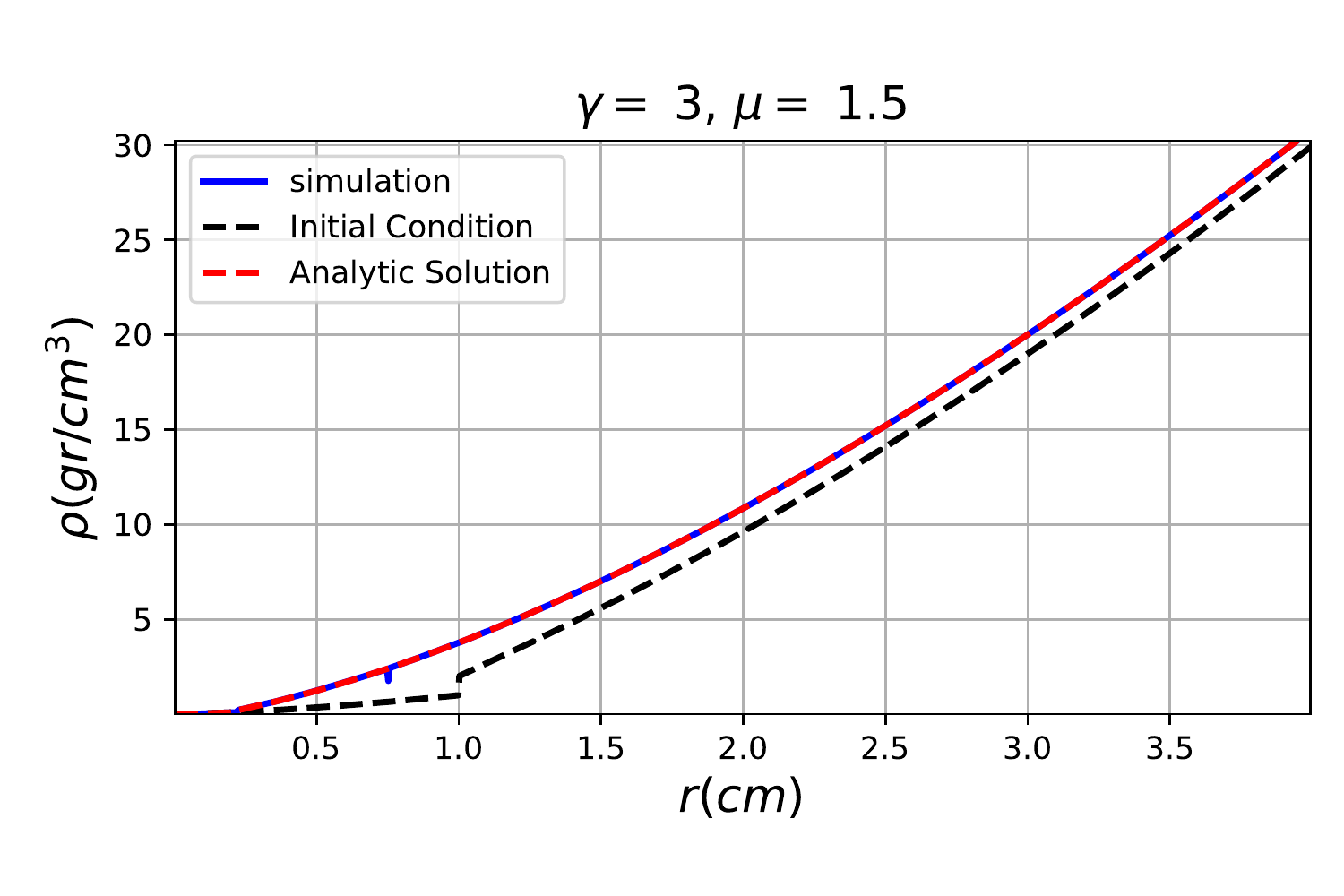}\includegraphics[scale=0.47]{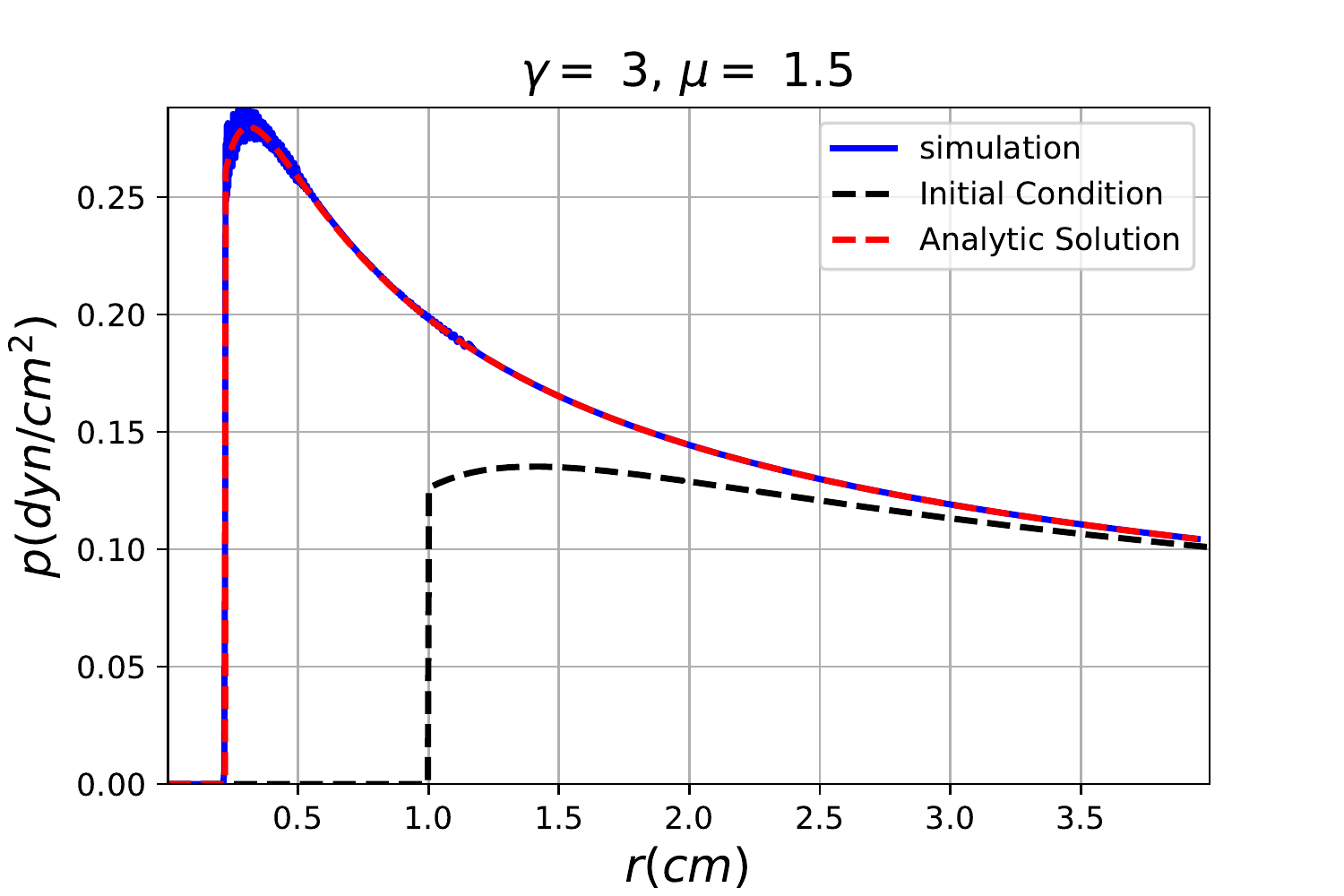}
\par\end{centering}
\begin{centering}
\includegraphics[scale=0.47]{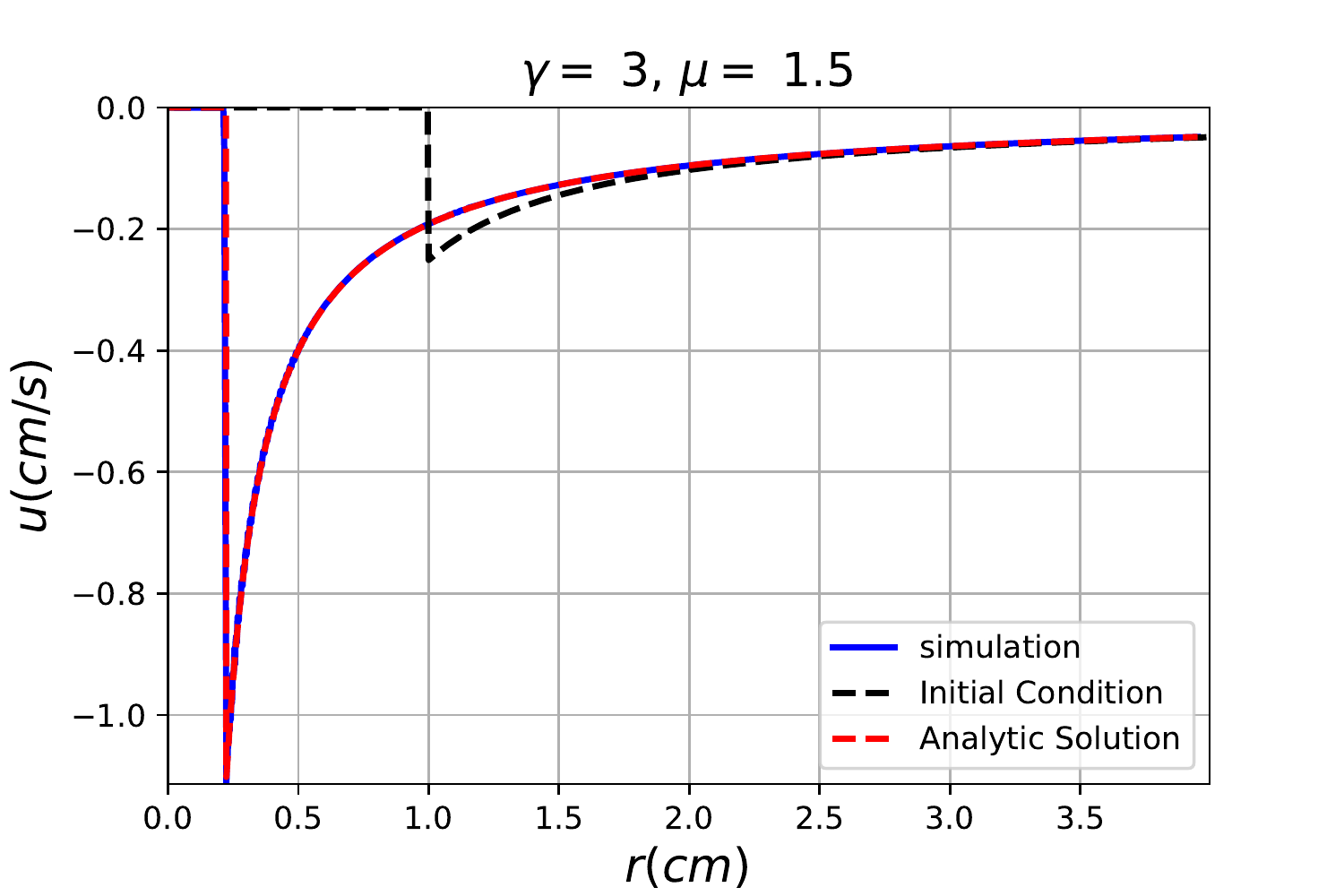}\includegraphics[scale=0.47]{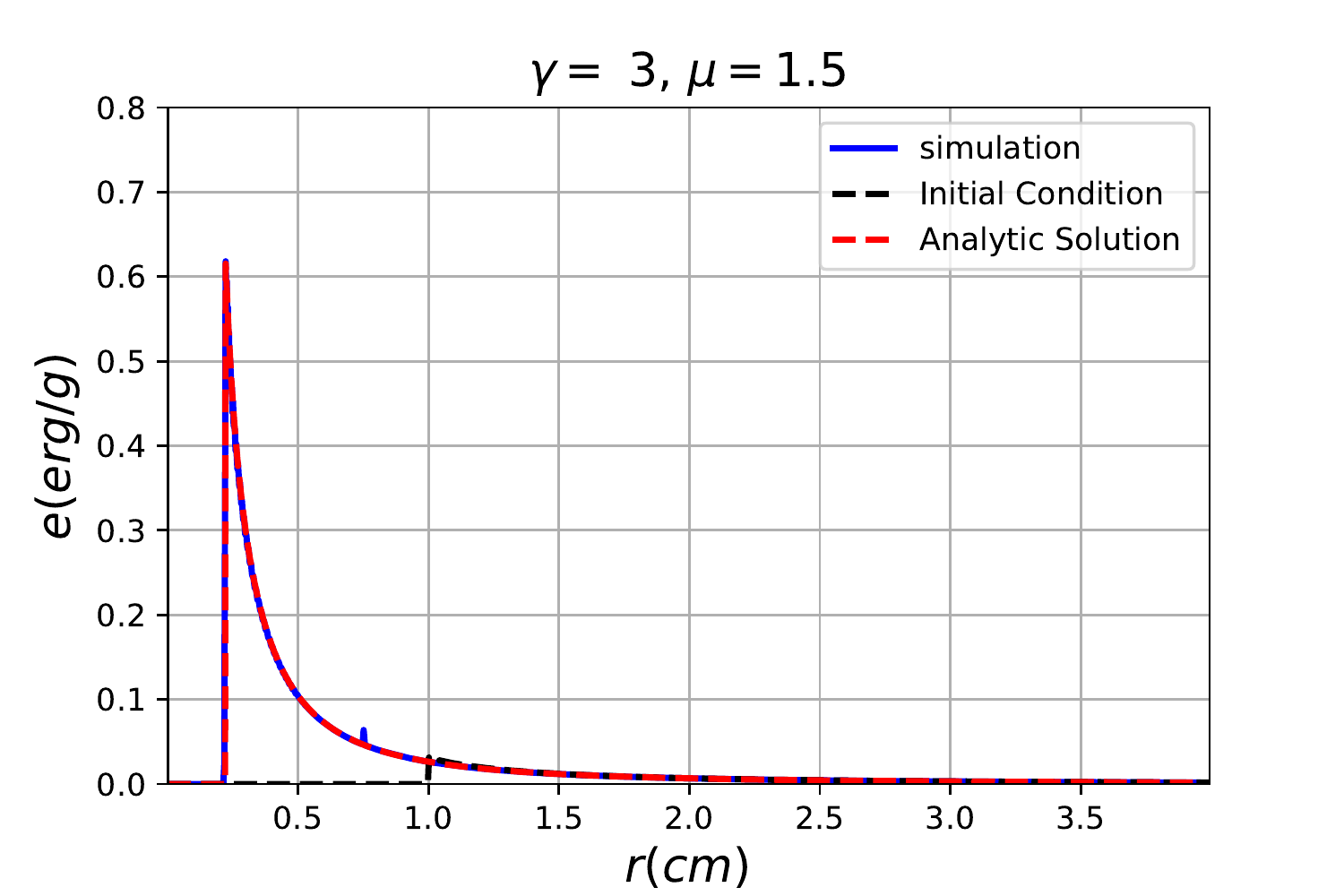}
\par\end{centering}
\caption{Same as Fig. \ref{fig:profiles}, for $\gamma=3,\ \mu=1.5$. \label{fig:profiles-1}}
\end{figure*}

\begin{figure*}[t]
\begin{centering}
\includegraphics[scale=0.47]{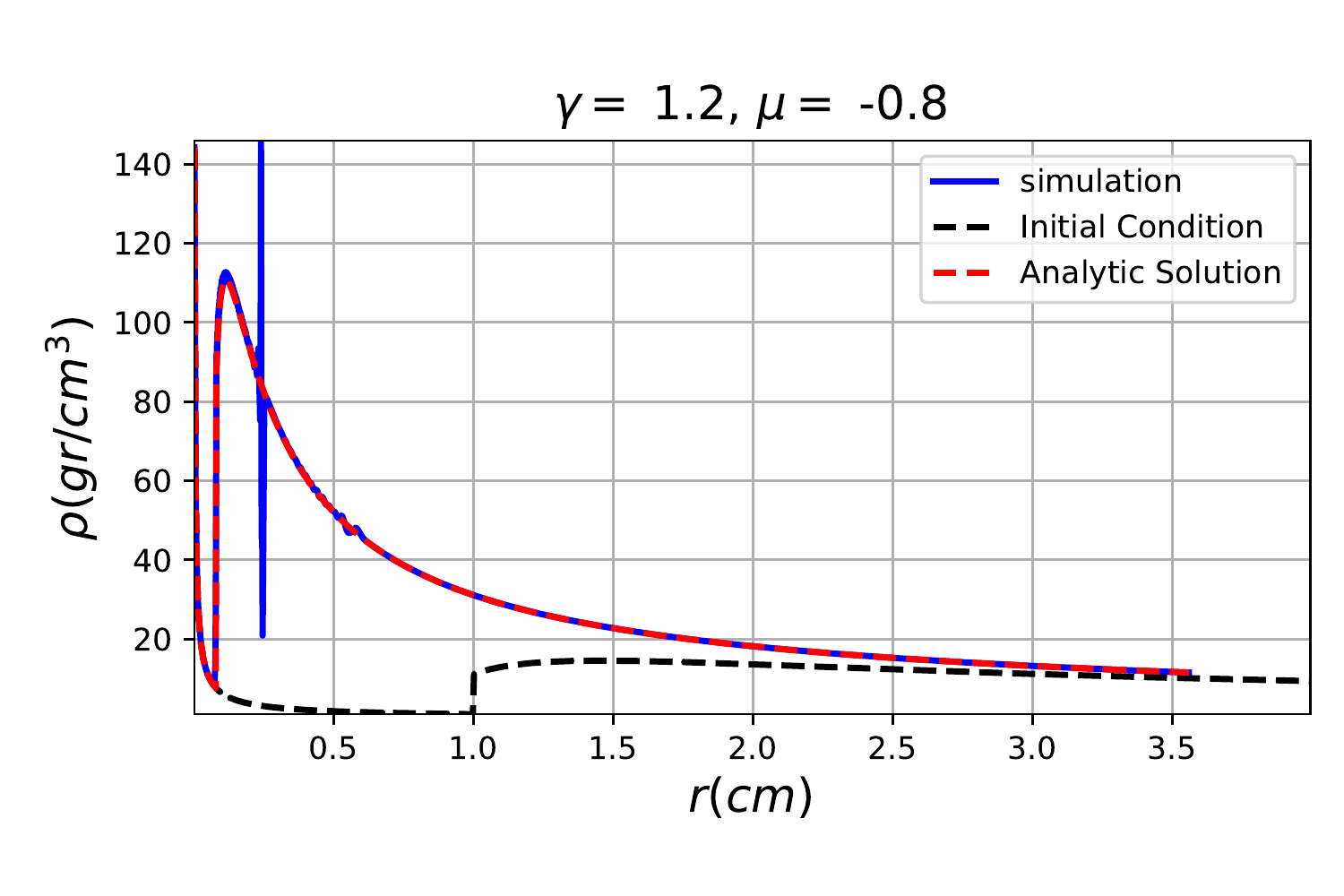}\includegraphics[scale=0.47]{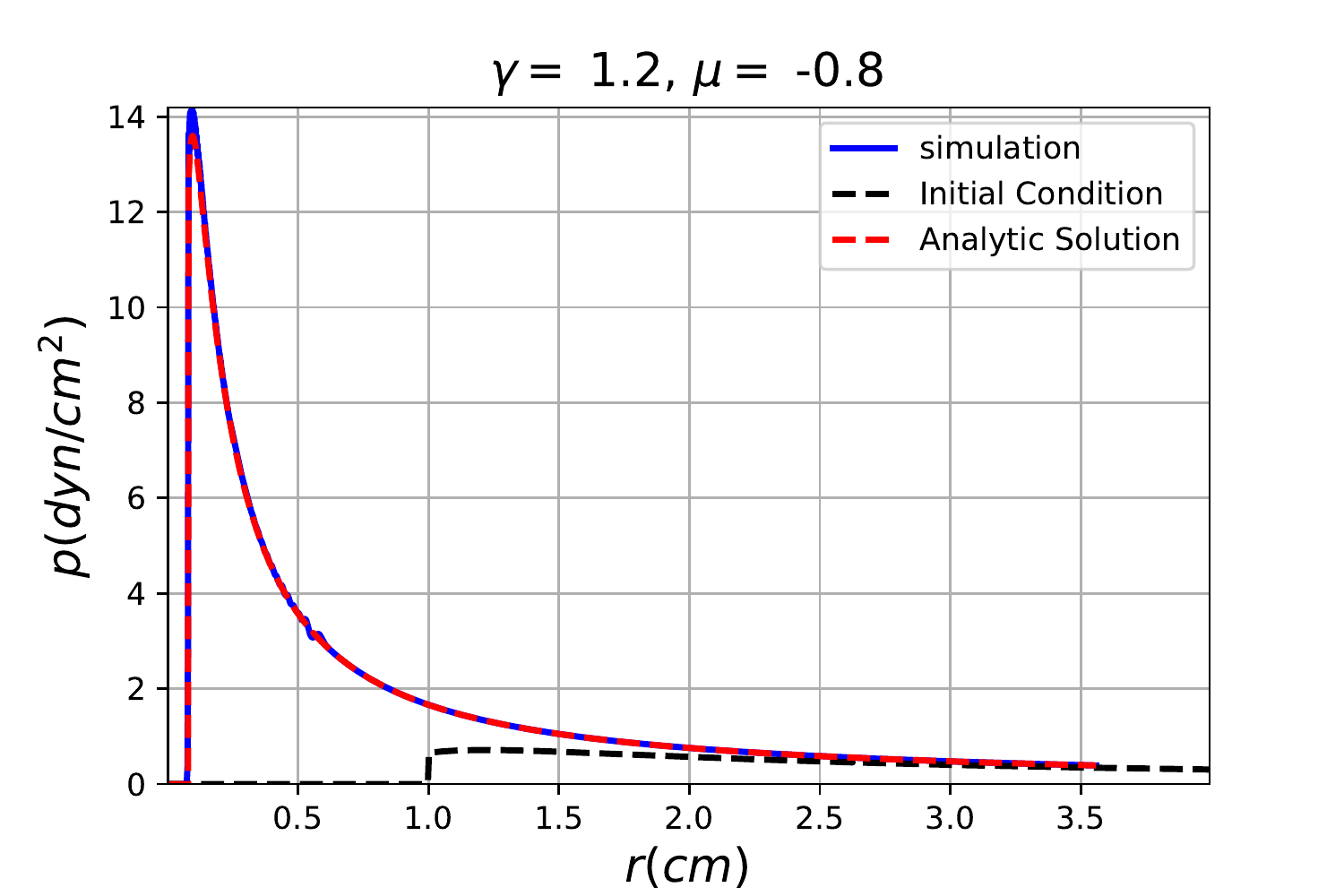}
\par\end{centering}
\begin{centering}
\includegraphics[scale=0.47]{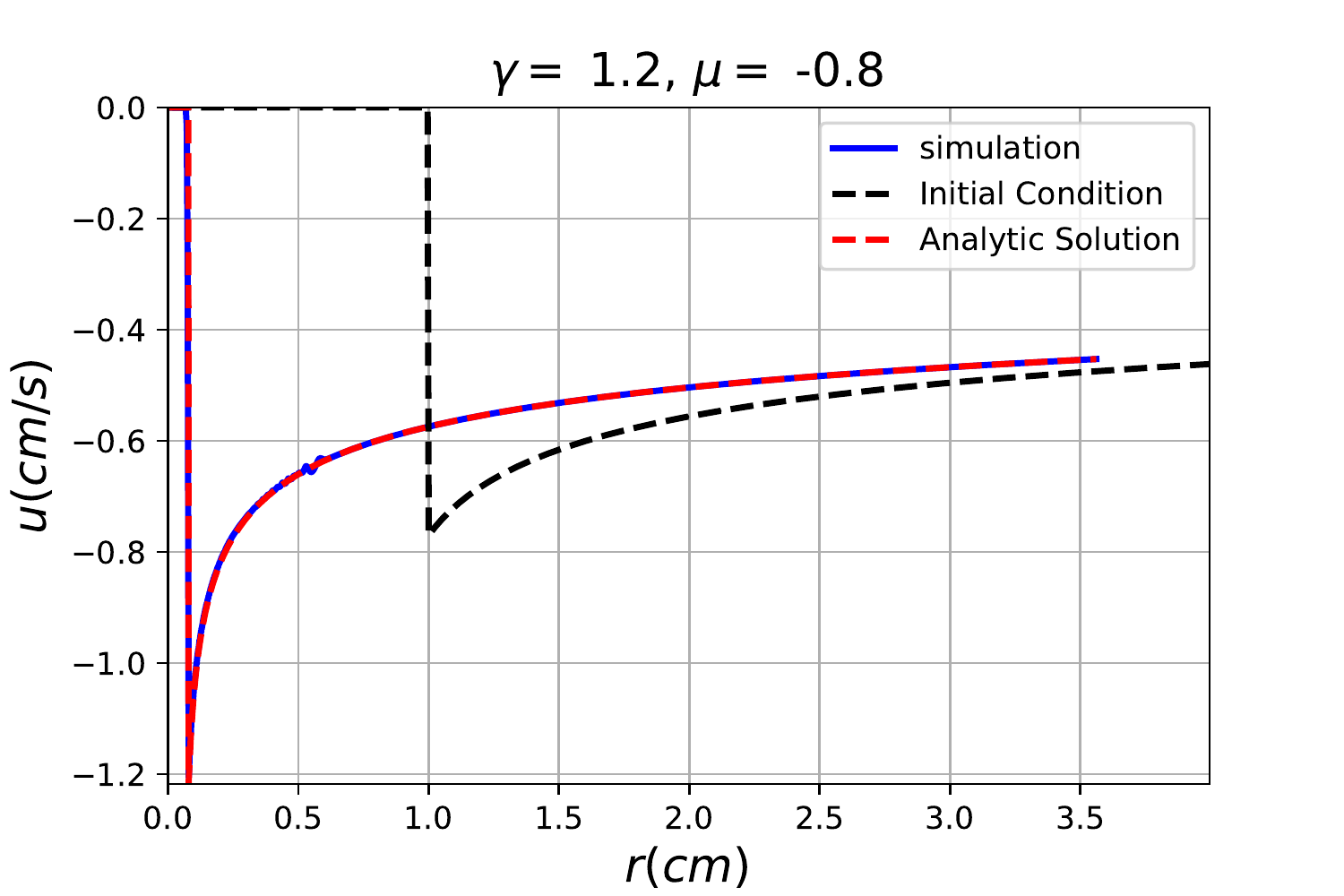}\includegraphics[scale=0.47]{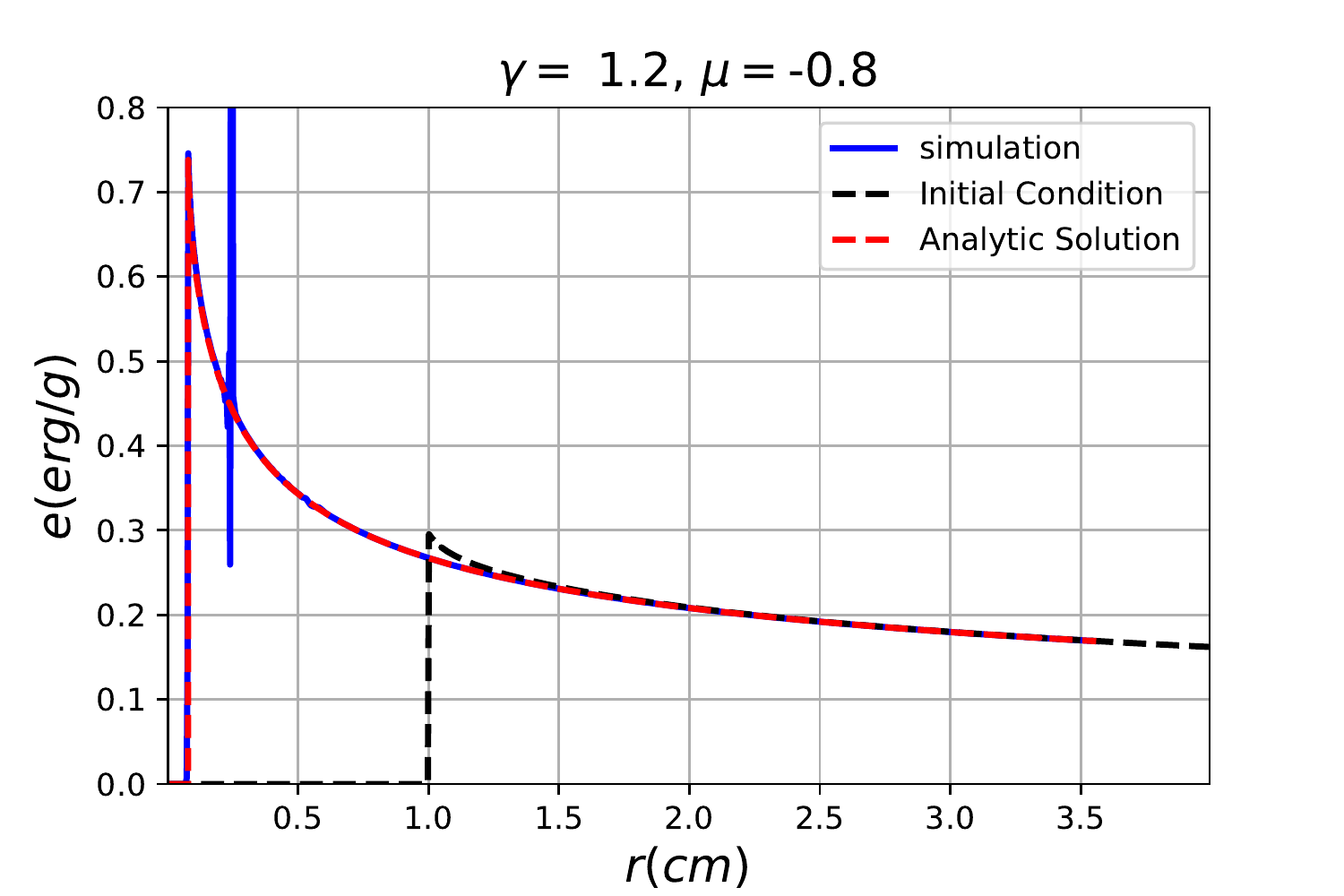}
\par\end{centering}
\caption{Same as Fig. \ref{fig:profiles}, for $\gamma=1.2,\ \mu=-0.8$ \label{fig:profiles-2}}
\end{figure*}

\begin{figure*}[t]
\begin{centering}
\includegraphics[scale=0.47]{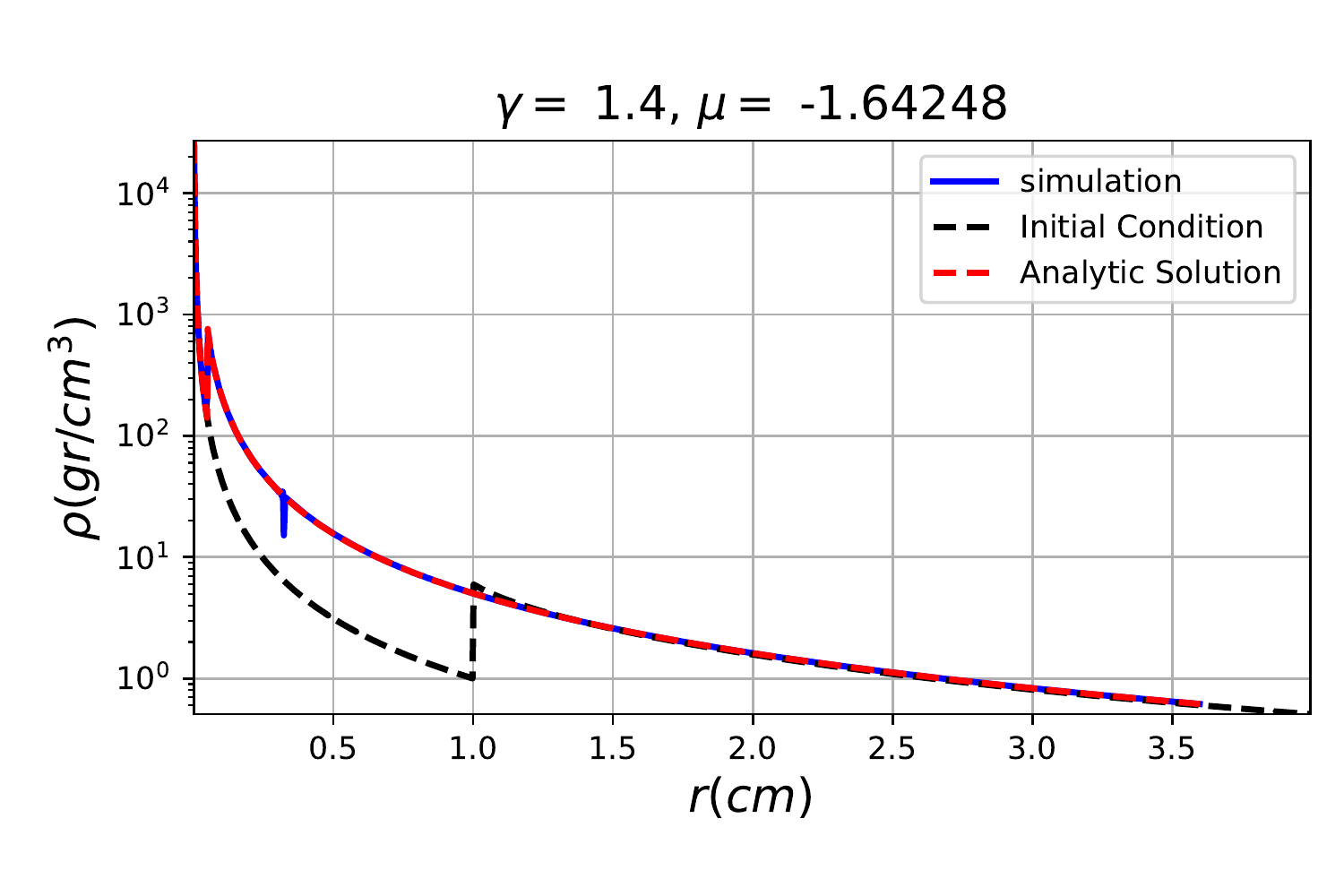}\includegraphics[scale=0.47]{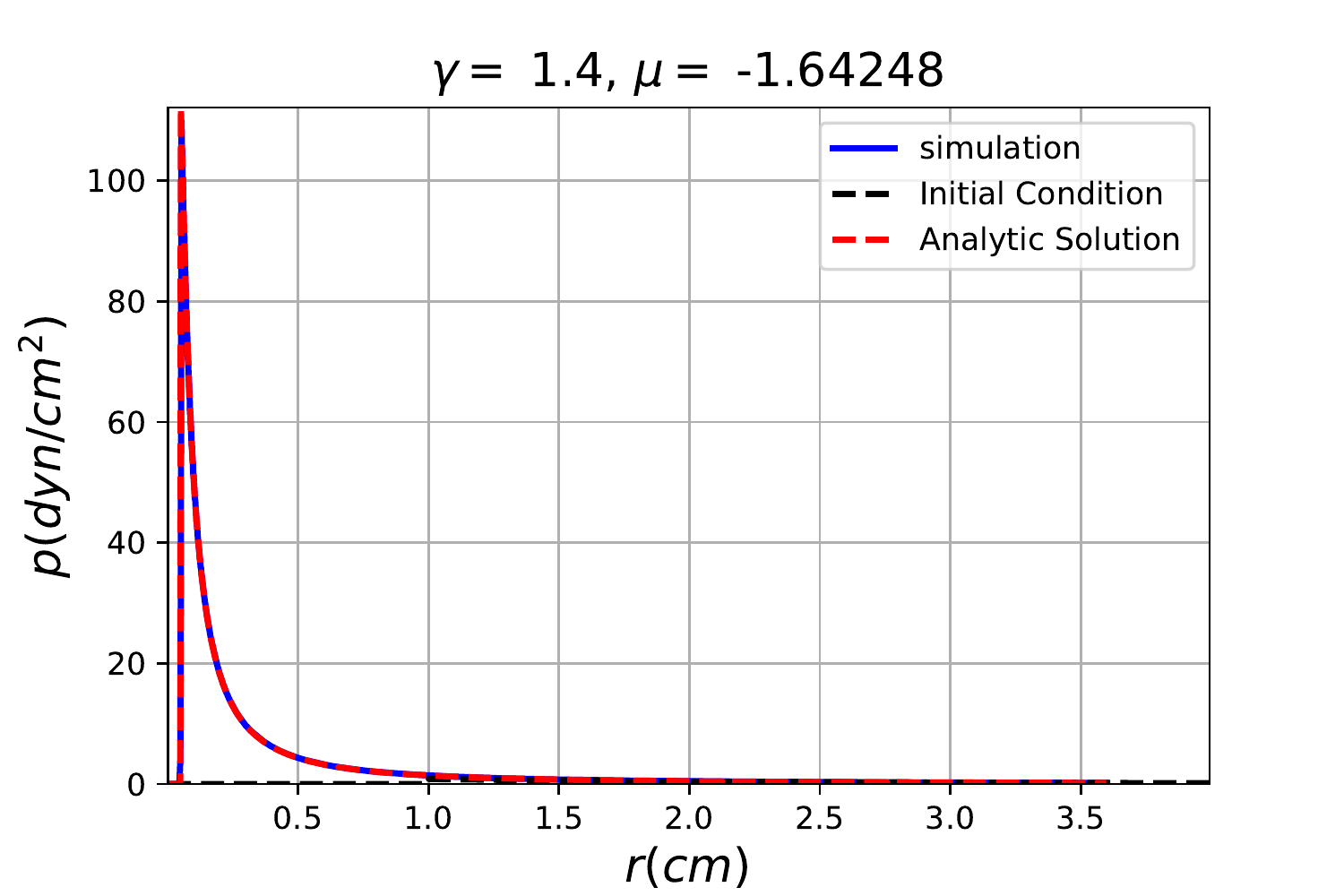}
\par\end{centering}
\begin{centering}
\includegraphics[scale=0.47]{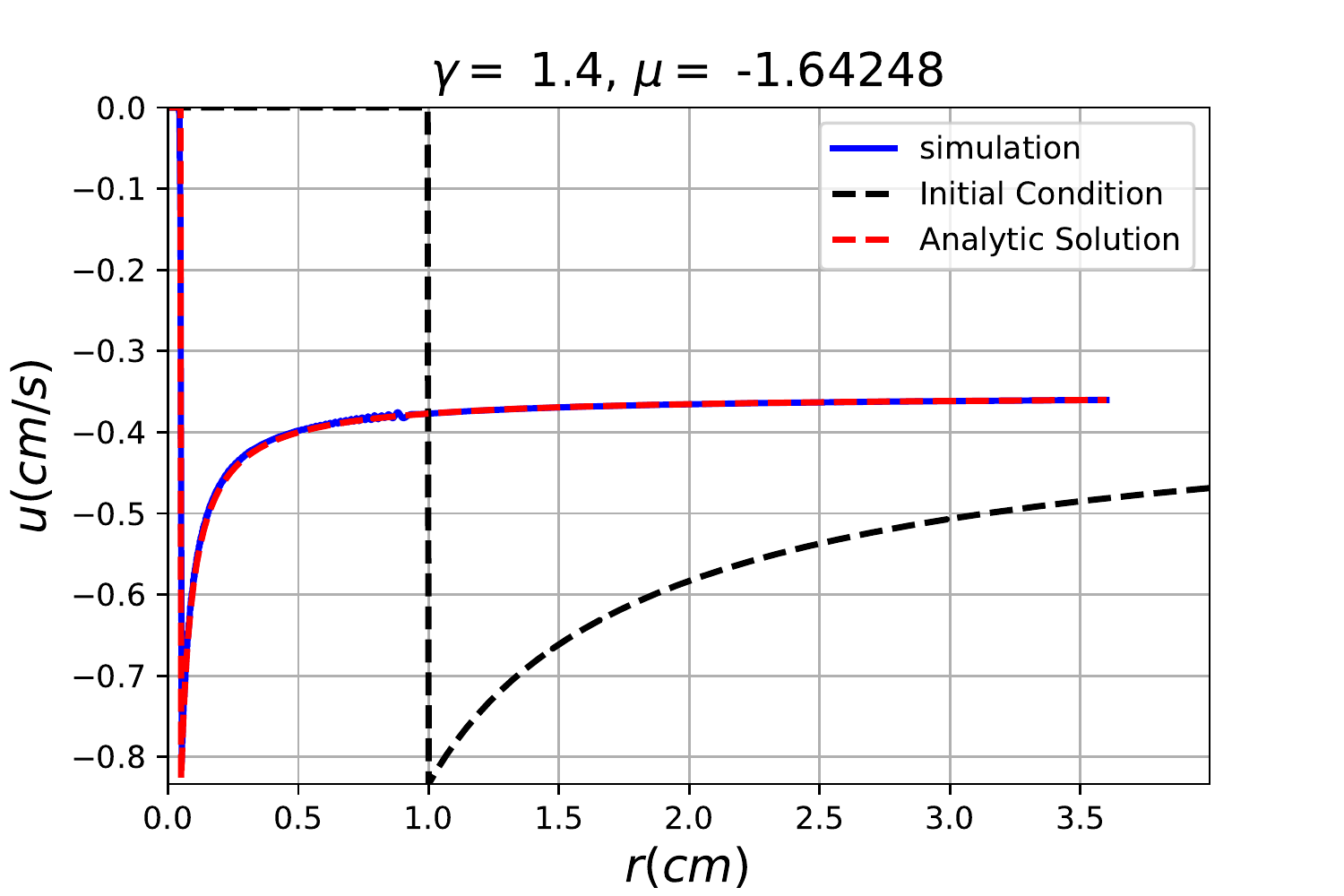}\includegraphics[scale=0.47]{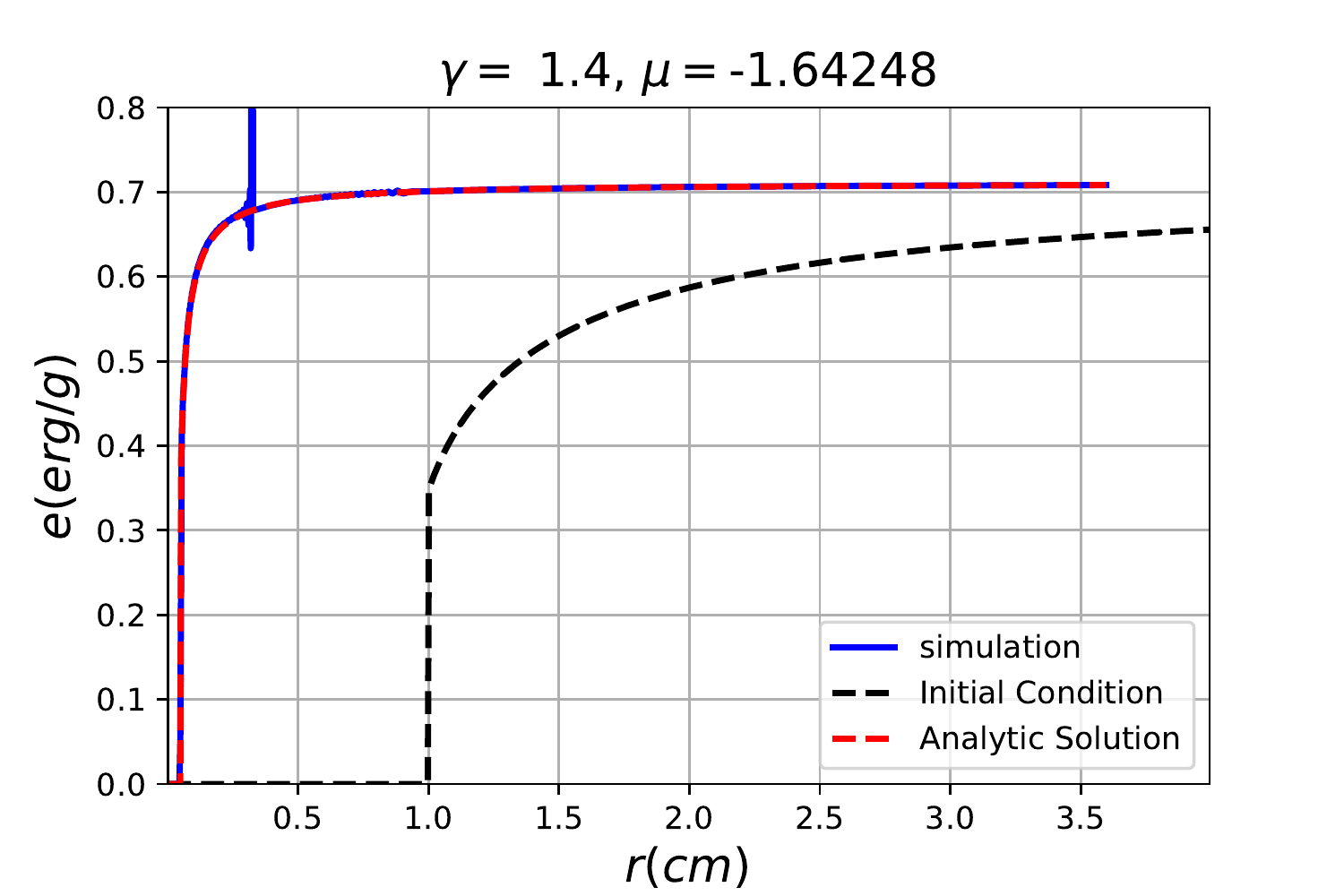}
\par\end{centering}
\caption{Same as Fig. \ref{fig:profiles}, for $\gamma=1.4,\ \mu=-1.64248$
\label{fig:profiles-3}}
\end{figure*}

\begin{figure*}[t]
\begin{centering}
\includegraphics[scale=0.47]{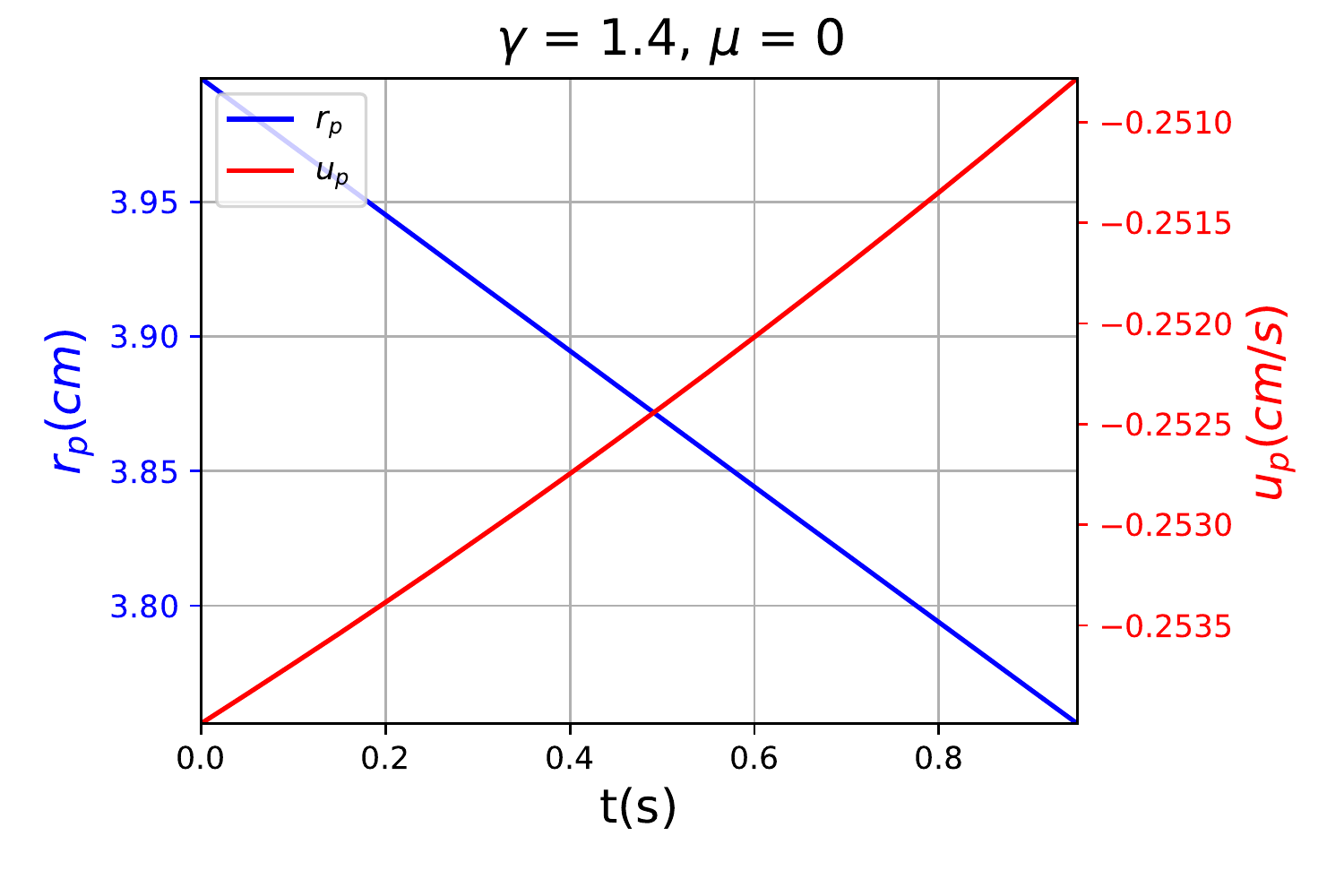}\includegraphics[scale=0.47]{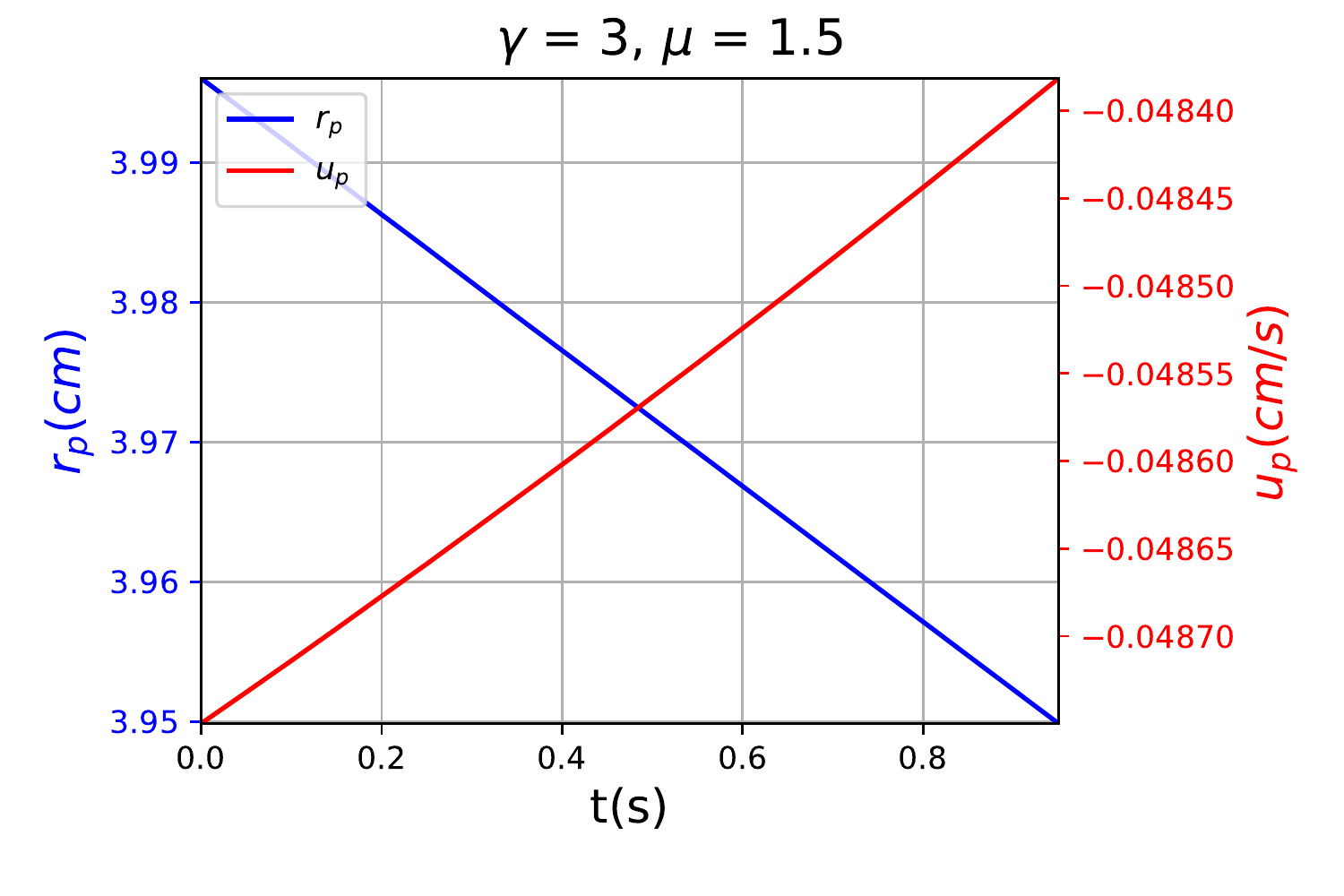}
\par\end{centering}
\begin{centering}
\includegraphics[scale=0.47]{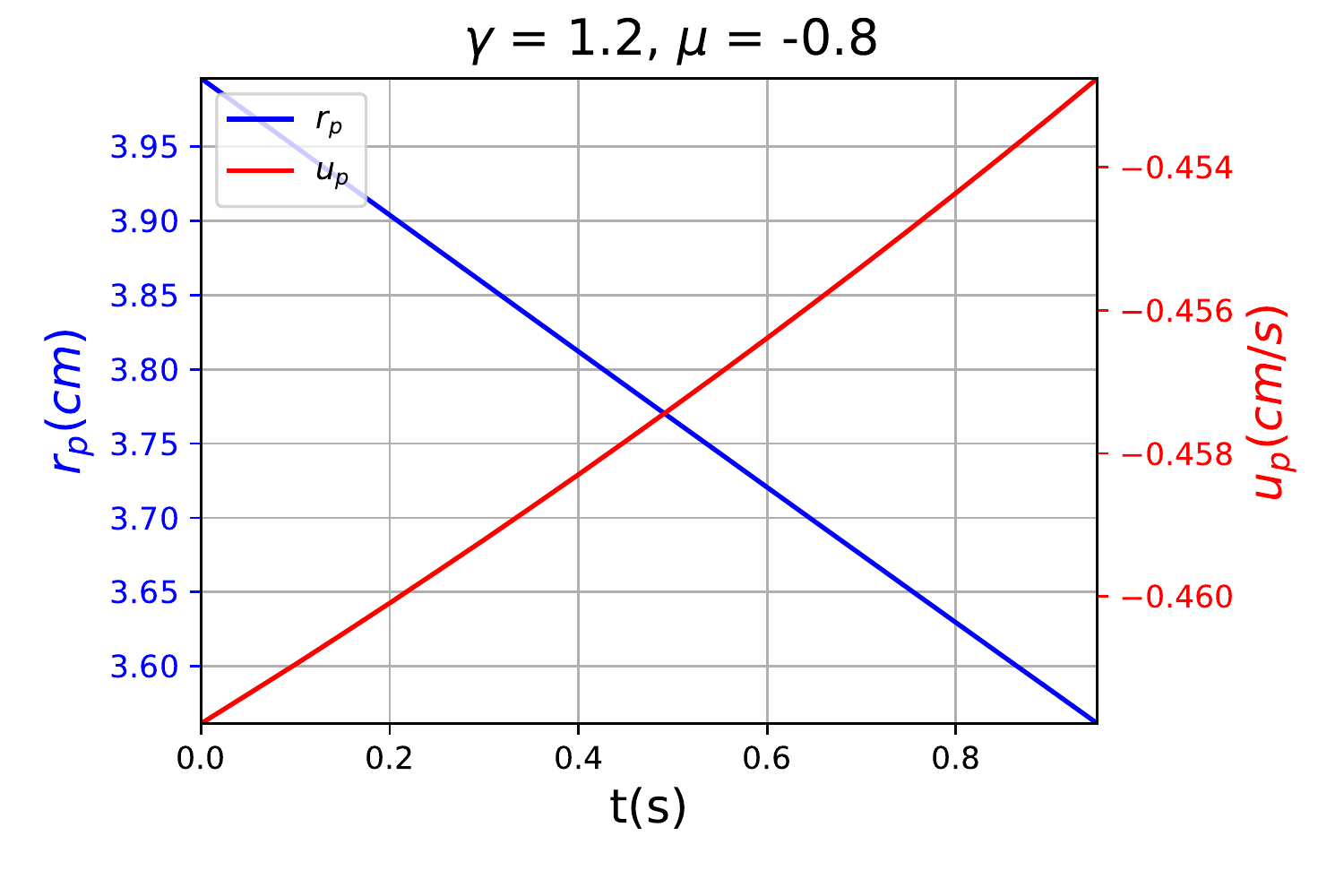}\includegraphics[scale=0.47]{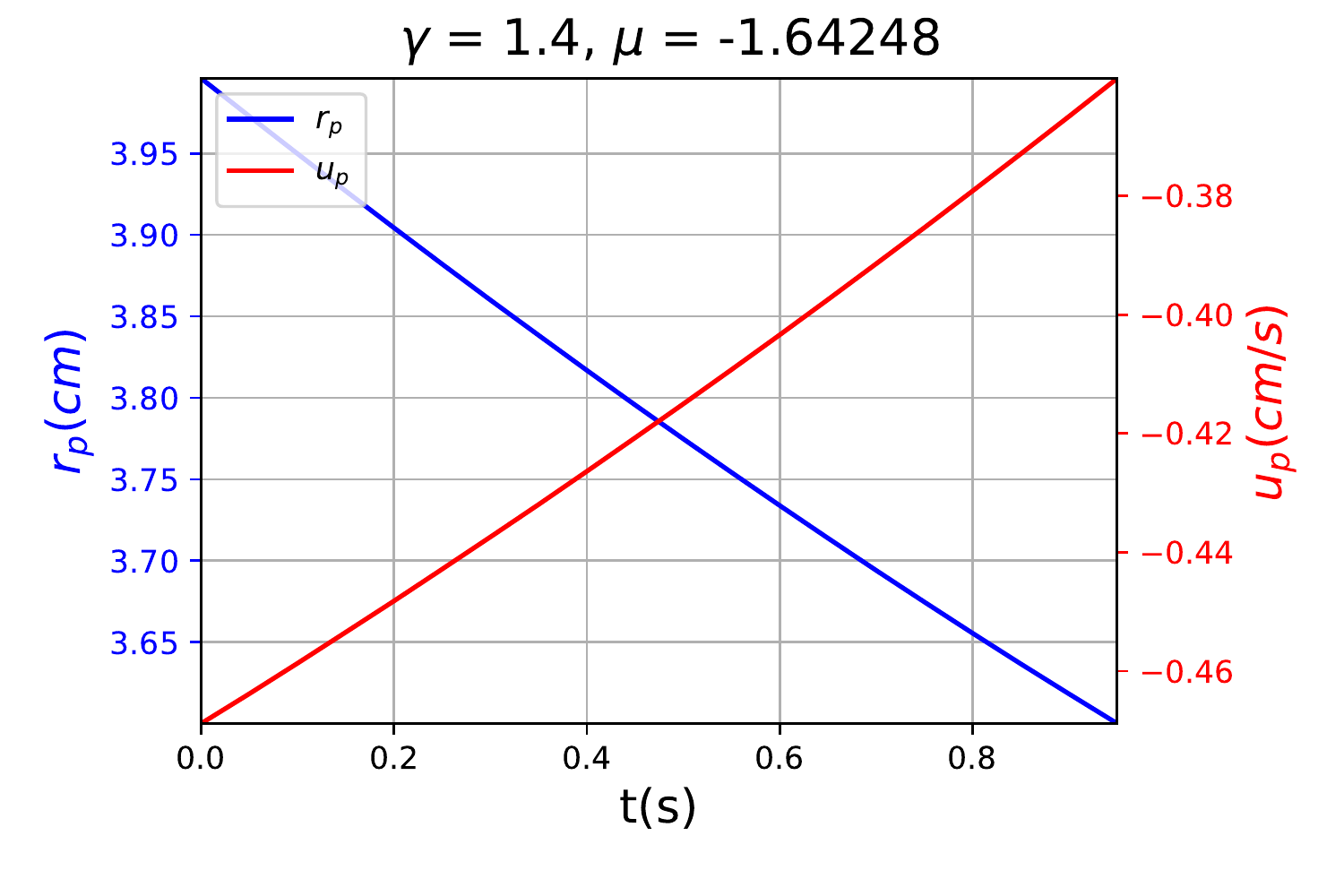}
\par\end{centering}
\caption{The piston boundary conditions as a function of time, for the four
cases detailed in the text (and also in figures \ref{fig:profiles}-\ref{fig:profiles-3}).
The piston position is given in blue on the left y axis and the piston
velocity is given in red on the right y axis.\label{fig:bc_piston}}
\end{figure*}

\begin{figure*}[t]
\begin{centering}
\includegraphics[scale=0.47]{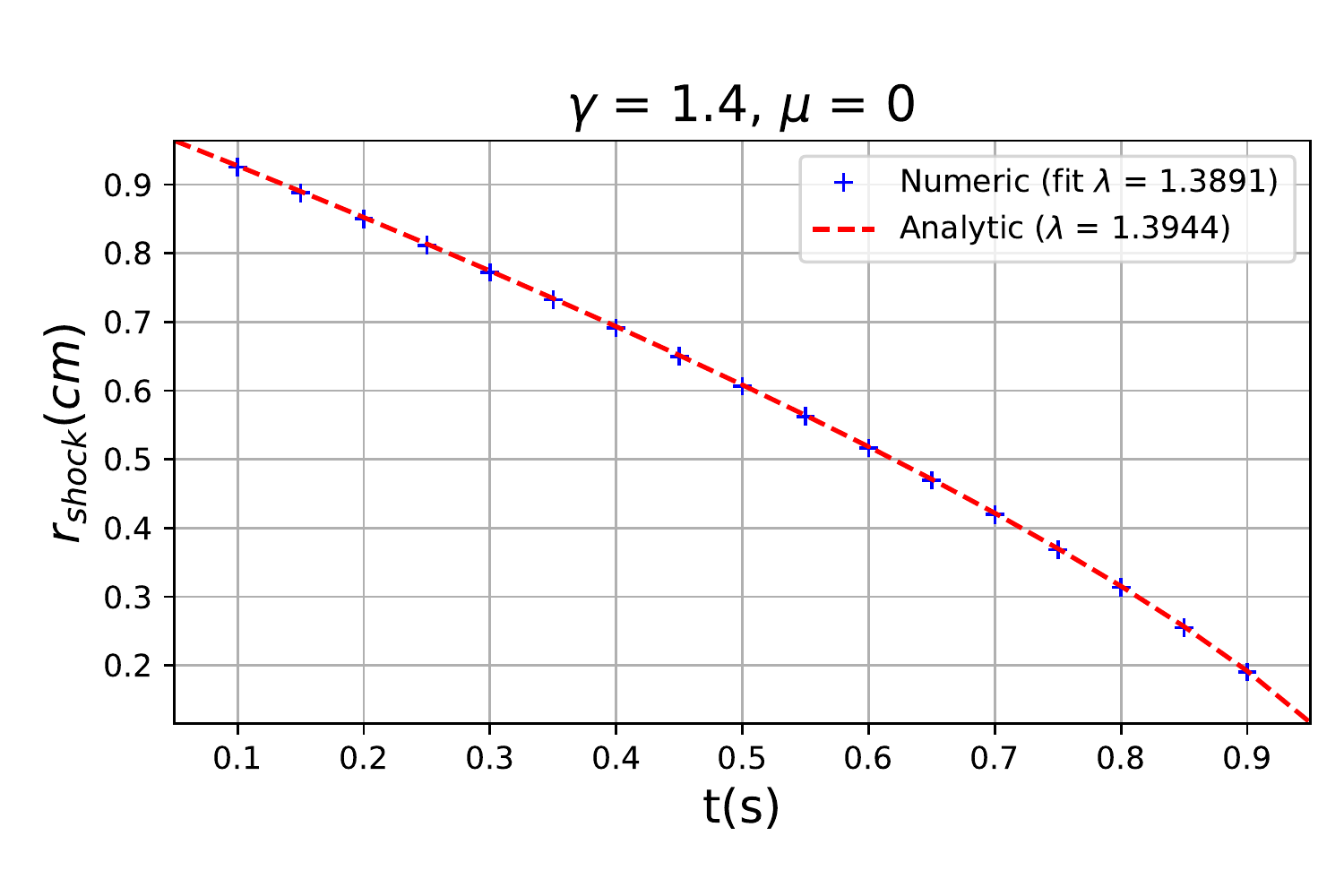}\includegraphics[scale=0.47]{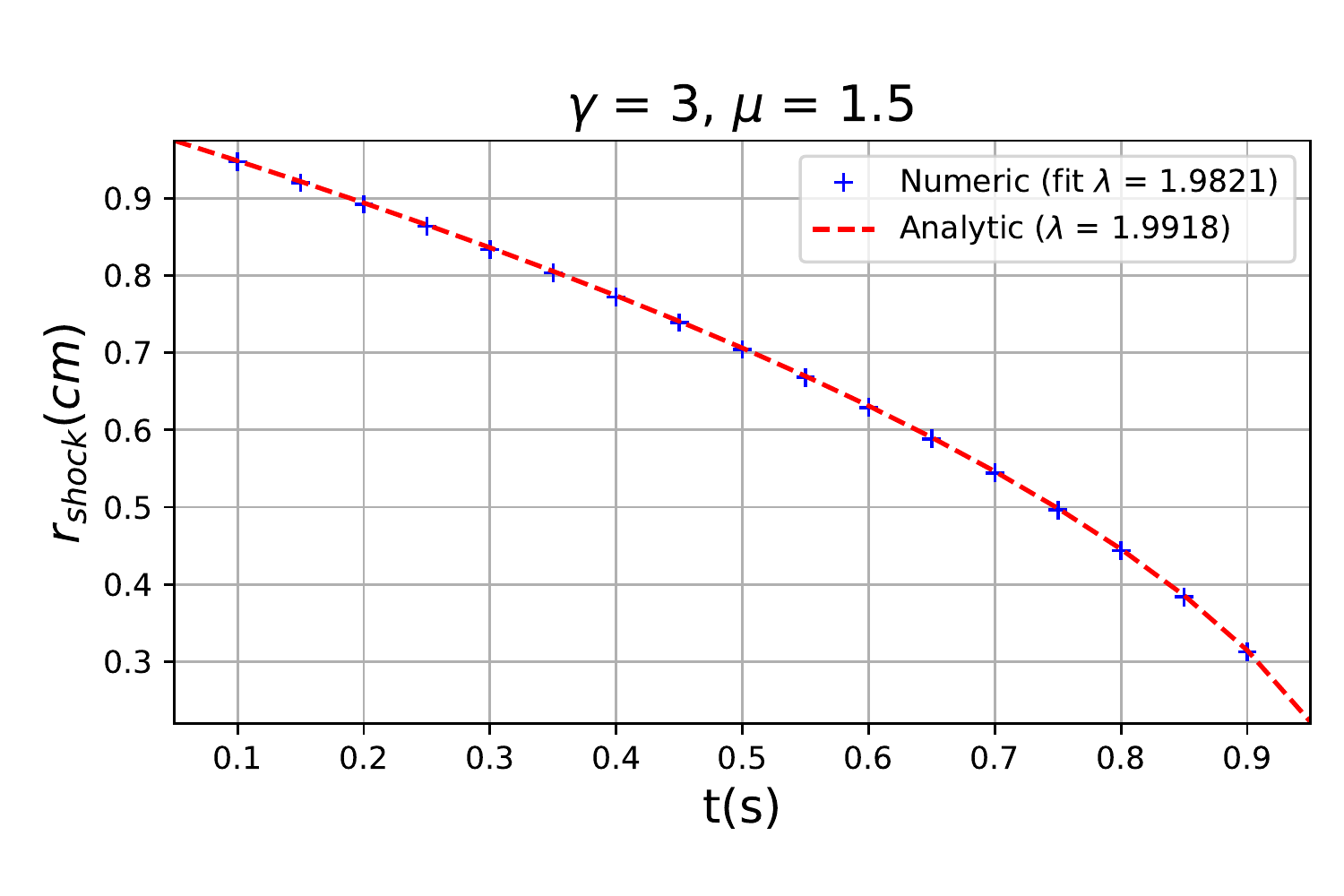}
\par\end{centering}
\begin{centering}
\includegraphics[scale=0.47]{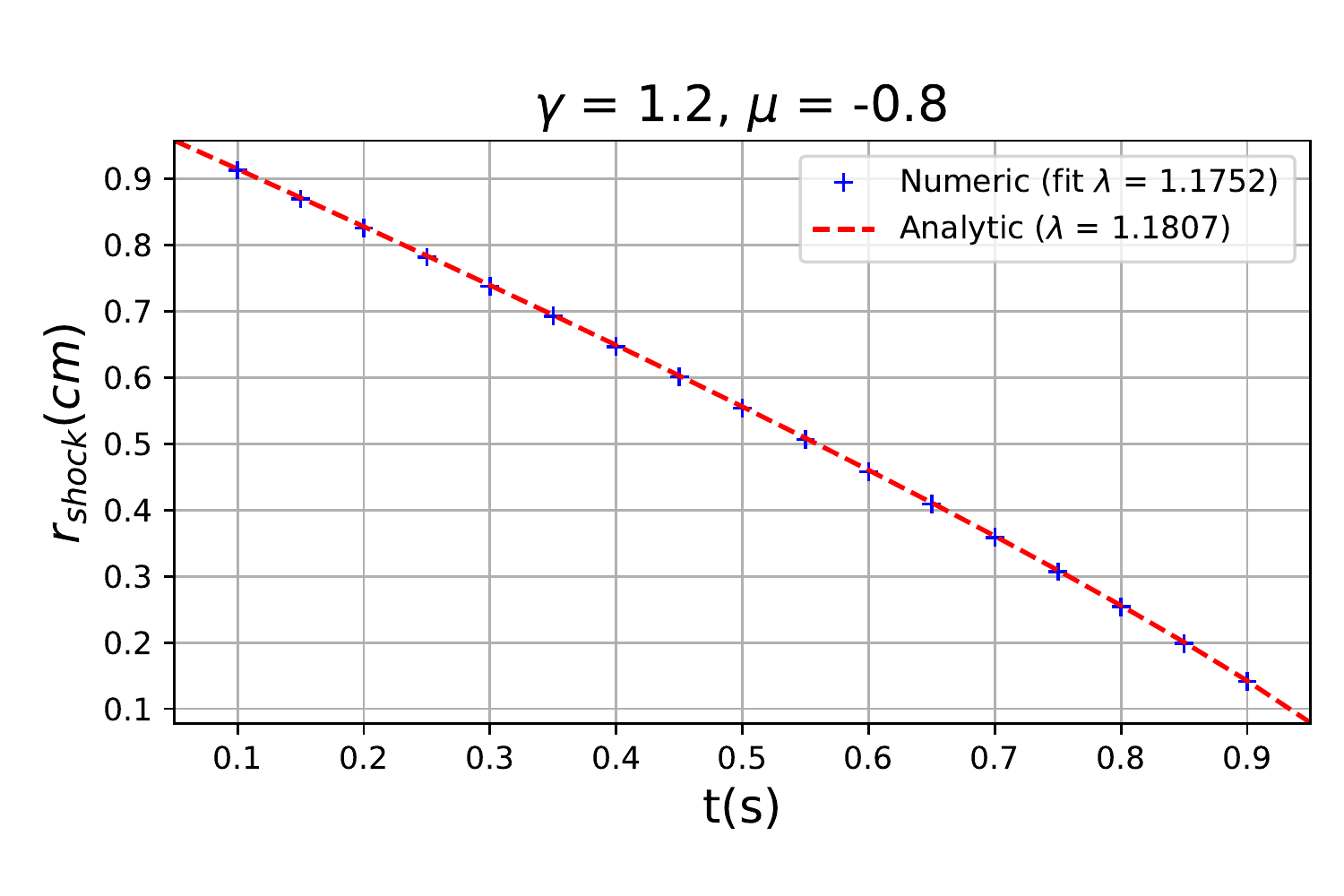}\includegraphics[scale=0.47]{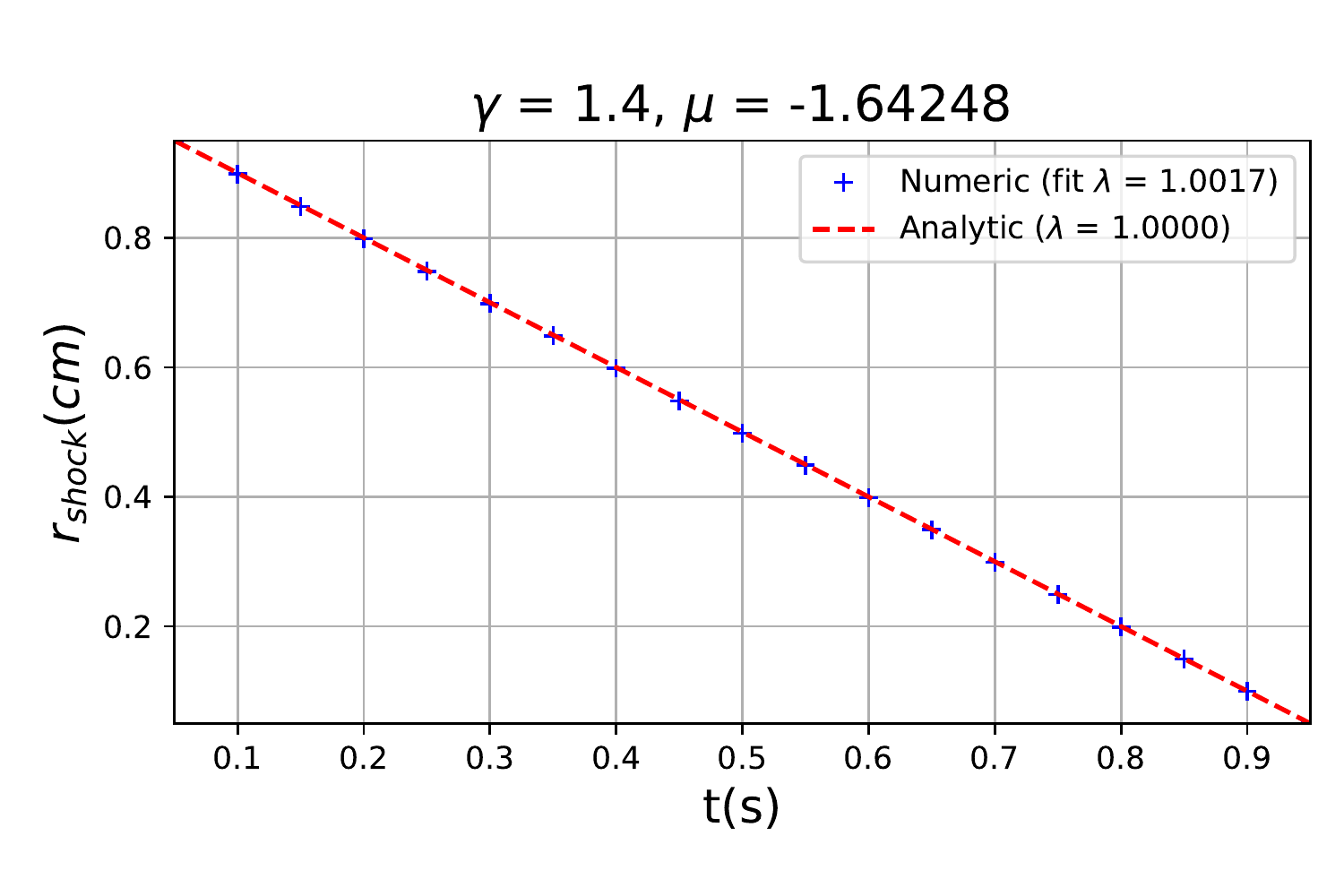}
\par\end{centering}
\caption{Comparison of the numerical (blue crosses) and analytical (red line)
shock positions as a function of time, for the four cases detailed
in the text (and also in figures \ref{fig:profiles}-\ref{fig:profiles-3}).
The numerical fitted value, and the analytic value of the similarity
exponent $\lambda$, are given in the legend.\label{fig:bc_shockpos-1}}
\end{figure*}

\begin{figure*}[t]
\begin{centering}
\includegraphics[scale=0.47]{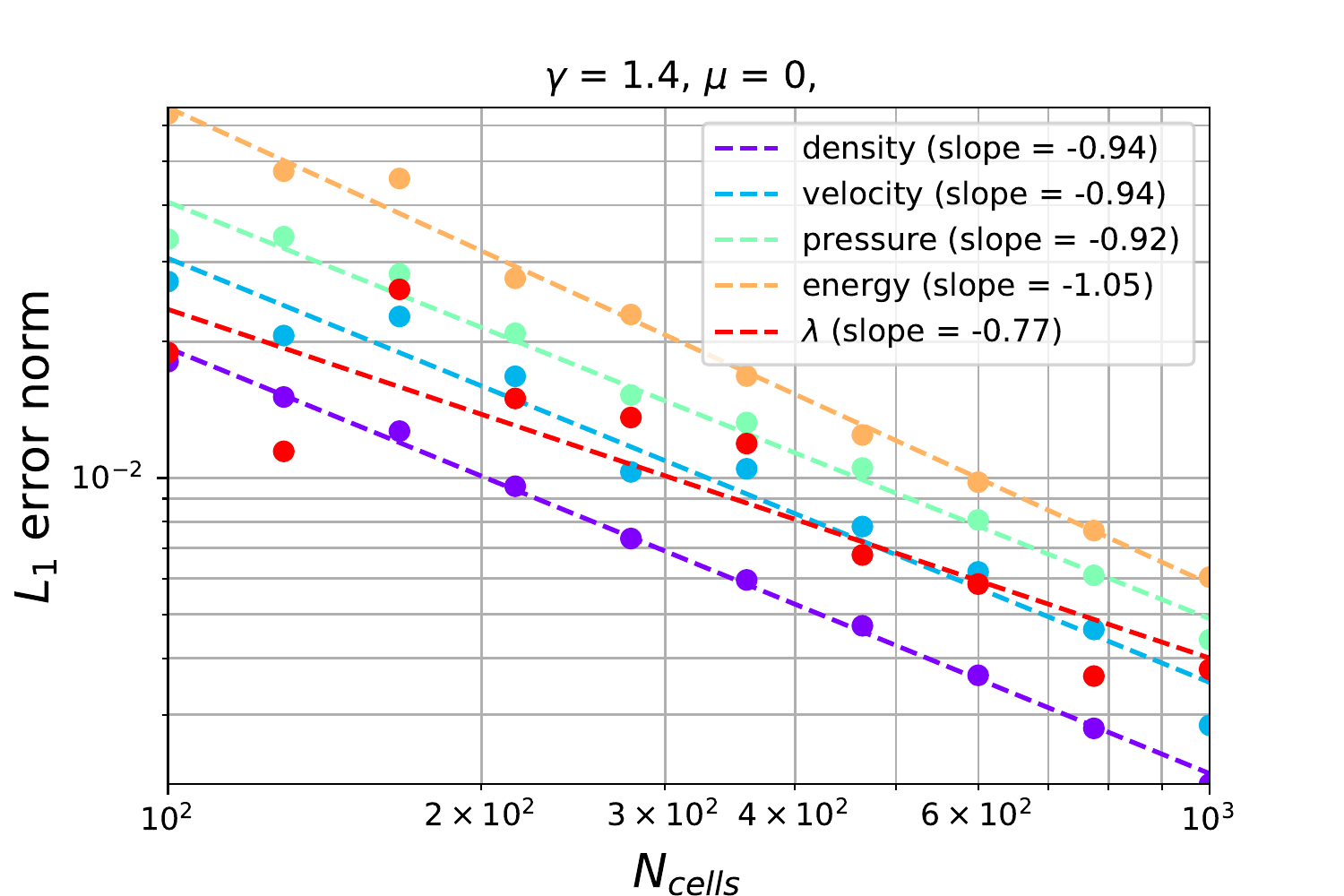}\includegraphics[scale=0.47]{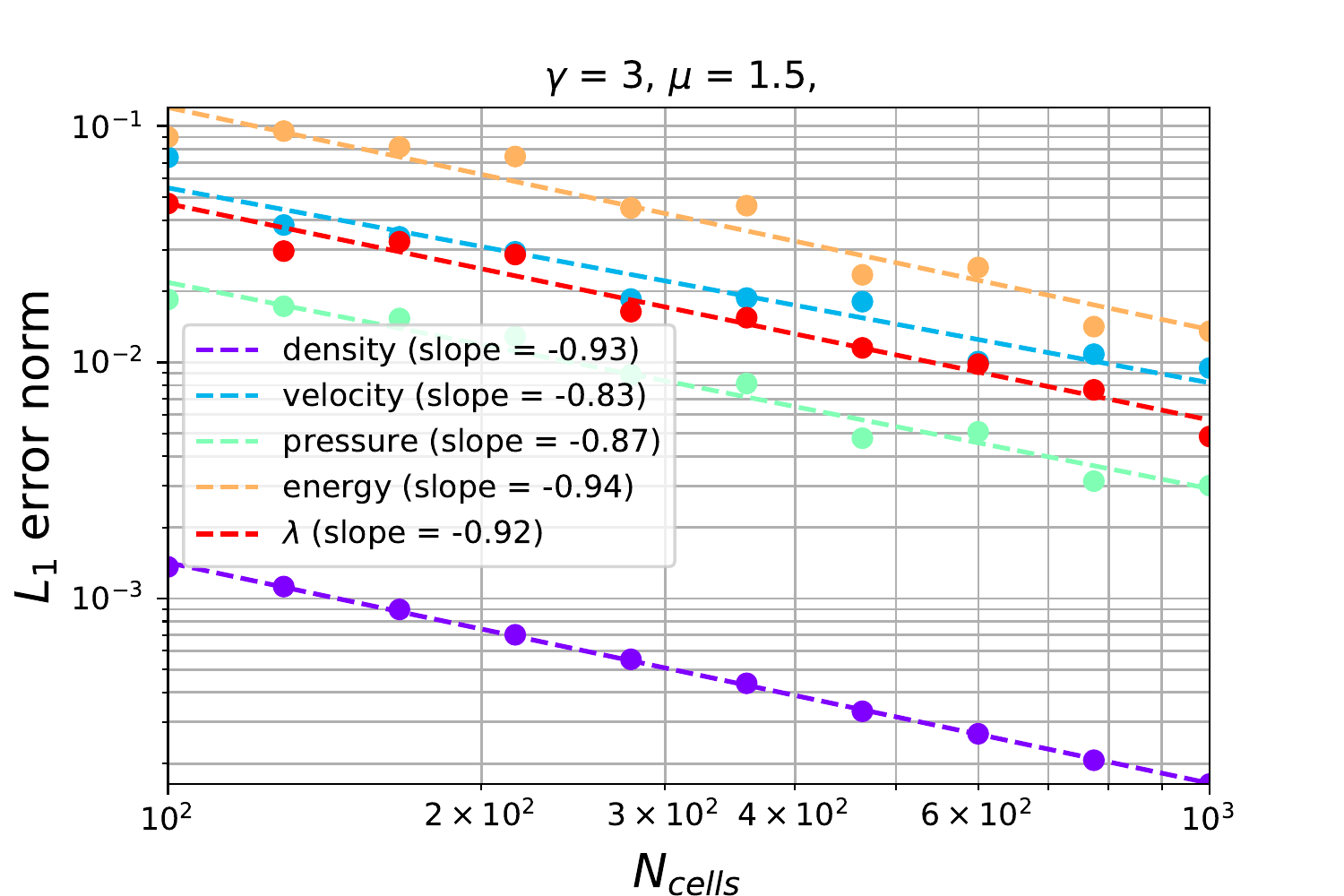}
\par\end{centering}
\begin{centering}
\includegraphics[scale=0.47]{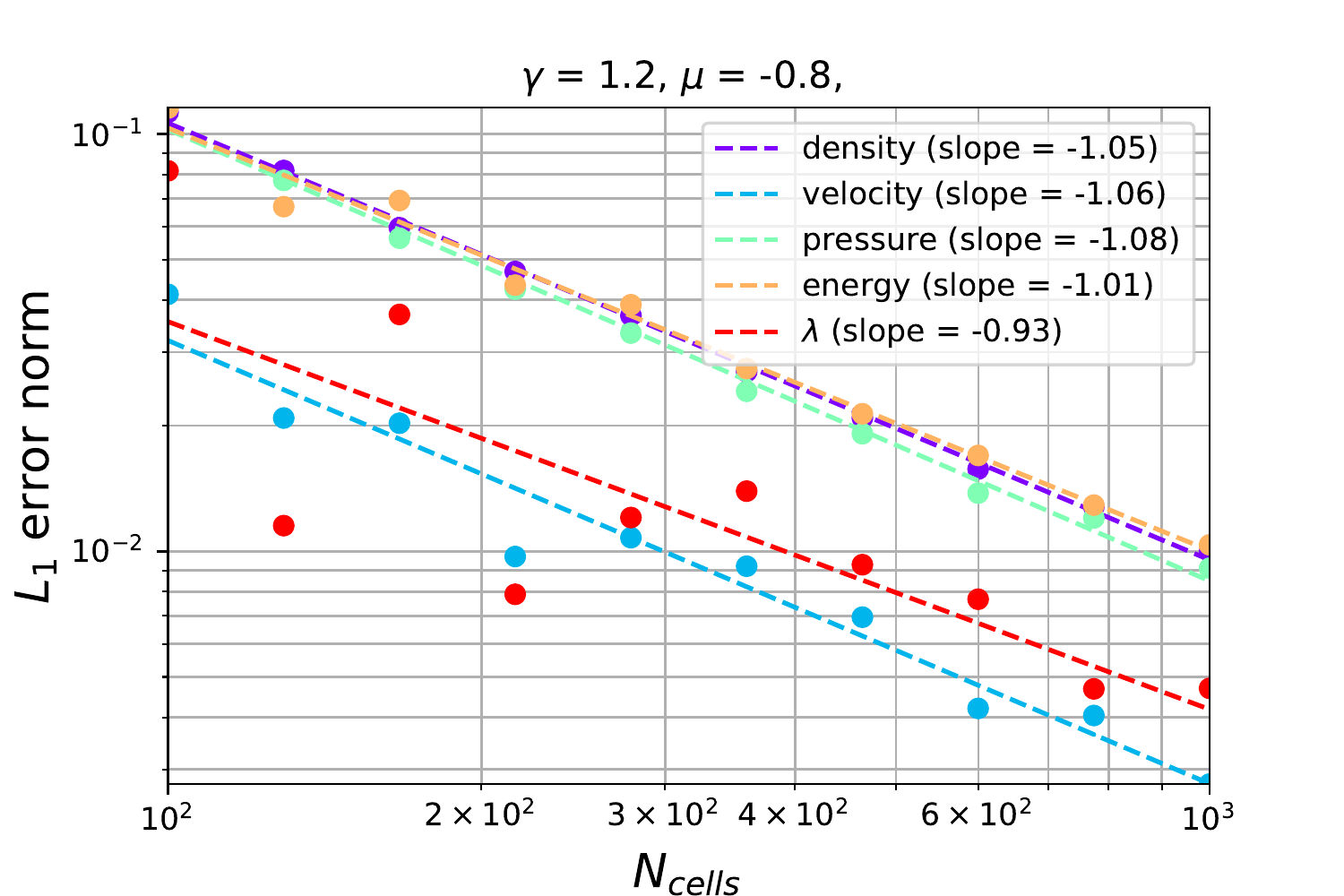}\includegraphics[scale=0.47]{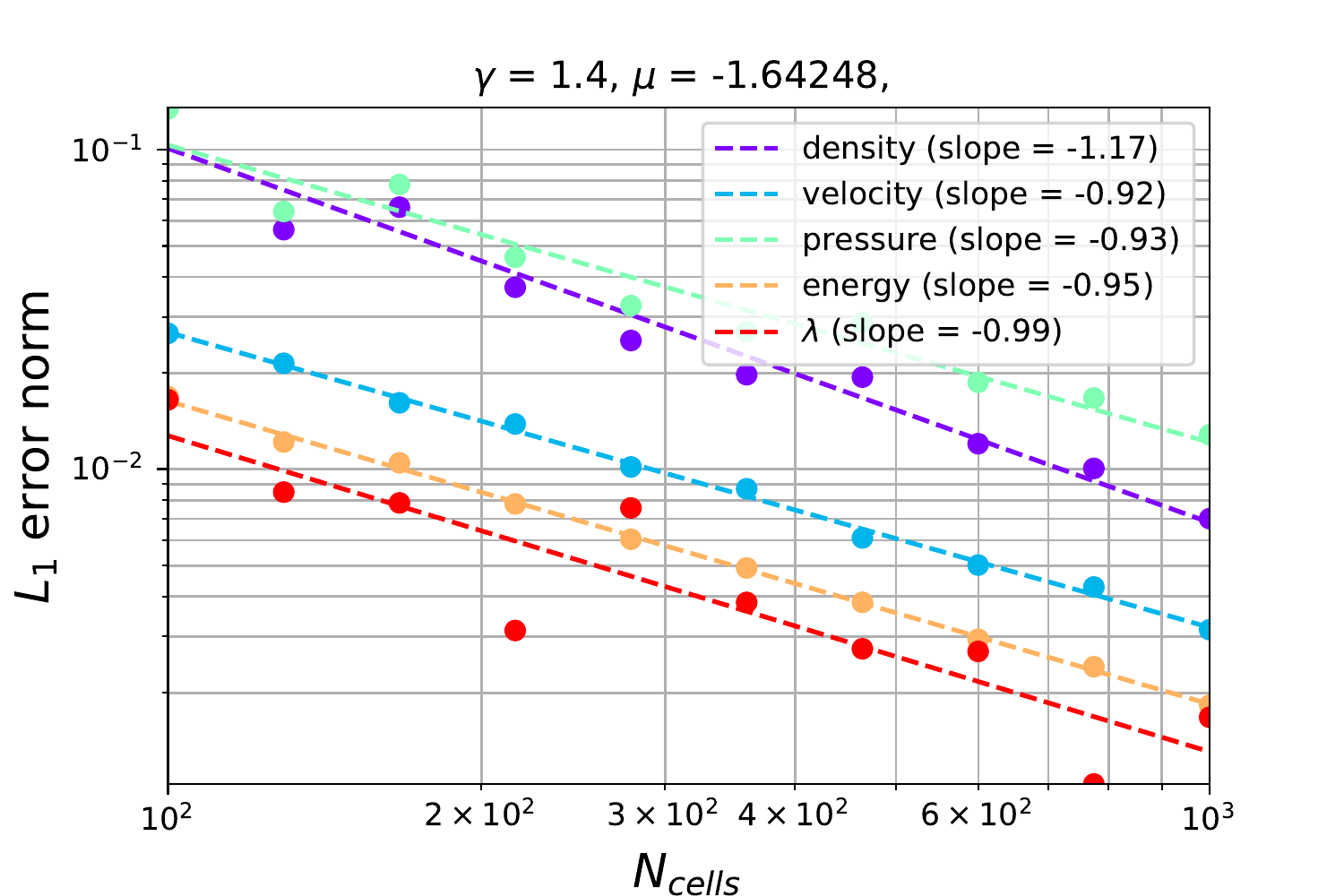}
\par\end{centering}
\caption{Convergence plots of the $L_{1}$ norm of the density (purple), velocity
(blue), pressure (green) and specific internal energy (orange), and
for the relative error in $\lambda$ (red), with respect to the analytical
solution. The convergence rates, which are obtained from the slope
of a linear fit (shown as a dashed line) on a log-log scale, are given
in the legend.\label{fig:convergence}}
\end{figure*}

We now turn to compare the semi-analytic solutions derived above with
one dimensional hydrodynamic simulations. Our goal is to demonstrate
that through a proper choice of initial and boundary conditions, the
known solutions of the imploding shock problem can serve as a test
problem for numerical codes simulating compressible flow.

\subsection{Initial and Boundary Conditions}

The Guderley problem is defined as a shock converging from infinity,
and application of a computational grid with a finite extent requires
some adaptation. In principle, the simulation can be initialized as
a Riemann problem at some large radius, with an inner region with
zero pressure and velocity, and an outer region with some positive
pressure. Indeed, such an initial setup will create a strong shock
converging towards the origin, and the flow will asymptotically approach
the analytic solution, but only up to the sonic point, where the flow
is truly self-similar (independent of the details of the initial conditions).

In order to accurately describe the flow in all points in space at
all times, we use a different approach to initialize the simulation.
This method was described by Ramsey et al. in \cite{ramsey2012guderley}
for an Eulerian hydrodynamic calculation and later in \cite{ramsey2017verification}
for a Lagrangian calculation. In this approach the simulation is initialized
with the hydrodynamic profiles (density, velocity and pressure), given
by the analytic solution for the entire spatial range at a specific
time $t_{0}$. We follow the Lagrangian method, in which a moving-piston
boundary condition at the outer radius is employed. The time dependent
velocity of the piston, $u_{p}(t)$, is given by

\begin{equation}
u_{p}\left(t\right)=u_{\text{analytic}}\left(r_{p}\left(t\right),t\right),\label{eq:pist_con}
\end{equation}
where $r_{p}\left(t\right)$ is the time-dependent position of the
piston, and $u_{\text{analytic}}$ is given by the analytic solution,
eq. \eqref{Similarity V}. Naturally, the numerical implementation
is the discrete analog of eq. \eqref{eq:pist_con}, applied at every
time step to the outermost vertex on the Lagrangian grid. 

\subsection{Results and Comparisons}

We implemented a standard one dimensional Lagrangian hydrodynamic
scheme, which is described for completeness in Appendix \ref{app:The-hydrodynamic-Scheme}.
As typical examples, in the following we present comparisons between
the simulations and the corresponding analytic solutions for the following
cases: 
\begin{itemize}
\item $\gamma=1.4,\ \mu=0$ - the standard Guderley problem, a converging
shock with a uniform initial density profile.
\item $\gamma=3,\ \mu=1.5$ - a case study for an initial density profile
that goes to zero towards the center. 
\item $\gamma=1.2,\ \mu=-0.8$ - a case study for an initial density profile
that diverges at the center. 
\item $\gamma=1.4,\ \mu=-1.64248$ - a particular choice for which $\lambda=1$,
corresponding to a constant shock velocity. 
\end{itemize}
All simulations where performed in spherical symmetry, using a spatial
grid of size $N=1000$, initialized using the analytical Guderley
profiles at time $t=-1$ (where the shock is at $r=1)$, and advanced
until the final time $t=-0.05$. The position and velocity of the
piston throughout the simulations for the four examples are also shown
in figure \ref{fig:bc_piston}.

We begin by visually comparing the numerical and analytical solutions.
Comparisons of the analytical and numerical results for the flow profiles
are presented for each case in figures \ref{fig:profiles}-\ref{fig:profiles-3}.
Shown are the hydrodynamic primitive variables: density, pressure,
velocity and specific internal energy, at the final time for the numerical
and analytic solutions in each of the four cases. A very good quantitative
agreement is achieved in all four examples. For comparison, the figures
also include the initial conditions for each of the physical quantities.
Notably, a sharp discontinuity appears in the profiles in all four
cases. The discontinuity exists over a single computational cell,
while the overall results of the simulations do correspond with the
analytic solution on either side of the discontinuity. This phenomena,
which occurs also in Eulerian simulations \cite{ramsey2012guderley},
is due to the discontinuity in of the initial profiles \cite{leveque2002finite},
and does not affect the overall stability and accuracy of the simulations.
A detailed analysis of such initialization errors is given in Ref.\textbf{
}\cite{ramsey2012guderley}. The shock position as a function of time
in the simulations is shown along with the analytic results for the
four examples, in figure \ref{fig:bc_shockpos-1}. An excellent fit
is found to exist throughout the simulations, confirming the self-similar
solutions derived in this work.

Several error measures can be applied to quantitatively assess the
accuracy of our derivations, as well the numerical convergence of
the simulations. Here we consider the relative $L_{1}$ error measure
for the profiles at the end of the simulations, defined by 
\begin{equation}
L_{1}=\frac{\sum_{k}\left|y_{k}-y_{k}^{\text{analytic}}\right|}{\frac{1}{2}\left(\sum_{k}\left|y_{k}\right|+\sum_{k}\left|y_{k}^{\text{analytic}}\right|\right)}\label{eq:L_1}
\end{equation}
where $y$ and $y^{\text{analytic}}$ denote the numerical and analytical
values, respectively, of the physical quantities. The index $k$ denotes
the cell or vertex in the simulation, for which the analytic solution
is derived through the physical location of the $k$-th cell/vertex.
The quality of numerical convergence in this measure for the four
simulations are shown in Fig. \ref{fig:convergence}, plotted as a
function of the number of numerical cells, ranging from $100$ to
 $1000$. While there exists some diversity in the accuracy of the
simulations regarding the different physical quantities, we observe
that a grid of 1000 cells will generally suffice to ensure an accuracy
of a few percent or even less than one percent in all physical quantities
at the end of the simulation. Accuracy is eroded, of course, for smaller
grids, increasing the error to the order of ten percent in the least
accurate quantities (usually the internal energy or the pressure).
Notably, the convergence rates (reduction of error as a function of
increase in grid size) is similar for all physical quantities in all
simulations. These rates are all of order unity (see explicitly in
the figures), as is to be expected for an artificial viscosity numerical
scheme, which is first order accurate in the presence of shocks.

Another integral measure of relative error is the difference between
the analytical value of similarity exponent and it numerical counterpart,
derived by fitting the numerical shock position as a function of time
to a power law of the form of eq. \eqref{eq:rshock}. This error measure,
$\left|\lambda_{\text{numerical}}/\lambda_{\text{analytic}}-1\right|$,
is also depicted in Fig. \ref{fig:convergence}, and we find that
it is generally as indicative of numerical convergence as the $L_{1}$
measure for the physical quantities.

\section{Summary \label{sec:summary}}

In this work we studied the Guderley problem of a strong shock imploding
in an ideal gas medium with an initial power-law density profile,
$\rho(r)\sim r^{\mu}$. We developed and reviewed the theoretical
framework required to construct self-similar solutions for the flow.
These solutions were systematically compared to numerical simulations
employing a one dimensional Lagrangian hydrodynamic code, and using
appropriate initial and boundary conditions.

From the physics stand point, we presented a first survey of the imploding
shock problem for a wide range of parameters, notably the adiabatic
constant $\gamma$ and the power of the initial density profile, $\mu$,
including $\mu<0$. Our results are in excellent agreement with previous
works with $\mu=0$, and with the few published cases which considered
$\mu>0$ only.

We demonstrated how the semi-analytic solution can be used to initialize
a nontrivial compressible flow problem which can serve for code verification.
In particular, we find that the numerical solution provides near-linear
convergence in terms of the $L_{1}$-norm, which bodes well for physical
problems where high-accuracy is required to asses the outcome, such
as double detonations in white dwarfs, sonoluminiscene and inertial
confinement fusion.

By expanding the Guderley problem to non-uniform density profiles,
we found a wide variability of the flow properties. This feature poses
the imploding shock problem as an attractive test for hydrodynamic
codes aimed at simulating compressible flow. Correspondingly, the
imploding strong shock offers a verification analysis for one-dimensional
compressible codes, as well as a starting point for two- and three-dimensional
codes.

Finally, we call attention to the complete Guderley problem covering
both the converging shock and the reflected shock which follows convergence
(for times $t>0$). The reflected shock and the entire flow profile
can also be solved semi-analytically in self-similar fashion. This
is well known to be the case for a uniform medium, and is also applicable
for non-zero values of $\mu$ covering both negative and positive
values. As discussed in detail by \cite{modelevsky2021revisiting},
a reflected shock actually exists only for a finite range $\mu_{-}<\mu<\mu_{+}$.
For $\mu<\mu_{-}$ the flow behind the converging shock stagnates
as the shock advances through the steep gradient towards the origin,
while for $\mu>\mu_{+}$ the diminishing density at the center causes
the pressure at the origin to vanish at convergence rather than to
become infinite (both $\mu_{-}$ and $\mu_{+}$, which are negative
and positive, respectively, depend on $\gamma$ and the geometry of
the flow). In cases where it does exist, the reflected shock marks
another advantage of the Guderley problem as a test case for hydrodynamic
codes, since it involves both converging and diverging flow \cite{ramsey2012guderley}.
In a follow-up work we will extend the physical and numerical analysis
of the reflected shock to relevant non-zero values of $\mu$, both
negative and positive, demonstrating the quality of the Guderley problem
as comprehensive test of hydrodynamic codes for compressible flow
through a non-uniform medium.

\subsection*{Availability of data}

The data that support the findings of this study are available from
the corresponding author upon request.

\bibliographystyle{plain}
\bibliography{datab}

\appendix

\section{Iterative algorithm for the calculation of the Similarity Exponent\label{app:Numerical-Calculation-of}}

In order to calculate the similarity exponent, $\lambda$, we construct
a scalar function, $f\left(\lambda\right)$, for which the correct
value of $\lambda$ is a root. As explained above in section \ref{subsec:The-Critical-Points},
only for the correct value of $\lambda$ does a numerical integration
of the ODE \eqref{eq:dCdV} intersects with the line $C=V+1$ at the
point $\left(V_{int},C_{int}\right)$ (as given by equations \eqref{eq:V_int},\eqref{eq:C_int})
and with with a slope $L_{int}$ (as given by eq. \eqref{eq:L_int}).
The function $f\left(\lambda\right)$ is defined below and solved
by iterations as follows: 
\begin{enumerate}
\item For a given guess for $\lambda$, calculate the values of the desired
intersection point $\left(V_{int},C_{int}\right)$ using equations
\eqref{eq:V_int}-\eqref{eq:C_int} and the slope of the $C(V)$ curve
at this point, using equations \eqref{eq:L_int}-\eqref{3dC}. 
\item Calculate the $C\left(V\right)$ curve, by numerically integrating
eq. \eqref{eq:dCdV} from the shock $\left(V_{s},C_{s}\right)$ (see
equations \eqref{eq:V_shock}-\eqref{eq:C_shock}) to the singular
point $V=V_{int}$. The resulting value of $C$ at that point is $C\left(V_{int}\right)\equiv C_{*}$.
In addition, the integration is explicitly evaluated at a point $V_{int}-\varepsilon$
close to the singular point, where $\varepsilon=\left|\frac{V_{s}-V_{int}}{V_{scale}}\right|$,
and the value of $C$ is denoted by $C\left(V_{int}-\epsilon\right)\equiv C_{*}^{-}$.
This gives the numerical value for the intersection point $\left(V_{int},C^{*}\right)$,
and for a point close to the intersection point, $\left(V_{int}-\varepsilon,C_{*}^{-}\right)$.
The ODE integration is performed via the LSODA integrator \cite{hindmarsh1983odepack}.
\item The numerical value for the slope near the intersection point, is
evaluated as: 
\begin{equation}
L_{*}=\frac{dC}{dV}\left(V_{int}-\varepsilon,C_{*}^{-}\right),
\end{equation}
where $dC/dV$ is calculated using eq. \eqref{eq:dCdV}. 
\item The value returned is given by: 
\begin{equation}
f\left(\lambda\right)=\begin{cases}
\left(C_{int}-C_{*}\right)\cdot10^{k+\alpha} & \gamma<\gamma_{crit}\\
\left(L_{int}-L_{*}\right)\cdot10^{k+\beta} & \gamma\ge\gamma_{crit}
\end{cases}
\end{equation}
where the calculation of $\gamma_{crit}$ is explained separately
below in Appendix \ref{app:gamma_crit}, and:
\begin{equation}
\alpha=\log\left|L_{*}-L_{int}\right|,
\end{equation}
\begin{equation}
\beta=\log\left|C_{*}-C_{int}\right|.
\end{equation}
\end{enumerate}
The quantities $V_{scale}$ and $10^{k}$ control the quality of convergence
and obviously must be chosen to be large. We find that $V_{scale}=5\cdot10^{4}$
and $k=5$ suffice to achieve accurate and fast convergence. The root
of $f\left(\lambda\right)$ is found using the bisection method, on
a range $\left[\lambda_{\text{min}},\lambda_{\text{max}}\right]$
such that $f\left(\lambda_{\text{min}}\right)\cdot f\left(\lambda_{\text{max}}\right)<0.$
In order to find such a range, we use a well known approximation for
the similarity exponent, denoted by $\lambda_{\text{approx}}$ (as
explained in Appendix \ref{app:Approximated-Calculation-of}). If
$f\left(\lambda_{\text{approx}}\right)>0$, we fix $\lambda_{\text{min}}=\lambda_{\text{approx}}$,
otherwise we take $\lambda_{\text{max}}=\lambda_{\text{approx}}$.

\section{Numerical Calculation of $\gamma_{crit}$ \label{app:gamma_crit}}

In this Appendix we describe the algorithm calculating $\gamma_{crit}$,
used in the calculation of the similarity exponent $\lambda$. For
a given $\gamma$, we integrate the ODE \eqref{eq:dCdV} from the
shock point $\left(V_{s},C_{s}\right)$ until it intersects with the
line $C=V+1$. In this integration, an approximated similarity exponent
$\lambda_{approx}$ is used, as detailed in Appendix \ref{app:Approximated-Calculation-of}.
Once the intersection is found, the distances in the $C-V$ plane
between the intersection point and the possible roots $\left(V_{\mp},C_{\mp}\right)$
are calculated. If the intersection point is closer to $\left(V_{-},C_{-}\right)$
we know that $\gamma<\gamma_{crit}$, otherwise $\gamma\geq\gamma_{crit}$.
This process results in a step function of $\gamma$ centered around
$\gamma_{crit}$, readily solvable via the bisection method.

Numerical results of $\gamma_{crit}$ for various values of $\mu$
are given for reference in table \ref{tab:gamma critical cylin}.
Specifically, we note that our results for $\mu=0$, $\gamma_{cirt}=1.9092$
and $\gamma_{crit}=1.8698$ for cylindrical and spherical symmetry
(respectively) are in excellent agreement with those found by Lazarus
\cite{lazarus1977similarity,lazarus1981self}. 
\begin{center}
\begin{table}[h]
\begin{centering}
\begin{tabular}{|c|c|c|}
\hline 
$n$  & $\mu$  & $\gamma_{crit}$\tabularnewline
\hline 
$2$  & $-1$  & $1$\tabularnewline
\hline 
$2$  & $-0.66$  & $1.08725$\tabularnewline
\hline 
$2$  & $-0.33$  & $1.39453$\tabularnewline
\hline 
$2$  & $0.0$  & $1.90920$\tabularnewline
\hline 
$2$  & $0.33$  & $2.66220$\tabularnewline
\hline 
$2$  & $0.66$  & $3.79614$\tabularnewline
\hline 
$2$  & $1.0$  & $5.74731$\tabularnewline
\hline 
\end{tabular}%
\begin{tabular}{|c|c|c|}
\hline 
$n$  & $\mu$  & $\gamma_{crit}$\tabularnewline
\hline 
$2$  & $1.33$  & $9.52816$\tabularnewline
\hline 
$2$  & $1.66$  & $20.6303$\tabularnewline
\hline 
$3$  & $-2$  & $1.00004$\tabularnewline
\hline 
$3$  & $-1.5$  & $1.04219$\tabularnewline
\hline 
$3$  & $-1$  & $1.19790$\tabularnewline
\hline 
$3$  & $-0.5$  & $1.47479$\tabularnewline
\hline 
$3$  & $0.0$  & $1.86976$\tabularnewline
\hline 
\end{tabular}%
\begin{tabular}{|c|c|c|}
\hline 
$n$  & $\mu$  & $\gamma_{crit}$\tabularnewline
\hline 
$3$  & $0.5$  & $2.40285$\tabularnewline
\hline 
$3$  & $1.0$  & $3.12706$\tabularnewline
\hline 
$3$  & $1.5$  & $4.14803$\tabularnewline
\hline 
$3$  & $2.0$  & $5.68258$\tabularnewline
\hline 
$3$  & $2.5$  & $8.24007$\tabularnewline
\hline 
$3$  & $3.0$  & $13.3511$\tabularnewline
\hline 
$3$  & $3.5$  & $28.6723$\tabularnewline
\hline 
\end{tabular}
\par\end{centering}
\caption{The values of $\gamma_{crit}$ for various value of $n$ and $\mu$
\label{tab:gamma critical cylin}}
\end{table}
\par\end{center}

\section{Calculation of an approximated similarity exponent $\lambda$ \label{app:Approximated-Calculation-of}}

For completeness, we review here the simple approximation for the
similarity exponent $\lambda$, developed by Chisnell for $\mu=0$
\cite{chisnell1998analytic}, and extended by Vishwakarama for $\mu\neq0$
\cite{vishwakarma2005analytic}.

The variables $\alpha=\frac{1}{\lambda}$ and $V_{0}$ are coupled
through a pair of equations:

\begin{equation}
\left(\frac{\alpha}{V_{0}}-1\right)^{2}=\frac{\gamma\left(\gamma-1\right)}{2}\left(\frac{\gamma-1}{\left(\gamma+1\right)\left(1-\frac{V_{0}}{\alpha}\right)}\right)^{\eta}\left(\frac{V_{0}+q}{\frac{2\alpha}{\gamma+1}+q}\right)^{B}\label{approx1}
\end{equation}

\begin{equation}
\left(n-1+\frac{\mu\alpha}{\gamma V_{0}}\right)\frac{\alpha}{1-\alpha}=\frac{1}{1-\frac{V_{0}}{\alpha}}+\frac{2\alpha}{\gamma V_{0}},\label{eq:approx2}
\end{equation}

where: 
\begin{equation}
\eta=\frac{2\left(1-\alpha\right)+\left(\gamma-1\right)\mu\alpha}{\alpha(n\gamma+\mu)-2\left(1-\alpha\right)},
\end{equation}

\begin{equation}
B=\eta+\left(n-1\right)\frac{2\left(\frac{\alpha}{V_{0}}-1\right)^{2}+\gamma-1}{1-n\left(\frac{\alpha}{V_{0}}-1\right)^{2}},
\end{equation}
\begin{equation}
q=-\frac{\alpha}{V_{0}\left(1-n\left(\frac{\alpha}{V_{0}}-1\right)^{2}\right)},
\end{equation}

The equations can be simplified by defining an auxiliary variable:
\begin{equation}
\xi=\frac{\alpha}{V_{0}}.
\end{equation}
Substituting $\xi$ and solving for $\alpha$ in eq. \eqref{eq:approx2},
we find: 
\begin{equation}
\alpha\left(\xi\right)=\frac{\frac{\xi}{\xi-1}+\frac{2}{\gamma}\xi}{n-1+\frac{\mu}{\gamma}\xi+\frac{\xi}{\xi-1}+\frac{2}{\gamma}\xi},\label{eq:alpha_xi}
\end{equation}
Inserting this result for $\alpha\left(\xi\right)$ into eq. \eqref{approx1},
is equivalent to finding the root of the function:

\begin{align}
f\left(\xi\right) & =\frac{\gamma\left(\gamma-1\right)}{2}\left(\frac{\gamma-1}{\left(\gamma+1\right)\left(1-\frac{1}{\xi}\right)}\right)^{\eta(\xi)}\left(\frac{\frac{\alpha\left(\xi\right)}{\xi}+q\left(\xi\right)}{\frac{2\alpha\left(\xi\right)}{\gamma+1}+q\left(\xi\right)}\right)^{B(\xi)}\nonumber \\
 & \ \ \ \ -\left(\xi-1\right)^{2},\label{eq:approx3}
\end{align}
where: 
\begin{equation}
q\left(\xi\right)=-\frac{\xi}{1-n\left(\xi-1\right)^{2}},
\end{equation}
\begin{equation}
\eta\left(\xi\right)=\frac{2\left(1-\alpha\left(\xi\right)\right)+\left(\gamma-1\right)\mu\alpha\left(\xi\right)}{\alpha\left(\xi\right)\left(n\gamma+\mu\right)-2\left(1-\alpha\left(\xi\right)\right)},
\end{equation}
\begin{equation}
B\left(\xi\right)=\eta\left(\xi\right)+\left(n-1\right)^{2}\frac{2\left(\xi-1\right)^{2}+\gamma-1}{1-n\left(\xi-1\right)^{2}}.
\end{equation}
The root of eq. \eqref{eq:approx3} is found numerically via the Newton-Raphson
method with the following initial value (see in \cite{chisnell1998analytic,vishwakarma2005analytic}):

\begin{equation}
\xi_{0}=\sqrt{\frac{\gamma\left(\gamma-1\right)}{2}}+1.
\end{equation}
Typically the root of eq. \eqref{eq:approx3} is found after three
or four Newton-Raphson iterations. After finding the root $\xi_{*}$
we substitute this root into eq. \eqref{eq:alpha_xi}, and obtain
the approximate value of the similarity exponent $\lambda_{approx}=\alpha^{-1}\left(\xi_{*}\right)$.

\section{The Hydrodynamic Scheme \label{app:The-hydrodynamic-Scheme}}

We used a standard one dimensional staggered grid Lagrangian hydrodynamic
scheme. The computational grid is divided into cells indexed $1...N$
and vertices indexed $\frac{1}{2},...N+\frac{1}{2}$. Thermodynamic
variables, $\rho_{i},p_{i},e_{i},c_{i}$, are defined on cells, while
kinematic variables $r_{i\pm\frac{1}{2}},u_{i\pm\frac{1}{2}},a_{i\pm\frac{1}{2}}$
(position, velocity and acceleration) are defined on vertices. The
mass of each cell, $m_{i}$, is time independent, given by the initial
combination of volumes, $V_{i}^{0}$, and densities $\rho_{i}^{0}$.

Denoting the number of the time step by an upper index $k$, the integration
timestep, $\Delta t^{k}=t^{k+1}-t^{k}$, is chosen according to the
CFL stability condition \cite{leveque2002finite}:

\begin{equation}
\Delta t^{k}=0.16\min_{i}\left(\frac{r_{i+\frac{1}{2}}^{k}-r_{i-\frac{1}{2}}^{k}}{c_{s,i}^{k}}\right),
\end{equation}
where the cell-related speed of sound is given by the ideal gas relation:
\[
c_{s,i}^{k}=\left(\frac{\gamma p_{i}^{k}}{\rho_{i}^{k}}\right)^{1/2}.
\]

We used a standard leap-frog temporal integration scheme, where the
next values of velocity, position, volume and density of each vertex
and cell are found through:

\begin{equation}
u_{i+\frac{1}{2}}^{k+\frac{1}{2}}=u_{i+\frac{1}{2}}^{k}+\frac{\Delta t^{k}}{2}a_{i+\frac{1}{2}}^{k},
\end{equation}

\begin{equation}
r_{i+\frac{1}{2}}^{k+1}=r_{i+\frac{1}{2}}^{k}+\Delta t^{k}u_{i+\frac{1}{2}}^{k+\frac{1}{2}},
\end{equation}

\begin{equation}
V_{i}^{k+1}=A\left(\left(r_{i+\frac{1}{2}}^{k+1}\right)^{n}-\left(r_{i-\frac{1}{2}}^{k+1}\right)^{n}\right),
\end{equation}

\begin{equation}
\rho_{i}^{k+1}=\frac{m_{i}}{V_{i}^{k+1}}.
\end{equation}
The geometry of the flow enters through the volume coefficient $A=1,\pi,\frac{4}{3}\pi$
and the dimensionality $n=1,2,3$ for planar, cylindrical and spherical
symmetries, respectively.

Numerical stability at the shock is achieved with the commonly used
Von-Neumann-Richtmyer \cite{vonneumann1950method} artificial viscosity:

\begin{equation}
q_{i}^{k+1}=\begin{cases}
0 & V_{i}^{k+1}\geq V_{i}^{k}\\
\sigma\rho_{i}^{k+1}\left(u_{i+\frac{1}{2}}^{k+\frac{1}{2}}-u_{i-\frac{1}{2}}^{k+\frac{1}{2}}\right)^{2}, & \text{else}
\end{cases}
\end{equation}
taking $\sigma=3$. Finally, the specific internal energy, pressure
and acceleration are solved according to:

\begin{equation}
e_{i}^{k+1}=\frac{e_{i}^{k}-\frac{1}{2}\left(p_{i}^{k}+q_{i}^{k+1}+q_{i}^{k}\right)\frac{V_{i}^{k+1}-V_{i}^{k}}{m_{i}}}{1+\frac{1}{2}\left(\gamma-1\right)\frac{V_{i}^{k+1}-V_{i}^{k}}{m_{i}}},
\end{equation}

\begin{equation}
p_{i}^{k+1}=\left(\gamma-1\right)e_{i}^{k+1}\rho_{i}^{k+1},
\end{equation}

\begin{equation}
a_{i+\frac{1}{2}}^{k+1}=-B\left(r_{i+\frac{1}{2}}^{k+1}\right)^{n-1}\frac{p_{i+1}^{k+1}+q_{i+1}^{k+1}-p_{i}^{k+1}-q_{i}^{k+1}}{\frac{1}{2}\left(m_{i+1}+m_{i}\right)},
\end{equation}
where the areal coefficient is $B=1,2\pi,4\pi$ for planar, cylindrical,
and spherical symmetries, respectively. 
\end{document}